\documentclass[11pt]{article}
\pdfoutput = 1

\usepackage[utf8]{inputenc}
\usepackage{color,graphicx}
\usepackage{epsfig}
\usepackage{amsmath}
\usepackage{verbatim}
\usepackage{caption}
\usepackage{amssymb}
\usepackage{mathabx}
\usepackage{dsfont}
\usepackage[mathcal]{eucal}
\usepackage{stmaryrd}
\usepackage{esint}
\usepackage{braket}
\usepackage{amsfonts}
\usepackage{cite}
\usepackage{array}
\usepackage{setspace}
\usepackage{float}
\usepackage{url}
\usepackage{mathtools}
\usepackage{bbold}
\usepackage{slashed}
\usepackage{tikz}
\usepackage{graphicx}
\usepackage{subcaption}
\usepackage{epstopdf}
\usepackage[labelfont=bf,font={sf}]{caption}
\usepackage{pgfplots}
\usepackage{mathrsfs}
\usepackage{fancybox}
\usepackage{eurosym}
\usepackage{tcolorbox}
\usepackage{tensor}
\allowdisplaybreaks
\usepackage{mdframed,lipsum}

\usepackage[margin = 2.2cm]{geometry}
\usepackage[ragged]{footmisc}
\usepackage[bookmarks=true,bookmarksnumbered=false,
hyperindex=true,bookmarksopen=true,hyperfigures=true,
colorlinks=true,linkcolor=webblue,citecolor=webgreen,urlcolor=webblue,breaklinks]{hyperref}
\definecolor{webgreen}{rgb}{0, 0.5, 0}
\definecolor{webblue}{rgb}{0, 0, 0.5}
\definecolor{webred}{rgb}{0.5, 0, 0}
\definecolor{darkgreen}{rgb}{0,0.5,0}

\renewcommand{\i}{\mathrm{i}}

\newcommand{\Tr}{\text{Tr}}

\def\ben{\begin{equation}}
	\def\een{\end{equation}}

     \let\r=v

\def\be{\begin{equation}}
	\def\ee{\end{equation}}
\def\ba{\begin{array}}
	\def\ea{\end{array}}

\def\dalemb#1#2{{\vbox{\hrule height .#2pt
			\hbox{\vrule width.#2pt height#1pt \kern#1pt
				\vrule width.#2pt}
			\hrule height.#2pt}}}

\newcommand{\bea}{\begin{eqnarray}}
	\newcommand{\eea}{\end{eqnarray}}

\def\mass{{{m_0}}}

\def\R{{{\mathbb{R}}}}

\let\tilde=\widetilde

\def\Imag{{\rm Im}\,}

\renewcommand{\i}{\mathrm{i}}

\numberwithin{equation}{section}

\title{}

\begin{document}
	
	\thispagestyle{empty}
	\begin{center}
		~\vspace{5mm}
		
		{\LARGE \bf 
		A clock is just a way to tell the time: gravitational algebras in cosmological spacetimes
		}
		
		\vspace{0.4in}
		
		{\bf  Chang-Han Chen and Geoff Penington}
		
		\vspace{0.4in}
		{Department of Physics, University of California, Berkeley, CA 94720, USA \\}
		\vspace{0.1in}
		
		{\tt changhanc@berkeley.edu, geoffp@berkeley.edu}
	\end{center}
\begin{abstract}
    We study algebras of observables for semiclassical quantum gravity in cosmological backgrounds, focusing on two key examples: slow-roll inflation and evaporating Schwarzschild-de Sitter black holes. In both cases, we demonstrate the existence of a nontrivial algebra of diffeomorphism-invariant observables \emph{without} the introduction of an explicit clock system or the presence of any asymptotic gravitational charges. Instead, the rolling inflaton field and the evaporating black hole act as physical clocks that allow a definition of gauge-invariant observables at $G = 0$. The resulting algebras are both Type II$_\infty$ factors, but neither is manifestly a crossed product algebra. For appropriate states, we establish a connection between the Type II entropy of these algebras and generalized entropies. Our work extends previous results on Type II gravitational algebras and highlights the crucial role of out-of-equilibrium dynamics when defining gauge-invariant observables in semiclassical canonically quantised gravity. We also briefly discuss the construction of gauge-invariant algebras for compact wedges bounded by extremal surfaces in generic spacetimes (i.e. in the absence of any Killing symmetry). In contrast to the inflaton and black hole cases, this algebra does end up being a simple crossed product. No clock or asymptotic charges are required because of the absence of any symmetry in the classical background.
\end{abstract}

 \tableofcontents
 \newpage
 \section{Introduction}
 Recent work has shown that the horizon entropy $A_{\rm hor}/4G$ associated both to black holes and to the cosmological horizon of an observer in de Sitter space can be accounted for by considering algebras of observables in the $G \to 0$ limit of canonical quantum gravity \cite{Leutheusser:2021frk,Leutheusser:2021qhd,Leutheusser:2022bgi,Witten_2022, CLPW,CPW, kudlerflam2024generalized, Jensen_2023,faulkner2024gravitational}. For a black hole, the relevant algebra consists of semiclassical observables at asymptotic spatial or null infinity; for de Sitter space, it is generated by observables localised along the observer's worldline. 

 In both cases, the algebras in question are Type II von Neumann factors.\footnote{See \cite{entropy1,entropy2,entropy3,entropy4} for other recent work on the role played by Type II factors in quantum gravity and quantum field theory.} Such algebras describe degrees of freedom that are infinitely entangled with their environment, reflecting the fact that horizon entropies diverge as $G \to 0$. However there is a natural notion of the renormalised entropy of a state on a Type II factor that is unique up to a state-independent additive constant. Roughly, this is because \emph{fluctuations} in the entanglement spectrum on a Type II algebra are finite, even though the total entanglement diverges.  For an appropriate class of states the entropy of the Type II algebra exactly matches the generalised entropy \cite{Bek, QES1, QES5}
 \begin{align}
 S_{\rm gen} =  A_{\rm hor}/4G + S_{\rm QFT}
 \end{align}
 of the observer's causal diamond.\footnote{Here $S_{\rm QFT}$ is the entropy of quantum fields in the causal diamond. This entropy is UV-divergent but the divergence can be absorbed into the renormalisation of Newton's constant $G$ so that the generalised entropy $S_{\rm gen}$ is UV-finite \cite{Susskind:1994sm}.}   The primary difference between the algebras for black holes and for de Sitter is that black hole algebras are Type II$_{\infty}$ whereas the de Sitter algebra is Type II$_1$. This reflects the fact that the entropy of a black hole is unbounded and can be increased arbitrarily by adding mass. On the other hand, the generalised entropy of the static patch is maximised by empty de Sitter space; any perturbation thereof can only decrease entropy.

 There is, nonetheless, an important subtlety here that is only present in the de Sitter context. Because Cauchy slices in de Sitter space are compact, the isometries of de Sitter space act as gauge constraints on the gravitational theory. In particular, gauge-invariant observables need to commute with the boost Hamiltonian $H$ that generates time translations of the observer's static patch. This is, in fact, the only reason that the algebra of a de Sitter observer in quantum gravity is different from that of the same observer in quantum field theory, even in the limit $G \to 0$.

 Unfortunately or otherwise, however, it is a well established fact about quantum field theory in de Sitter space that any operator that is both a) localised to the static patch and b) time translation invariant is necessarily a $c$-number, i.e. a multiple of the identity. As a result, the observer's algebra is apparently trivial. One way to see why this is the case is the following. To construct a gauge-invariant observable that is independent of coordinate time, the obvious approach is to integrate a time-dependent observable $a(t)$ over the observer's worldline to obtain an apparently boost-invariant observable
 \begin{align}\label{eq:introtildeadef}
     \tilde a \stackrel{?}{=} \int dt \,a(t)
 \end{align}
 However, any state of quantum fields in de Sitter space thermalises on the static patch at sufficiently early and late times. Since the thermal state is separating (i.e. it is not annihilated by any nonzero operator $a$) the integral in \eqref{eq:introtildeadef} diverges when acting on any state $\ket{\Phi}$. As a result, the operator $\tilde a$ does not exist.

 This explanation also makes clear why the physically relevant algebra of observables should \emph{not} be trivial. It is true that physical observations cannot depend on a choice of coordinate time. However, neither do they require averaging an observation over all times from past to future infinity. Instead, in practice, physical observations occur at a particular ``clock time'', defined with respect to the state of a physical system or clock. In this context, a clock can mean anything from a literal device used to tell the time to the state of the observer's brain or the state of some external dynamical system. Crucially, since the universe is quantum mechanical, the clock, whatever it might be, should itself be a quantum system.
 
 In \cite{CLPW}, a minimal model of a quantum clock was used to define the algebra of an observer in de Sitter space. To enable nontrivial observables, the clock needed to be able to measure arbitrarily long times $\Delta t \gg \ell_{dS}$ (in units of the de Sitter radius $\ell_{dS}$) using finite energy. This required it to have a continuous spectrum in order to prevent Poincar\'{e} recurrences.\footnote{More precisely, it requires the typical spectral gap $\delta E$ to satisfy $\delta E \ll 1/\ell_{dS}$ as $G\to 0$. This is easy enough to engineer, even with the size of the clock remaining much smaller than the de Sitter scale, because the typical spectral gap of a quantum system is exponentially small in its entropy.} Additionally, it was assumed on physical grounds that the energy of the clock was expected to be bounded from below.

 The simplest model of a clock satisfying those properties is the Hilbert space $L^2(\R^+)$ of wavefunctions with position $x \geq 0$ and Hamiltonian $H_{\rm clock} = x$. Assuming no coupling between the clock and the background quantum fields, so that the boost gauge constraint is simply $H+x = 0$, the inclusion of this clock was shown in \cite{CLPW} to lead to a Type II$_1$ algebra of observables with the properties described above.

 It is important to emphasize that the lesson of \cite{CLPW} (or, at the very least, the lesson as interpreted by one of its authors) was not that in practice physical observers are equipped with clocks of exactly the form just described. Instead, it is that a clock of that form provides a simple model with features that all good physical clocks should have. The hope was that the lessons drawn from studying such a clock, and in particular the conclusion that the entropy of the observer's algebra includes a contribution from the cosmological horizon area, would generalise fairly universally to any reasonable physical system that can act as a clock.

 In this paper, we provide strong evidence that this is indeed the case, by considering two natural classes of ``clocks'' that are not associated to the observer themselves, but rather to a dynamical quantum evolution of the observer's environment that remains out-of-equilibrium for parametrically long times. The first such example is quasi-de Sitter space in the presence of a slow-rolling inflaton field. The second is an evaporating Schwarzschild-de Sitter black hole. 

 The technical details of how the algebra works in each case is somewhat different, but the overall conclusions end up being very similar. Firstly, the out-of-equilibrium nature of the solutions mean that there exist nontrivial gauge-invariant observables even if we don't equip the observer with an explicit clock; the dynamics of the quantum system already provides one for us. In the inflaton case, the scalar field wants to roll downhill as we approach asymptotic future (or past) infinity. The value of the inflaton field therefore acts as the physical clock distinguishing different times. This claim will of course come as no surprise to cosmologists, since treated the inflaton value as a clock is standard practice in inflationary cosmology. Similarly, because the temperature of the black hole is higher than the cosmological horizon, a Schwarzschild-de Sitter black hole will tend to evaporate as we approach past or future infinity. The energy of the black hole can therefore be used to distinguish different times.

 Secondly, in both cases the algebra becomes a Type II$_{\infty}$ von Neumann factor. This is because, unlike for perturbations around pure de Sitter space, the entropy of these solutions can be arbitrarily increased simply by simply shifting the state further along the road to equilibrium. For example, given any state of the inflaton field we can produce a state with smaller cosmological constant, and hence larger generalised entropy, by shifting the inflaton wavefunction down its potential gradient. Similarly, we can arbitrarily increase generalised entropy in the presence of a Schwarzschild-de Sitter black hole by decreasing the energy of the black hole and hence increasing the area of the cosmological horizon. Of course, at finite $G$, neither process can be continued indefinitely: eventually an inflaton will reach the bottom of its potential and a black hole will reach zero mass. But, as $G \to 0$, the increase in generalised entropy ends up diverging parametrically sooner than either of these events will occur.

 An interesting feature of our constructions is that -- to the best of our current understanding -- neither algebra has a description as a crossed product of a Type III von Neumann factor by a modular automorphism group. (The black hole algebra does involve such a crossed product as an intermediate step in its construction.) This is in contrast to the  Type II algebras that describe de Sitter space in the presence of a worldline clock or semiclassical black holes in asymptotically flat/AdS spacetimes and is instead more similar to the boundary algebras describing quantum JT gravity (beyond the semiclassical limit) \cite{penington2023algebras, Kolchmeyer:2023gwa}.

 However, like in the de Sitter algebra introduced in \cite{CLPW}, the traces on both algebras we will define do both have a natural interpretation as expectation values of operators in a no-boundary Hartle-Hawking state. This is consistent with the proposal of \cite{witten2023background} that the no-boundary Hartle-Hawking state should be viewed as a maximum entropy state in background-independent quantum gravity.

Like previous gravitational Type II algebras, the algebras we construct have traces that are proportional to the number of horizon microstates $\exp(A_{\rm hor}/4G)$. In particular, the inflaton algebra has a symmetry that shifts the inflaton field by a constant. We show that this symmetry rescales the trace by $\exp(\delta A_{\rm hor}/4G)$ where $\delta A_{\rm hor}$ is the perturbative change in cosmological horizon area induced by the shift in the inflaton potential. Similarly, perturbatively increasing the mass of  a Schwarzschild-de Sitter black hole rescales the trace on its gravitational algebra by $\exp((\delta A_{BH} + \delta A_{CH})/4G)$ where $\delta A_{BH} > 0$ and $\delta A_{CH} < 0$ are the respective changes in the black hole and cosmological horizon areas.

We also obtain a precise match between entropies on our algebras and generalised entropies of semiclassical states that have been formally averaged over the gauge group. In the inflaton case, this correspondence holds for any QFT state. For Schwarzschild-de Sitter black holes, there is a timeshift mode associated to travelling in a nontrivial spacelike cycle through the black hole and cosmological horizons that cannot be described by perturbative quantum field theory. The correspondence between the entropy of the Type II algebra and generalised entropy holds for states with sufficiently small fluctuations in this timeshift.

 The structure of this paper is as follows. In Section \ref{sec:inflaton}, we describe the semiclassical gravitational algebra associated to a slow-rolling inflaton in quasi de Sitter space. In Section \ref{sec:blackholes}, we describe the algebras for a Schwarzschild-de Sitter black hole, after first pausing briefly to discuss asymptotically flat black holes. Finally in Section \ref{sec:generic}, we briefly discuss the gravitational algebra of a wedge bounded  by an extremal surface in a generic cosmological spacetime (i.e. in the absence of any background symmetry). In Appendices, we provide some technical details on Tomita-Takesaki theory and higher-dimensional spherical harmonics. We also include a detailed glossary of notation used throughout the paper.

\textbf{\underline{Note added}:} The submission of this paper to arXiv has been coordinated with the submission of \cite{KLS}, which contains related discussion of von Neumann algebras in the presence of a slow-rolling inflaton field. A significant difference between our work and \cite{KLS} is that \cite{KLS} makes use of an explicit gravitating clock, while we do not.

 \section{A slow rolling inflaton in quasi-de Sitter space} \label{sec:inflaton}
 We consider a single scalar inflaton field $\phi$ with potential $V(\phi)$. This is a well-studied inflation model that has produced successful predictions for primordial curvature perturbations \cite{inflation0,inflation,inflation1,inflation2,inflation3}. See \cite{Maldacena_2003,maldacena2024comments} for excellent reviews. For simplicity, we assume four spacetime dimensions, although the generalisation to other dimensions is straightforward. In units where $\hbar = 1$, the slow-roll approximation assumes that
 \begin{align}
     \varepsilon = \frac{G}{2} \left(\frac{V'}{V}\right)^2 \ll 1 ~~~~\text{and}~~~~|\eta| = G \frac{|V''|}{V} \ll 1.
 \end{align}
 In this limit, symmetric vacuum solutions to the Einstein equations can be well approximated by de Sitter space with cosmological constant $\Lambda = 8 \pi G V$ over scales that are large compared to the de Sitter length $\ell_{dS} = \sqrt{3/\Lambda}$.

 In cosmology, one usually takes $\varepsilon$ and $|\eta|$ to be small but finite so that the exit from inflation occurs at finite time. Mathematically, however, giving a precise algebraic description of gravitational observables at finite $\varepsilon,|\eta|$, or indeed even in a formal perturbative expansion in $\varepsilon,|\eta|$, would be very challenging. It is much easier to take a strict $G \to 0$ limit where $V \sim O(1/G)$, $V' \sim O(1)$ and $V'' \sim o(1)$. In this limit, the cosmological constant $\Lambda$ remains finite while dynamical backreaction from the matter fields goes to zero. Meanwhile -- if we include the cosmological constant in the Einstein-Hilbert action -- the inflaton action becomes
 \begin{align}\label{eq:inflatonaction}
     S= -\int d^4 x\sqrt{-g}\; \bigg[\frac{1}{2}\nabla_a\phi\nabla^a\phi +V' \phi \bigg].
 \end{align}
 which is just the action for massless scalar field with a potential $V(\phi) = V' \phi$ that has constant linear slope $V'$. Without loss of generality we can assume $V' > 0$. There also exist graviton fluctuations that are described by a free massless spin-2 quantum field theory, along with quantum perturbations of any other quantum fields that we choose to include in our theory. We assume that, at least in the limit $G \to 0$, all such fields are either free or described by a UV-complete interacting theory. In the $G \to 0$ limit, the full gravitational theory then reduces to a continuum quantum field theory in a fixed background de Sitter spacetime.

\begin{figure}
  \hspace{4.7 cm}
\begin{tikzpicture}[scale =1]
     \pgftext{\includegraphics[scale =.5]{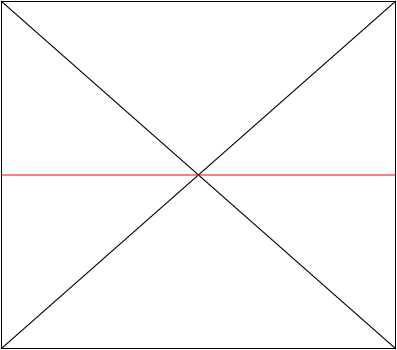} 
     \draw (-1.4,4.2) node {\LARGE $\mathcal{A}$};
     \draw (-6.6,4.2) node {\LARGE $\mathcal{A}'$};
     \draw (-1.8,2.7) node {\color{red} $\tau =0$};
     \draw (-.5,2.2) node { $\chi =0$};
     \draw (-8.9,2.2) node { $\chi =\pi$};
}
 \end{tikzpicture}

 \caption{\label{fig:ds} A Penrose diagram for global de Sitter space. The compact $\tau=0$ Cauchy slice is shown in red. We consider the QFT algebra $\mathcal{A}$ of operators localised in the static patch centered around $\chi = 0$. The commutant algebra $\mathcal{A}'$ describes operators in the opposite static patch.}
 \end{figure}
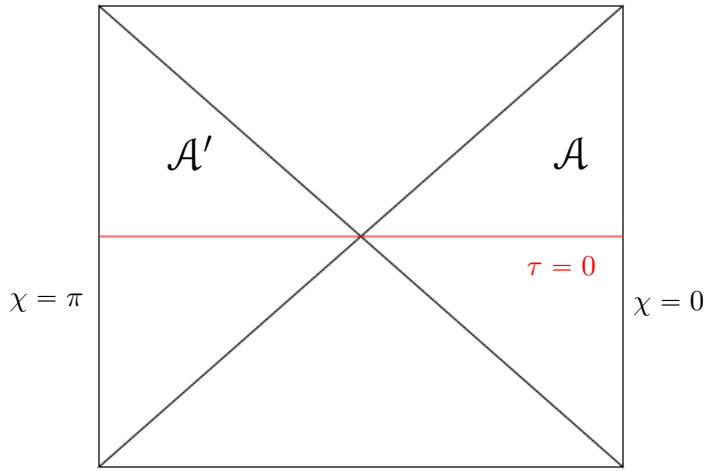

  In global coordinates, the metric for $dS_4$ is
 \begin{align}\label{ds}
     ds^2 =  -\ell_{dS}^2\big(d\tau^2+ \cosh^2(\tau) d\Omega_3^2\big),
 \end{align} where $\Omega_3 = (\chi,\theta,\phi)$ describes angular coordinates on $S^3$. The Penrose diagram is shown in Figure \ref{fig:ds}. 
 For notational convenience, we will mostly set $\ell_{dS} = 1$ for the remainder of this section. Since global de Sitter space has compact spatial Cauchy slices, there exists a unique natural Hilbert space $\mathcal{H}$ describing any set of free quantum fields propagating within it; this result is expected, but is not known, to extend to UV-complete interacting theories \cite{Wald:1995yp,witten2022does}. Essentially, this is because, on a compact spatial slice, there are a finite number of long-wavelength modes, which consequently have a unique representation by the Stone-von Neumann theorem, together with an asymptotically large number of high-energy short-wavelength modes that have an unambiguous decomposition into positive and negative frequencies because locally the spacetime looks like Minkowski space. Our goal in this section will be to study this Hilbert space for the slow rolling inflaton (and other fields) and show that it contains a Type II$_\infty$ von Neumann subalgebra describing gauge-invariant observables in any given static patch.

 Working with global de Sitter space is again slightly unusual in inflationary cosmology, where typically only the inflating patch, covered by a flat slicing of de Sitter, is considered. We make this choice for two reasons. First, at a technical level, quantising an inflaton field in a flat slicing of de Sitter space leads to a number of mathematical difficulties related to the fact that the inflationary patch does not contain compact Cauchy slices. In particular, in the flat slicing of de Sitter,  there exist, at any finite time, infinitely many long wavelength modes that have already exited the cosmological horizon. These contribute divergent fluctuations to the value of the inflaton at sub-horizon scales. Cosmologists typically get around this issue by treating such modes classically, which works well for predicting cosmological observables since only a finite number of these IR modes have so far reentered our particle horizon. However, to describe the mathematical structure of the inflaton algebra of observables, we find it simplest to avoid IR divergences in the first place by working in global de Sitter space, where Cauchy slices are compact and hence (angular) momentum is quantised.

 A second, somewhat related, point is that the Hartle-Hawking proposal \cite{Hartle:1983ai,HH2,hh3,hh4} for the no-boundary state in quantum gravity defines a natural choice of wavefunctional at finite $G$ for spatially compact spacetimes like global de Sitter. An excellent recent review of the Hartle-Hawking state and its (incorrect) predictions for inflationary cosmology, was given in \cite{maldacena2024comments}.
 The Hartle-Hawking state will play a very important role in our story: indeed, the trace that we construct to show that our algebra is Type II$_\infty$ is essentially the $G \to 0$ limit of the Hartle-Hawking state. Again, all of this is simplest to describe when working in global de Sitter space.

 \subsection{The Bunch-Davies state for a light scalar field}\label{sec:bdmassive}

For a massive scalar field theory (and more generally for any gapped QFT) in de Sitter space, there exists a unique normalisable state in $\mathcal{H}$, called the Bunch-Davies state, that is invariant under the isometry group $SO(4,1)$.

\begin{figure}
  \centering
\begin{tikzpicture}[scale =1]
     \pgftext{\includegraphics[scale =.5]{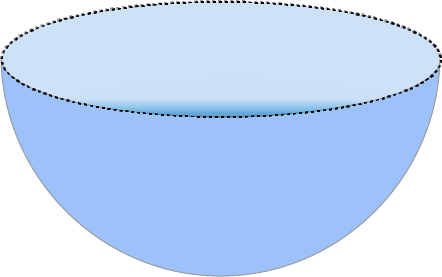}
        \draw (- 4.5,5.5) node {$\phi(\Omega_3)$};
}
 \end{tikzpicture}
 \caption{\label{fig:bd} The Bunch-Davies wavefunctional $\Psi_{BD}(\phi(\Omega_3))$ is defined by a path integral on a Euclidean hemisphere with Dirichlet boundary conditions $\phi(\Omega_3)$.}
 \end{figure}

 One way to construct the Bunch-Davies state is via a path integral on a Euclidean hemisphere as shown in Figure \ref{fig:bd}.
  The Bunch-Davies wavefunctional $\Psi(\phi(\Omega_3))$ is proportional to the value of the path integral with boundary conditions $\phi(\Omega_3)$.

 For free theories, there is a very useful trick where the path integral can be evaluated by considering the classical action of a single saddle point solution with the correct boundary conditions. Consider e.g. a single harmonic oscillator, with Euclidean action
 \begin{align}
     S =  \int dt \,\frac{\mass}{2} \dot{x}^2 - \frac{1}{2} \mass\omega^2 x^2.
 \end{align}
 The vacuum wavefunction $\psi(x_0)$ for the harmonic oscillator can be evaluated using a Euclidean path integral on the half-line $ it < 0$ with boundary conditions $x(it = -\infty) = 0$ and $x(it = 0) = x$. The classical solution with these boundary conditions is $x(it) =x e^{\omega it}$ which leads to a saddle-point contribution to the partition function of
 \begin{align} \label{eq:qho}
     \psi(x) \sim \exp(i S) \sim \exp(-\frac{1}{2} \mass\omega x^2).
 \end{align}
 Crucially, because the theory is free, this answer is exact. The one-loop determinant depends only on the second derivatives of the action around the classical saddle, which are constant because the action is quadratic in $x$. And, for the same reason, no higher-order corrections to \eqref{eq:qho} exist.
    
Let us recall how to implement the same procedure for a massive scalar field in the de Sitter space. Partially for technical simplicity, and partially because it means that our result will be useful later when we come to consider an inflaton field, we assume that the mass $\mass \ll \ell_{dS}$ of the scalar field is parametrically small and hence can be treated perturbatively. The action for this field is
 \begin{align}\label{eq:massivescalaraction}
     S=-\int d^4 x\sqrt{-g}\;\frac{1}{2}\,\nabla_a\phi\nabla^a\phi + \frac{1}{2}\mass^2 \phi^2.
 \end{align} In the limit where the mass $\mass$ is small, this leads to a classical equation of motion
 \begin{align}
     \frac{1}{\sqrt{-g}}\partial_a\big(\sqrt{-g}g^{ab}\partial_b\phi\big) = 0,\qquad
     \sqrt{-g} = \cosh^3\tau\cdot\sin^2\chi\cdot\sin\theta.
 \end{align} With respect to the metric (\ref{ds}), this becomes
 \begin{align}\label{eq:eommasslessscalar}
     \frac{1}{\cosh^3\tau}\partial_\tau\big(\cosh^3\tau\partial_\tau\phi\big)-\frac{\Delta_{S^3}\phi}{\cosh^2\tau}=0,
 \end{align} where $\Delta_{S^3}$ is the Laplacian for the unit $3-$sphere. This can be solved using separation of variables,
 \begin{align}
     \phi(\tau,\Omega_3) = A(\tau)B(\Omega_3).
 \end{align} 

The solutions for $B$ are the standard spherical harmonics on $S^3$, namely $Y^{k\ell m}(\Omega_3)$ with $k\ge\ell\ge 0$ and $\ell\ge m\ge -\ell;$ see Appx.\ref{sh} for details. The two independent solutions for $A$ with non-zero angular momentum, $k\neq0,$ are known \cite{maldacena2024comments,Bousso_2002}\footnote{Compared to \cite{maldacena2024comments}, we normalize the modes with respect to the Klein-Gordon inner product
\begin{align}
    \cosh^{3}\tau(\dot{A}_kA_k^\ast-\dot{A}_k^\ast A_k)\overset{!}{=}-\i\nonumber.
\end{align} This is not obvious but can checked straightforwardly for (\ref{sol}). As usual, the normalization is guaranteed to be a time-independent constant by the equation of motion},

    \begin{align}\label{sol}
    A_k(\tau)& = (-1)^{\frac{k+1}{2}}\frac{\Gamma(k+3)\,\Gamma(k+\frac{3}{2})}{\Gamma(\frac{3}{2})\,\Gamma(2k+3)}\frac{1}{\sqrt{2k(k+1)(k+2)}}\; e^{3\tau}\,(1+e^{2\tau})^k\;{}_2F_1\big(\frac{3}{2}+k,3+k,3+2k;1+e^{2\tau}\big),
\end{align} and its complex conjugate
\begin{align} \label{eq:sol2}
    A_k^\ast(\tau)& = A_k(-\tau).
\end{align}
Taking a formal limit $k \to 0$, \eqref{sol} and \eqref{eq:sol2} become constant, which is one of the two independent solutions to \eqref{eq:eommasslessscalar} with $k=0$. The other is
\begin{align}\label{zero}
    A_0(\tau) = \frac{1}{2}\left[\frac{\sinh\tau}{\cosh^2\tau}+\tan^{-1}(\sinh \tau)\right].
\end{align}
The most general real solution to \eqref{eq:eommasslessscalar} is therefore
\begin{align}
     \phi &= \phi_0 \frac{1}{\sqrt{2}\pi} +\pi_0 A_0(\tau) \frac{1}{\sqrt{2}\pi}+ \sum \phi_{k\ell m} Y_{k\ell m}\label{modesperfect}\\
    & = \phi_0\frac{1}{\sqrt{2}\pi} +\pi_0 A_0(\tau) \frac{1}{\sqrt{2}\pi}+ \sum \bigg[a_{k\ell m}A_k(\tau)+(-1)^m a^\ast_{k\ell- m}A_k^\ast(\tau)\bigg] Y_{k\ell m}\\
     &= \phi_0\frac{1}{\sqrt{2}\pi} +\pi_0\,A_0(\tau)\frac{1}{\sqrt{2}\pi}+\sum a_{k\ell m}A_k Y_{k\ell m}+a_{k\ell m}^\ast A_k^\ast Y^\ast_{k\ell m},\label{eq:gensoln}
 \end{align}
 for arbitrary real coefficients $\phi_0$, $\pi_0$ and complex coefficients $a_{k\ell m}$. The conjugate momentum field is
 \begin{align}
     \pi_\phi &= \cosh^3\tau\, \partial_\tau\phi.\\
     &=\pi_0\frac{1}{\sqrt{2} \pi }+\sum_{k\ell m} \pi_{k\ell m} Y_{k\ell m}; \label{eq:momentummodes}
 \end{align} the second line defines the modes for conjugate momentum field. Clearly, $\pi_0$ is the conjugate momentum of $\phi_0$. 

 To compute the wavefunctional $\Psi(\phi(\Omega_3))$ of the Bunch-Davies state, we need to look for (complex) solutions to the classical equations of motion with boundary conditions $\phi(0,\Omega_3) = \phi(\Omega_3)$ that are regular at the Euclidean south pole $\tau = i\pi/2$. It turns out that analytic continuations of the solutions \eqref{sol} and the constant solution are regular, while \eqref{eq:sol2} and \eqref{zero} are not. Since
 \begin{align}
     A_k(\tau=0)=-\sqrt{\frac{k+1}{2 \, k\big(k+2\big)}}.
 \end{align} 
 the regular solution with boundary conditions
 \begin{align}
\phi(0,\Omega_3) = \phi_0 \frac{1}{\sqrt{2}\pi} + \sum \phi_{k\ell m} Y_{k\ell m}(\Omega_3)
 \end{align}
 is therefore
 \begin{align}\label{phiE}
\phi(\tau,\Omega_3) = \phi_0 \frac{1}{\sqrt{2}\pi} - \sum \sqrt{\frac{2 \, k\big(k+2\big)}{k+1}}\phi_{k\ell m} A_k (\tau) Y_{k\ell m}(\Omega_3).
 \end{align} 
 Plugging \eqref{phiE} back into the action \eqref{eq:massivescalaraction}, we have
 \begin{align}\label{eq:dsaction}
     \i S &= \i\int d\tau\cosh^3\tau \Bigg\{-\frac{1}{2}m_0^2\,\phi_0^2+\frac{k(k+2)}{k+1}\bigg[ \sum (\dot{A_k})^2 \phi_{k\ell m}\phi_{k\ell -m}-\frac{1}{\cosh^2\tau}\sum k(k+2)\,A_k^2 \phi_{k\ell m}\phi_{k\ell -m}\bigg]\Bigg\}\\
     &=\i\sum \phi_{k\ell -m}\bigg[\int d\tau \frac{k(k+2)}{k+1}\bigg((\dot{A_k})^2\cosh^3\tau-k(k+2)A_k^2\cosh\tau\bigg)\bigg]\phi_{k\ell m}- \i\int d\tau\cosh^3\tau \frac{1}{2}m_0^2\,\phi_0^2,
 \end{align} and we are left with the time integral inside the large bracket to evaluate. To find the Bunch-Davies wavefunctional, we integrate $\tau$ along the imaginary axis from $\i \frac{\pi}{2}$ to $0$ to obtain

 \begin{align} \label{eq:massiveBDstate}
\Psi_{BD}[\phi(\Omega_3)]\sim\exp\Bigg\{- \frac{1}{3}\mass^2 \phi_0^2 -\sum_{\substack{ k\neq 0}}\frac{k\big(k+2\big)}{2\big(k+1)}\phi_{k\ell -m}\phi_{k\ell m}\Bigg\}.
 \end{align} 

\subsection{The Bunch-Davies weight for an inflaton field}\label{sec:bdinflaton}
In the limit $\mass \to 0$, the state \eqref{eq:massiveBDstate} stops being normalisable and if therefore no longer contained in the Hilbert space $\mathcal{H}$. However the expectation values of a dense set of observables on the unnormalized state \eqref{eq:massiveBDstate} converge to finite limits as $\mass \to 0$. For example, consider the smeared field operator
\begin{align}
    \phi_f = \int d^4 x f(x) \phi(x) 
\end{align}
for some suitable smearing function $f(x)$. This is a densely defined unbounded self-adjoint operator, and so we can define the bounded operator $F(\phi_f)$ for any bounded function $F$. If $F$ vanishes sufficiently fast as $\phi_f \to \pm \infty$, e.g.
\begin{align}\label{eq:exp-phi_f^2}
    F(\phi_f) = \exp(-\phi_f^2)
\end{align}
the expectation value $\braket{\Psi_{BD}|F(\phi_f)|\Psi_{BD}}$ will remain finite $\mass \to 0$. As a result, the $\mass \to 0$ limit of \eqref{eq:massiveBDstate} defines a faithful, normal semifinite weight on $\mathcal{H}$, which we will call the Bunch-Davies weight. Importantly, since in the argument above $\phi_f$ could be localised on an arbitrarily small region of spacetime, the Bunch-Davies weight is semifinite on the von Neumann subalgebra $\mathcal{A} \subseteq \mathcal{B}(\mathcal{H})$ associated to any causal diamond.

Our goal is to construct the analogous Bunch-Davies weight for the inflaton action \eqref{eq:inflatonaction}. We do so by an identical procedure to the one above. We first define the unnormalizable Bunch-Davies wavefunctional by a path integral over the Euclidean hemisphere with appropriate boundary conditions and then use the fact that the action \eqref{eq:inflatonaction} is free to evaluate this path integral by evaluating the action of the regular Euclidean solution to the classical equation of motion with appropriate boundary conditions. In this case, the classical equations of motion are
\begin{align}\label{eq:inflatoneom}
\frac{1}{\cosh^3\tau}\partial_\tau\big(\cosh^3\tau\partial_\tau\phi\big)-\frac{\Delta_{S^3}\phi}{\cosh^2\tau}= V',
 \end{align}
 which is a linear inhomogeneous equation with homogeneous part equal to \eqref{eq:eommasslessscalar}. A solution to \eqref{eq:inflatoneom} is
 \begin{align}\label{eq:particularsoln}
     \phi(\tau, \Omega_3) = C_0(\tau) = \frac{V'}{3} \left(\frac{1}{\cosh^2 \tau} - \log\left[2\cosh\tau\right]\right).
 \end{align}
 In the limit $\tau \to \pm \infty$, this becomes the standard solution $C_0(\tau) = - V' |\tau|/3$ for a slow-rolling inflaton field in a flat slicing with Hubble constant $H = \mathrm{sgn}(\tau)$. The inflaton field naturally rolls towards lower value of the potential, at a terminal velocity determined by ``friction'' from the expansion of space in the asymptotic past and future.
 
 The most general real solution to $\eqref{eq:eommasslessscalar}$ is a sum of \eqref{eq:particularsoln} and the general solution \eqref{eq:gensoln} to the homogeneous equation \eqref{eq:eommasslessscalar}. Note that there is no freedom in rescale the solution \eqref{eq:particularsoln}; its coefficient is fixed by the slope of the linear potential.

 As in Section \ref{sec:bdmassive}, to solve for the Bunch-Davies wavefunctional we need to look for solutions with given boundary conditions at $\tau = 0$ that are regular at the Euclidean south pole $\tau = i \pi/2$. The regular solution with boundary conditions $\phi(0,\Omega_3) = 0$ is
\begin{align}\label{eq:zerobcsregular}
     \phi(\tau,\Omega_3) = C_0(\tau) +  \frac{V'}{3} \left[2 iA_0(\tau) + (\log 2 - 1)\right] = \frac{V'}{3}\Bigg(\frac{1+\i \sinh\tau}{\cosh^2\tau}-\log[1-\i\sinh\tau]-1\Bigg).
 \end{align}
 The regular solution with general boundary conditions is given by a linear combination of \eqref{eq:zerobcsregular} and the solution \eqref{phiE} to the homogeneous equation \eqref{eq:eommasslessscalar} with those boundary conditions.

 Plugging this solution into the action \eqref{eq:inflatonaction}, we obtain
 \begin{align}
     \i S &= \i\int d\tau\cosh^3\tau \Bigg\{\frac{k(k+2)}{k+1}\bigg[ \sum (\dot{A_k})^2 \phi_{k\ell m}\phi_{k\ell -m}-\frac{1}{\cosh^2\tau}\sum k(k+2)\,A_k^2 \phi_{k\ell m}\phi_{k\ell -m} \bigg] +2\pi^2\frac{1}{\sqrt{2}\pi} V' \phi_0\Bigg\}.
 \end{align}
 The Bunch-Davies weight is therefore
 \begin{align}\label{eq:inflatonBDweight}
\Psi_{BD}[\phi(\Omega_3)]\sim\exp\Bigg\{- \frac{4\pi}{3\sqrt{2}} V' \phi_0 -\sum_{\substack{ k\neq 0}}\frac{k\big(k+2\big)}{2\big(k+1)}\phi_{k\ell -m}\phi_{k\ell m}\Bigg\}.
  \end{align}

So far, we have considered only the inflaton field and not any other fields, including e.g. graviton excitations that may be present in the spacetime. However this is easy to rectify. Because the different fields do not interact, the full Bunch-Davies weight for a theory containing an inflaton and additional quantum fields is simply a tensor product of the Bunch-Davies weights for the individual decoupled sectors. For example, in the presence of additional light scalar field the Bunch-Davies weight will be a tensor product of \eqref{eq:inflatonBDweight} with \eqref{eq:massiveBDstate}.

The formula \eqref{eq:inflatonBDweight} was derived purely using quantum field theory. However, because gravitational effects decouple in the limit we have taken, one obtains the same result by taking the $G \to 0$ limit of the no-boundary Hartle-Hawking state in quantum gravity. More precisely, given an observable that e.g. projects onto an $O(1)$ range of inflaton values with potential $V =\Lambda/8\pi G$, the $G\to 0$ limit of the expectation value of that observable for the Hartle-Hawking state agrees with the corresponding expectation value for the Bunch-Davies weight, up to an overall divergent factor. In particular, the exponential factor proportional to $V'$ in \eqref{eq:inflatonBDweight} means that the weight is dominated by field values with a small inflaton potential. This is the same phenomenon reviewed for the Hartle-Hawking state in \cite{maldacena2024comments}: the inflaton wants to already start in equilibrium at the bottom of the potential rather than rolling down for a long time.

 An important check of our results is that the Bunch-Davies weight is invariant under the $SO(4,1)$ isometry group of de Sitter space. Rotation invariance is easy to check, and follows from the fact that \eqref{eq:inflatonBDweight} only depends explicitly on $k$ and not on $\ell$ or $m$. To demonstrate full de Sitter invariance it therefore suffices to check invariance under a single boost generator, which for simplicity we take to be the generator that preserves the quantum numbers $\ell$ and $m$. 
 
To find the boost generator, it is easiest to use the embedding coordinate $\{X_0,X_1,\cdots,X_4\},$ in which the de Sitter space is a hyperboloid that satisfies $-X_0^2+\sum_{i=1}^4 X_i^2=1.$ The boost generator in embedding coordinate in the $i$-th direction is
 \begin{align}
     K_i = X_i\frac{\partial}{\partial X_0}+ X_0\frac{\partial}{\partial X_i}.
 \end{align} The global slicing is given by
 \begin{align}\label{eq:embedglobal}
     X_0 &= \sinh{\tau},\\
     X_i&= \cosh{\tau}z_i,
 \end{align} where $\sum_{i=1}^{4} z_i=1,$ parametrizing the three-sphere. With this parametrization, the boost generator becomes
 \begin{align}
     K_i = z_i \,\partial_\tau+\sum_{j=1}^{4}\tanh{\tau}\,\Big(\delta_{ij}-z_iz_j\Big)\partial_j.
 \end{align} In quantum theory, the boost Hamiltonian can be obtained from the boost generator by substituting in the stress energy tensor and integrating over a Cauchy slice so that
 \begin{align}
   H_i = \int d^3 x \sqrt{-h} z_i \,T^0{}_0+\sum_{j=1}^{4}\tanh{\tau}\,\Big(\delta_{ij}-z_iz_j\Big)T^0{}_j.
 \end{align} 
 with $d^3 x \sqrt{-h}$ the volume form on the Cauchy slice.
 
 Without loss of generality, we can set $z_1 =\cos\chi$ and choose to focus on the static patch centered at $\chi=0.$ Plugging in the expression for the energy density $T^0{}_0$ and evaluating $H$ on the $\tau = 0$ slice, the corresponding boost Hamiltonian \footnote{The boost Hamiltonian $H$ should not be confused with the global Hamiltonian $H_{\rm gl}(\tau)$, which generates translations in $\tau$. The latter is explicitly time dependent since global time translations are not a symmetry of de Sitter space.}  is \begin{align}
     H&=  \int \frac{1}{2}\,\cos\chi\bigg((\partial_\tau\phi)^2+(\nabla_{S^3}\phi)^2 +V' \phi \bigg)\\
     &=\int \frac{1}{2}\,\cos\chi\bigg(\pi_\phi^2+(\nabla_{S^3}\phi)^2 +V' \phi \bigg).
 \end{align} 
 In the second line we replaced $\partial_\tau \phi$  by the conjugate momentum $\pi_\phi$. 
Expanding the boost Hamiltonian in terms of angular momentum modes for the field $\phi$ as in \eqref{modesperfect} and the conjugate momentum field $\pi$ as in \eqref{eq:momentummodes}, this becomes
\begin{align}\label{eq:boost}
     H =&-\frac{1}{2}\Bigg[\pi_0\pi_{100}+\sum_{k>0}\sqrt{\frac{(k+\ell+2)(k-\ell+1)}{(k+1)(k+2)}}\bigg(\pi_{k\ell m}\pi_{k+1\ell -m }+k(k+3)\phi_{k\ell m}\phi_{k+1\ell -m}\bigg)\Bigg] -\frac{\pi}{\sqrt{2}}V_\ast'\phi_{100},
 \end{align} where we have used the orthogonality relations of the spherical harmonics \eqref{ortho}.

In canonical quantization, the conjugate momenta get promoted to the functional derivative of fields,
\begin{align}
    \pi_{k\ell m} \rightarrow -\i \frac{\delta}{\delta\phi_{k\ell m}}.
\end{align} The boost Hamiltonian then becomes
\begin{align} \label{eq:boostcanonicalaction}
    H=&\frac{1}{2}\Bigg[\frac{\delta^2}{\delta\phi_0\delta\phi_{100}}+\sum_{k>0}\sqrt{\frac{(k+\ell+2)(k-\ell+1)}{(k+1)(k+2)}}\bigg(\frac{\delta^2}{\delta\phi_{k\ell m}\delta\phi_{k+1\ell-m}}-k(k+3)\phi_{k\ell m}\phi_{k+1\ell -m}\bigg)\Bigg]-\frac{\pi}{\sqrt{2}}V_\ast'\phi_{100}.
\end{align} Applying \eqref{eq:boostcanonicalaction} to \eqref{eq:inflatonBDweight}, we find that 
\begin{align}
   H \Psi_{BD} = \Bigg(\bigg[\frac{1}{2}&\left(-\frac{4\pi V'}{\sqrt{2}\, 3}\right)\left(-\frac{1(1+2)}{1+1}\right)-\frac{\pi}{\sqrt{2}}V'\bigg]\phi_{100} \nonumber\\&+ \sum_{k>0} \bigg[\frac{k\big(k+2\big)}{k+1}\frac{\big(k+1\big)\big(k+3\big)}{k+2}-k(k+3)\bigg]\phi_{k\ell m}\phi_{k+1\ell{-m}} \Bigg)\Psi_{BD}= 0,
\end{align}  so the Bunch-Davies weight is indeed boost invariant.
 
 \subsection{The algebra of observables}\label{sec:inflatonalgebra}
So far all of our discussion has not really been about quantum gravity at all, but about quantum field theory in de Sitter space. However, despite the fact that we took a strict $G \to 0$ limit, there continue to exist residual effects of quantum gravity that go beyond the existence of a free massless spin-2 field. Specifically, the isometry group of the background de Sitter space acts not as the physical symmetries of a quantum field theory but as quantum gravity gauge constraints. 

Our goal in this section is to explain how those constraints impact the physical algebra of observables $\widetilde{\mathcal{A}}$ associated to a single static patch $P$. In pure de Sitter space, there is no gauge-invariant way to pick out a particular static patch $P$. Instead, we will assume that $P$ is defined in a gauge-invariant way relative to the location of the observer. This observer could either be present during inflationary era or, perhaps more physically given the role inflation seems to have played in our universe, they could exist only at some asymptotically late time. Regardless, the location and velocity of the observer can then be used to define the patch $P$. Crucially, unlike in \cite{CLPW} we will only make use of the observer's (classical) location to define $P$ and will not additionally equip them with any quantum Hilbert space to act as their clock.

The choice of $P$ breaks the isometric group of de Sitter space down to the group $\R \times SO(3)$ of the boost- and rotation-isometries of the patch $P$. The algebra $\widetilde{\mathcal{A}}$ is then the boost- and rotation-invariant subalgebra of the QFT static patch algebra $\mathcal{A}$. 

While there exist many rotation-invariant observables in the algebra $\mathcal{A}$, an important part of the story of \cite{CLPW} is that for quantum fields in de Sitter space with a potential bounded from below the only boost-invariant operators in the algebra $\mathcal{A}$ are $c$-numbers. This led to the need to include an explicit model of a clock in order to obtain a nontrivial algebra $\widetilde{\mathcal{A}}$.

This is not true, however, in the presence of an inflaton field. As we saw in Section \ref{sec:bdinflaton}, a classical solution to the inflaton equations of motion will tend to roll linearly down the potential gradient at early and late times as described by \eqref{eq:particularsoln}. Perturbations to this solution are described by \eqref{eq:gensoln} and have finite limits as $\tau \to \pm \infty$. So this behaviour is universal.

The late- and early-time behaviour of the quantum theory is related but slightly more complicated. At arbitrarily late times short wavelength modes with nonzero vacuum fluctuations continue to exit the horizon and contribute to the effective classical background $\phi_{\rm cl}$ -- the sum over all modes that have exited the horizon -- within the static patch. As a result, the evolution of $\phi_{\rm cl}$ at long times becomes a Wiener process, with a drift proportional to $V'$ and stochastic noise from modes exiting the horizon. The probability distribution $p(\phi_{\rm cl}, t)$ as a function of the time $t$ is therefore described by the Fokker-Planck equation  \cite{chaos1,chaos2,chaos3}
\begin{align}\label{eq:fokker}
    \frac{\partial}{\partial t} p(\phi_{\rm cl}) = \frac{V'}{3} \frac{\partial}{\partial \phi}p(\phi_{\rm cl},\tau) + \frac{1}{8\pi^2}\frac{\partial^2}{\partial \phi^2}p(\phi_{\rm cl},\tau).
\end{align}
It follows from \eqref{eq:fokker} by integrating by parts that the expectation value $\langle\phi_{\rm cl}\rangle$ satisfies \begin{align}
    \frac{\partial}{\partial t} \langle\phi_{\rm cl}\rangle = - \frac{V'}{3}
\end{align} 
while the variance $\delta \phi_{\rm cl}^2$ satisfies 
\begin{align}
    \frac{\partial}{\partial t} \delta \phi_{\rm cl}^2 = \frac{\partial}{\partial t} \left(\langle\phi_{\rm cl}^2\rangle - \langle\phi_{\rm cl}\rangle^2\right) = \frac{1}{4 \pi^2}.
\end{align}
Since the late-time probability distribution is Gaussian by the central limit theorem, this means that the probability $p(\phi_{\rm cl} > \phi_0)$ that the contribution to the inflaton field from modes that have exited the horizon remains above fixed finite value $\phi_0$ decays exponentially at late times. By a time-reversed version of the same argument, this is also true at early times. In particular, this means that our argument in the introduction that the operator \eqref{eq:introtildeadef} did not exist is no longer valid in the presence of an inflaton. In the inflaton quantum field theory, the static patch never equilibrates and so there is no reason that the integral over time in \eqref{eq:introtildeadef} needs to diverge when acting on a normalisable state. 

Naively, the evolution of $\phi_{\rm cl}$ seems in tension with the boost-invariance of the Bunch-Davies weight. However, the above argument applies only to normalisable probability distributions $p(\phi_{\rm cl})$ where we can integrate by parts without running into boundary terms. In contrast, under the field redefinition $\phi \to \phi + \kappa$  for some constant $\kappa$ (or equivalently under $\phi_0 \to \phi_0 + \sqrt{2}\pi\kappa$), we see from \eqref{eq:inflatonBDweight} that
\begin{align}
    \ket{\Psi_{BD}} \to \exp\left(-\frac{4 \pi^2}{3}\kappa\right) \ket{\Psi_{BD}}.
\end{align}
It follows that the Bunch-Davies weight must have
\begin{align} \label{eq:invariantprob}
    p(\phi_{\rm cl}) \sim \exp(-\frac{8 \pi^2 V'}{3} \phi_{\rm cl}),
\end{align}
which is of course unnormalisable. It is easy to check that \eqref{eq:invariantprob} is in fact preserved by the Fokker-Planck equation \eqref{eq:fokker}, consistent with the boost invariance of $\ket{\Psi_{BD}}$.

Suppose we consider a QFT operator $a \in \mathcal{A}$ that decays to zero faster than exponentially as $\phi_{\rm cl} \to -\infty$ (assuming modes with subhorizon wavelength are in the Bunch-Davies state). For example, $a$ might be the operator $\exp(-\phi_f^2)$ defined in \eqref{eq:exp-phi_f^2}. Let $a(t) = e^{iHt} a e^{-iHt}$ be the same operator but boosted by time $t$. Then for any fixed state $\ket{\Phi} \in \mathcal{H}$, we have $\lVert a(t) \ket{\Phi}\rVert \to 0$ exponentially fast as $|t| \to \infty$. This means that the operator
\begin{align} \label{eq:timeintegral}
    \widetilde{a} = \int dt \, a(t)
\end{align}
is densely defined, boost invariant and has nontrivial action on the Hilbert space $\mathcal{H}$. It will also have finite expectation value $\braket{\Psi_{BD}|a|\Psi_{BD}}$ in the Bunch-Davies weight. Since operators that behave like $a$ are dense in $\mathcal{A}$ with respect to the strong operator topology, this leads to a large algebra $\widetilde{\mathcal{A}} \subseteq \mathcal{A}$ of boost- (and rotation-) invariant operators.

We will now argue that the Bunch-Davies weight can be used to construct a normal semifinite trace for the algebra $\widetilde{\mathcal{A}}$. Suppose we tried to construct a formal density matrix for the Bunch-Davies weight on $\mathcal{A}$. To do so, we would take the Bunch-Davies wavefunctional and partial trace over fields at $\chi > \pi/2$. This leads to the path integral shown in Figure \ref{fig:rhobd},
 \begin{figure}
  \hspace{4.5 cm}
\begin{tikzpicture}[scale =1]
     \pgftext{\includegraphics[scale =.5]{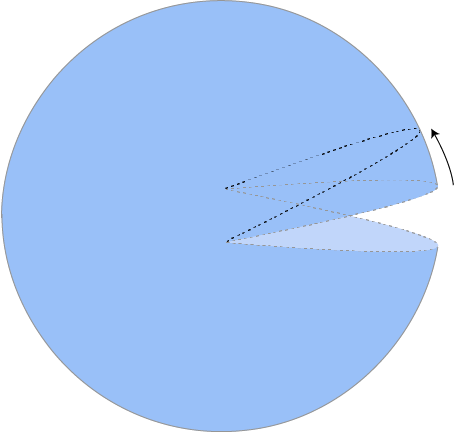}
        \draw (0,5.3) node {\LARGE $e^{-\varepsilon \boldsymbol{h}}$};
         \draw (0,3) node { $\chi=0$};
         \draw (-5,3.) node { $\chi=\pi/2$};
          \draw (-10,3.) node { $\chi=\pi$};
}
 \end{tikzpicture}
 \caption{\label{fig:rhobd} The formal density matrix of the Bunch-Davies weight on the algebra $\mathcal{A}$ is computed by a Euclidean path integral over the full sphere, with fixed boundary conditions on either side of a cut at the static patch centered on $\chi = 0$. This can be reinterpreted as a sequence of operators $\exp(-\varepsilon \boldsymbol{h})$ generating Euclidean rotations (i.e. the analytic continuation of Lorentzian boosts). The full density matrix describes (up to normalisation) a Euclidean rotation by $2\pi$.}
 \end{figure}
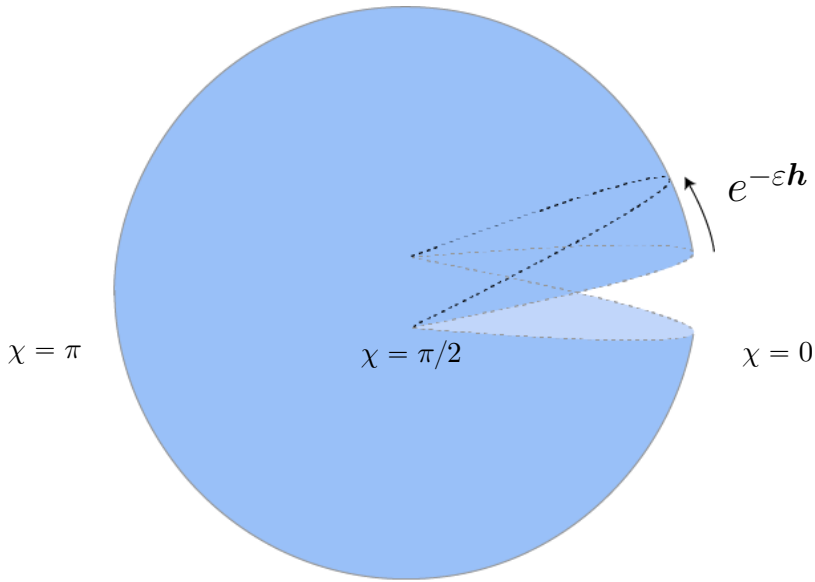 which can be reinterpreted as the path integral for a boost of the static patch by Euclidean time $2\pi$. In other words we have 
 \begin{align}\label{eq:rho}
     \boldsymbol{\rho} \sim \exp(-2\pi \boldsymbol{h} + {\rm const})
 \end{align} 
 where 
 \begin{align}\label{eq:Sec2formalHamsplit} 
     H = \boldsymbol{h} - \boldsymbol{h'}
 \end{align}
 is a formal splitting of the boost Hamiltonian $H$ into singular Hamiltonians $\boldsymbol{h}$ and $\boldsymbol{h'}$ on the static patch and its commutant respectively. A constant, that turns out to be divergent, has been included because the density matrix needs to be normalised to have trace one. 
 
 The splitting \eqref{eq:Sec2formalHamsplit} certainly exists with $\boldsymbol{h}$ and $\boldsymbol{h}'$ defined as integrals over the stress-energy tensor. We can therefore also define $\boldsymbol{h}$ and $\boldsymbol{h}'$ as sesquilinear forms mapping dense subsets of bras and kets to their correspoding matrix elements. However $\boldsymbol{h}$ and $\boldsymbol{h}'$ do not exist as densely defined operators on $\mathcal{H}$ because $\boldsymbol{h} \ket{\Phi}$ is not normalisable for a dense set of states $\ket{\Phi} \in \mathcal{H}$.\footnote{If a sesquilinear form $\boldsymbol{h}$ is not actually an operator, it might be unclear what it means to say that it is localised in the static patch $P$. What we mean by this is that $\boldsymbol{h}$ is invariant under conjugation by any unitary $U_\ell \in \mathcal{A}_{\ell}$. Equivalently, $\boldsymbol{h}$ defines a linear functional on a dense set of states on $\mathcal{A}_r$. (Recall that states on $\mathcal{A}$ are really themselves linear functionals on the algebra $\mathcal{A}$. So $\boldsymbol{h}$ is a linear functional on the space of linear functionals.} To indicate this, we write them in bold font. As a result, it does not make sense, except at a very formal level, to talk about nonlinear functions of $\boldsymbol{h}$. However, one can consider the effect of the action generated by the full boost Hamiltonian $H$  on the algebra $\mathcal{A}$. It turns out that $H$ generates an (outer) automorphism of $\mathcal{A}$, which is another sense in which $H$ morally splits as \eqref{eq:Sec2formalHamsplit}, without literally doing so.

In a similar manner to \eqref{eq:rho}, the density matrix $\boldsymbol{\rho'}$ on the commutant static patch can be written as 
\begin{align}\label{eq:rho'}
    \boldsymbol{\rho'} \sim \exp(- 2\pi \boldsymbol{h'} +{\rm const}).
\end{align}
By symmetry, we expect the divergent constants in \eqref{eq:rho} and \eqref{eq:rho'}  should be the same. While the density matrices $\boldsymbol{\rho}$ and $\boldsymbol{\rho'}$ do not actually exist as we explained above, the formal combination $\Delta_{\Psi} = \boldsymbol{\rho} \otimes \boldsymbol{\rho'}^{-1}$ describes a densely defined operator called the modular operator for the semifinite weight $\ket{\Psi_{BD}}$. This operator can be rigorously defined using Tomita-Takesaki theory, briefly reviewed in Appendix \ref{appx:a}. By the formal manipulations above, we have 
\begin{align}\label{eq:deltapsiinflaton}
    \Delta_\Psi = \exp(-2 \pi H).
\end{align} 
More rigorously, one can prove \eqref{eq:deltapsiinflaton} using KMS condition. Given $a,b \in \mathcal{A}$, the KMS condition says that
\begin{align}\label{eq:KMS}
    \braket{\Psi_{BD}|a(t) b|\Psi_{BD}} = \braket{\Psi_{BD}|b\, a(t + 2\pi i)|\Psi_{BD}},
\end{align}
where the right-hand side is defined by analytic continuation. \eqref{eq:KMS} follows immediately from the definition of $\ket{\Psi_{BD}}$ via a Euclidean path integral because the left- and right-hand sides of \eqref{eq:KMS} are defined by the same path integral. Consider a path integral on the Euclidean sphere with operators $b$ and $a$ inserted separated by a Euclidean angle $0 > it > -2\pi$. By analytic continuation, this correlation function can be extended to the strip $0 > \Imag(t) > -2\pi$. In the limits $\Imag(t) \to 0$ and $\Imag(t) \to -2\pi$, the difference in Euclidean time between the two insertions goes to zero. But, as $\Imag(t) \to 0$, the operator $b$ is inserted slightly before $a$ in Euclidean time and the correlation function limits to the left-hand side of \eqref{eq:KMS}. On the other hand, as $\Imag(t) \to -2\pi$ the operator $a$ is inserted slightly prior to $b$ and the correlation function limits to the right-hand side of \eqref{eq:KMS}. But, by Theorem 16 in \cite{Sorce_2023}, \eqref{eq:deltapsiinflaton} follows from the four conditions that a) $\ket{\Psi_{BD}}$ is cyclic-separating on $\mathcal{A}$, b) $H\ket{\Psi_{BD}} = 0$, c) $H$ generates an automorphism of $\mathcal{A}$ and d) \eqref{eq:KMS}.

Modulo a minor subtlety about normalisation, the Bunch-Davies weight will act as a trace on the boost-invariant algebra $\widetilde{\mathcal{A}}$ because, for boost-invariant states, the KMS condition \eqref{eq:KMS} is identical to the tracial condition 
\begin{align}
\Tr[\tilde a \tilde b] = \Tr[\tilde b \tilde a].
\end{align}

More precisely, suppose we have an operator $\widetilde{a} \in \widetilde{\mathcal{A}}$ that can be written as \eqref{eq:timeintegral}. We define the trace
\begin{align}\label{eq:inflaton1sttracedef}
    \Tr(\widetilde{a}) = \braket{\Psi_{BD}|a(t)|\Psi_{BD}}
\end{align}
Since the Bunch-Davies weight is boost invariant, this is independent of the time $t$ that we pick. Furthermore, suppose we have the operator equality
\begin{align}
    \widetilde{a} = \int dt\, a_0(t) = \int dt\, a_1(t).
\end{align}
If
\begin{align} \label{eq:equivalencetraceinflaton}
    0 = \int dt\,\braket{\Psi_{BD}|(a_0(t) - a_1(t)) |\Psi_{BD}},
\end{align}
then the integrand must itself vanish because the Bunch-Davies weight is boost invariant. Therefore \eqref{eq:inflaton1sttracedef} is not only independent of $t$ but also of the choice of expansion \eqref{eq:timeintegral}.

We can write the product of two operators $\widetilde{a}$ and $\widetilde{b}$ as
\begin{align}
    \widetilde{a}\widetilde{b} = \int dt_1 dt_2\, a(t_1) b(t_2) = \int dt \int dt' \,a(t) b(t + t').
\end{align}
But then
\begin{align}
\Tr(\widetilde{a}\widetilde{b}) &= \int dt' \braket{\Psi_{BD}|a(t) b(t + t')|\Psi_{BD}} \\&= \int dt' \braket{\Psi_{BD}|b(t + t' - 2 \pi i) a(t) |\Psi_{BD}} \\&= \int dt'\braket{\Psi_{BD}|b(t - t'') a(t) |\Psi_{BD}}
\\&= \int dt'' \braket{\Psi_{BD}|b(t) a(t + t'') |\Psi_{BD}} \\&= \Tr(\widetilde{b}\widetilde{a}).
\end{align}
In the second equality we used the KMS condition, while  in the third equality we substituted $t'' = 2 \pi i - t'$ and then shifted the contour of integration to real $t''$. An interesting feature of this proof is that we never needed to assume rotation invariance of $\tilde a$ or $a(t)$. It is only the boost symmetry, i.e. the noncompact part of the gauge group, that plays a nontrivial role. 

It would be nice to express $\Tr(\tilde a)$ directly as a linear functional of the operator $\tilde a$ rather than indirectly via $a(t)$. The problem is that the boost invariance of $\ket{\Psi_{BD}}$ means that the integral over $t$ in
\begin{align}
\braket{\Psi_{BD}|\tilde a|\Psi_{BD}} = \int dt \braket{\Psi_{BD}|a(t) |\Psi_{BD}}
\end{align}
will always diverge. Since this divergent factor is independent of the operator $\tilde a$, we could hope remove it by a formal infinite rescaling of $\ket{\Psi_{BD}}$. Indeed, this is exactly what the definition \eqref{eq:inflaton1sttracedef} achieves. 

However, we can also achieve the same result by regulating $\ket{\Psi_{BD}}$ to produce a normalisable state and then dividing through by a regulator-dependent factor that goes to infinity as the regulator is removed. Explicitly let $P_{\phi_0 >  \phi_{\rm min}}$ be the projector onto the spatial zero mode $\phi_0 >  \phi_{\rm min}$ for some constant $\phi_{\rm min}$. $P_{\phi_0 >  \phi_{\rm min}} \ket{\Psi_{BD}}$ is then a normalisable state. We obviously have
\begin{align}
    \lim_{\phi_{\rm min} \to- \infty} \braket{\Psi_{BD}|P_{\phi_0 >  \phi_{\rm min}}\,a(t)\,P_{\phi_0 >  \phi_{\rm min}}|\Psi_{BD}} = \braket{\Psi_{BD}|a(t)|\Psi_{BD}}
\end{align}
for any $a \in \mathcal{A}$. However, for any finite $\phi_{\rm min}$, the expectation values $\braket{\Psi_{BD}|P_{\phi_0 >  \phi_{\rm min}}a(t) P_{\phi_0 >  \phi_{\rm min}}|\Psi_{BD}}$ and $\braket{\Psi_{BD}|a(t)|\Psi_{BD}}$ will be very different whenever $|t|$ is sufficiently large. Since we are only interested in the behaviour of the inflaton field at very late times, the relevant physics is captured by the Fokker-Planck equation \eqref{eq:fokker}. A standard property of Wiener processes, which can be obtained by solving the Fokker-Planck equation with a delta-function initial condition, says that the conditional probability distribution
\begin{align}\label{eq:condprobinflaton}
    p(\phi_{\rm cl},t|\phi_{{\rm cl},0}) = \frac{(2 \pi)^{1/2}}{t}\exp(-\frac{2 \pi^2 (\phi_{\rm cl} - \phi_{{\rm cl},0} + V' t/3)^2}{t}),
\end{align}
where the initial probability distribution is $p(\phi_{{\rm cl},0})$. For the state $P_{\phi_0 >  \phi_{\rm min}} \ket{\Psi_{BD}}$ we have
\begin{align}
p(\phi_{{\rm cl},0}) = \exp(- \frac{8 \pi^2 V'}{3} \phi_{{\rm cl},0}) \theta (\phi_{{\rm cl},0} - \phi_{\rm min}).
\end{align}
We therefore obtain
\begin{align}
    p(\phi_{\rm cl},t) &= \int d \phi_{{\rm cl},0}\, p(\phi_{\rm cl},t|\phi_{{\rm cl},0}) p(\phi_{{\rm cl},0}) \nonumber
    \\&=\exp(- \frac{8 \pi^2 V'}{3} \phi_{\rm cl})\int d \phi_{{\rm cl},0} \frac{(2 \pi)^{1/2}}{t} \exp(-\frac{2 \pi^2 (\phi_{\rm cl} - \phi_{{\rm cl},0} - V' t/3)^2}{t}) \theta (\phi_{{\rm cl},0} - \phi_{\rm min})\nonumber
    \\&=\exp(- \frac{8 \pi^2 V'}{3} \phi_{\rm cl}) \,\Phi(\frac{2 \pi(\phi_{\rm cl} - V' t/3 - \phi_{\rm min})}{\sqrt{t}}) \label{eq:phicllatetimeevolution}
\end{align}
where $\Phi(z) = \int_{-\infty}^z dt \exp(-t^2/2)/\sqrt{2\pi} $ is the cumulative probability distribution for the standard Gaussian. In other words, we see that for any fixed $\phi_{\rm cl}$ and large negative $\phi_{\rm min}$, the probability distribution $p(\phi_{\rm cl},t)$ looks like \eqref{eq:invariantprob} for $t \lesssim 3 |\phi_{\rm min}|/V'$ and then quickly goes to zero over a comparatively short time $\Delta t \sim t^{1/2}$. By symmetry, the same behaviour also occurs at very early times. We therefore have 
\begin{align} \label{eq:inflatontrace2}
    \Tr(\widetilde{a}) = \lim_{\phi_{\rm min} \to -\infty} \frac{V'}{6 \,|\phi_{\rm min}|} \braket{\Psi_{BD}|P_{\phi_0 >  \phi_{\rm min}}\widetilde{a} P_{\phi_0 >  \phi_{\rm min}}|\Psi_{BD}},
\end{align}
which gives our desired definition of $\Tr(\tilde{a})$ as an explicit linear functional of $\tilde a$. Note that \eqref{eq:inflatontrace2} only defines a semifinite normal weight on $\widetilde{\mathcal{A}}$; unlike the Bunch-Davies weight $\ket{\Psi_{BD}}$ it does not form a semifinite, normal weight on $\mathcal{A}$.

On general grounds, we expect that the algebra $\widetilde{\mathcal{A}}$ does not contain any nontrivial centre and so is a von Neumann factor. Since it has a trace, it cannot be Type III. Since the trace of the identity is infinite (as can be easily verified from \eqref{eq:inflatontrace2}), $\widetilde{\mathcal{A}}$ is not finite-dimensional or Type II$_1$. The remaining possibilities are Type I$_\infty$ or Type II$_\infty$. Finally, we observe that the boost Hamiltonian $H$ is invariant under the field transformation $\phi \to \phi + \kappa$ for constant $\kappa$. So $\phi \to \phi + \kappa$ generates an automorphism of the algebra $\widetilde{\mathcal{A}}$. As we already saw, that automorphism rescales the trace $\Tr$ by a factor of $\exp(-8 \pi^2 V' \kappa/3)$, which is inconsistent with the existence of a minimal trace for projectors and hence with a Type I$_\infty$ factor. We conclude that $\widetilde{\mathcal{A}}$ is Type II$_\infty$.

We have so far not talked about imposing the gauge constraints on the commutant algebra $\mathcal{A}'$. However it is obvious by symmetry that this works in exactly the same way. The gauge-invariant algebras $\widetilde{\mathcal{A}} \subseteq \mathcal{A}$ and $\widetilde{\mathcal{A}}' \subseteq \mathcal{A}'$ commute because they are contained in commuting algebras. However, as algebras acting on the Hilbert space $\mathcal{H}$, they are not commutants, because e.g. gauge transformations are nontrivial operators on $\mathcal{H}$ that commute with both.

The example of gauge transformations makes it clear that the reason (or at least one reason) $\widetilde{\mathcal{A}}$ and $\widetilde{\mathcal{A}}'$ are not commutants is that we have imposed the gauge constraints on the algebras but not on the Hilbert space $\mathcal{H}$. Because the gauge group is not compact, the constraints on $\mathcal{H}$ need to be imposed  using the method of coinvariants; see e.g. Appendix B of \cite{CLPW} for a review. This leads to a gauge-invariant Hilbert space $\widetilde{\mathcal{H}}$ with natural actions of $\widetilde{\mathcal{A}}$ and $\widetilde{\mathcal{A}}'$.\footnote{We again assume that only the symmetries of the static patch $P$ need to be imposed as constraints, because the remaining symmetries of de Sitter space are broken by the presence of an observer.} It is easy to check that these actions still commute. We would like to conjecture that in fact $\widetilde{\mathcal{A}}$ and $\widetilde{\mathcal{A}}'$ are commutants when acting on $\widetilde{\mathcal{H}}$. We certainly don't know of any operators that commute with both. But it is not true in general that if two algebras $\mathcal{A}$ and $\mathcal{A}'$ are commutants on a Hilbert space $\mathcal{H}$ and we gauge an automorphism group of those algebras that the invariant subalgebras $\widetilde{\mathcal{A}}\subseteq \mathcal{A}$ and $\widetilde{\mathcal{A}}' \subseteq \mathcal{A}'$ act as commutants on the coinvariant Hilbert space $\widetilde{\mathcal{H}}$. A counterexample is given by stable quantum fields in de Sitter space, where $\widetilde{\mathcal{A}}$ and $\widetilde{\mathcal{A}}'$ are trivial but the coinvariant Hilbert space $\widetilde{\mathcal{H}}$ and hence the algebra of operators $\mathcal{B}(\widetilde{\mathcal{H}})$ are not.

 \subsection{Density matrices and entropies} \label{sec:bddensity}
In previous work, the trace of gravitational Type II von Neumann algebras behaves as if there is a (divergent) density of states proportional to $\exp(A_{\rm hor}/4G)$ with $A_{\rm hor}$ the area of the black hole or cosmological horizon. We find the same result for the algebra $\widetilde{\mathcal{A}}$. 

We have already seen that shifting $\phi \to \phi + \kappa$ rescales $\Tr$ by $\exp(-8 \pi^2 V' \kappa/3)$. Changes in the de Sitter horizon area $A_{\rm hor}$ are related to the changes in the cosmological constant $\Lambda = 3/\ell_{dS}^2$ by 
\begin{align}
\delta A_{\rm hor} = 8 \pi \delta \ell_{dS} = - \frac{4 \pi}{3}  \delta \Lambda = - \frac{32 \pi^2}{3} G \delta V.
\end{align}
To absorb the constant shift in the QFT Lagrangian into the Einstein-Hilbert action, and leave the total action invariant under the field redefinition, we need to shift the cosmological constant $\Lambda$ by
\begin{align}
\delta \Lambda = 8 \pi G \delta V = 8 \pi G V' \kappa.
\end{align}
So the trace will be rescaled by
\begin{align}
    \exp(-8 \pi^2 V' \kappa/3) = \exp(- \delta A_{\rm hor}/4G),
\end{align}
consistent with our claim above.

Furthermore, as we explained in Section \ref{sec:bdinflaton}, the Bunch-Davies weight can be thought of as a $G \to 0$ limit of the no boundary Hartle-Hawking state (up to an overall normalisation factor). Our results are therefore consistent with the proposal of \cite{witten2023background} that, beyond the strict $G \to 0$ limit, the no-boundary Hartle-Hawking state acts as a trace on the gauge-invariant algebra associated to an observer.
 
To obtain a more precise relationship between the Type II algebra $\widetilde{\mathcal{A}}$ and the generalised entropy of its static patch, we need to find the density matrix $\rho_\Phi$ associated with some state $\ket{\Phi}$ for the gauge-invariant algebra $\widetilde{\mathcal{A}}$. 

We assume for simplicity that the state $\ket{\Phi}$ is already rotation-invariant. If not, then one can produce a rotation invariant state with the same expectation values for all operators in $\widetilde{\mathcal{A}}$ by simply integrating the state (i.e. the positive linear functional) on $\mathcal{A}$ defined by $\ket{\Phi}$ over the rotation group and dividing by its volume, which is finite because the rotation group is compact. This defines a normalised rotation-invariant positive linear functional on $\mathcal{A}$ which can then always be purified (for a Type III$_1$ algebra) to a rotation-invariant pure state on $\mathcal{H}$. 

By definition, $\rho_\Phi$ is the unique operator affiliated to the algebra $\widetilde{\mathcal{A}}$ such that
\begin{align}\label{eq:condition}
    \bra{\Phi} \widetilde{a}\ket{\Phi} = \Tr(\rho_\Phi \,\widetilde{a}),\quad \forall\, \widetilde{a}\in \widetilde{\mathcal{A}}.
\end{align}
We claim that 
\begin{align}
    \rho_\Phi = \int d t\; e^{(\pi-\i t) H}\,\Delta_{\Phi|\Psi}\,e^{(\pi+\i t) H} = \int d t\; \Delta_\Psi^{-\frac{1}{2}-\i \frac{ t}{2\pi}}\,\Delta_{\Phi|\Psi}\,\Delta_\Psi^{-\frac{1}{2}+\i \frac{ t}{2 \pi}},
\end{align}
where $\Delta_{\Phi|\Psi}$ is the relative modular operator of $\ket{\Phi}$ relative to $\ket{\Psi_{BD}}$ on the algebra $\mathcal{A}$. See Appendix \ref{appx:a} for a brief review. Given an operator $a'$ in the commutant algebra $\mathcal{A}'$, we have
\begin{align}
    [\log \Delta_{\Phi|\Psi}, a'] = [\log \Delta_\Psi, a'] = - 2 \pi [H,a']
\end{align}
It follows immediately that $[\rho_\Phi, a'] =0 $ and hence $\rho_{\Phi}$ affiliated to $\mathcal{A}$. Since
\begin{align}
[H, \rho_\Phi] = i \int d t\; \frac{\partial}{\partial t}\left[e^{(\pi-\i t) H}\,\Delta_{\Phi|\Psi}\,e^{(\pi+\i t) H}\right] = 0,
\end{align}
$\rho_\Phi$ is also affiliated to the boost-invariant algebra $\widetilde{\mathcal{A}}$. Finally, we have
\begin{align}
    \Tr(\rho_\Phi a) &= \braket{\Psi_{BD}|e^{\pi H}\Delta_{\Phi|\Psi} e^{\pi H } a|\Psi_{BD}}
    \\& = \braket{\Psi_{BD}|\Delta_{\Phi|\Psi} a|\Psi_{BD}}
    \\& = \braket{\Phi|a |\Phi}.
\end{align}
In the second equality we used $[H,a] = 0$ and $H \ket{\Psi_{BD}} = 0$, and in the last step we used (\ref{eq:appxeq}).

In Appendix \ref{appx:a}, we review proofs of two other standard properties of the relative modular operator that we will need: a) given a reference system $\mathcal{H}_R$ and a state $\ket{\Phi} = \ket{\Phi_0} \ket{0} + \ket{\Phi_1} \ket{1} \in \mathcal{H} \otimes \mathcal{H}_R$, the relative modular operator $\Delta_{\Phi|\Psi} = \Delta_{\Phi_0|\Psi} + \Delta_{\Phi_1|\Psi}$ and b) if we define the state $\ket{\Phi(s)} = \Delta_\Psi^{-is} \ket{\Phi}$, then
\begin{align}
    \Delta_{\Phi(s)|\Psi} = \Delta_{\Psi}^{-is}\Delta_{\Phi|\Psi} \Delta_{\Psi}^{is}.
\end{align}
We therefore have
\begin{align}
\rho_\Phi = \int dt e^{\pi H} \Delta_{\Phi(t)|\Psi} e^{\pi H} = e^{\pi H} \Delta_{\tilde\Phi|\Psi} e^{\pi H},
\end{align}
where $\ket{\Phi(t)} = e^{-it H} \ket{\Phi}$ and the weight $\ket{\tilde\Phi}$ is defined such that
\begin{align} \label{eq:hatphi}
    \braket{\tilde \Phi| a |\tilde \Phi} = \int dt \braket{\Phi(t) | a|\Phi(t)} \int dt \braket{\Phi | a(t)|\Phi},
\end{align}
The functional \eqref{eq:hatphi} is finite whenever $a$ defines a boost-invariant operator $a$ by \eqref{eq:timeintegral}. We already argued that this was true for a dense set of operators $a$, which means the weight $\ket{\tilde\Phi}$ is semifinite. 

Since the expectation value \eqref{eq:hatphi} is boost invariant, we have $[\Delta_{\tilde\Phi|\Psi} , H] = 0$. We therefore have
\begin{align} \label{eq:inflatonlogrho}
    \log \rho_{\Phi} = \log \Delta_{\tilde\Phi|\Psi} - \log \Delta_\Psi = \log \Delta_{\tilde\Phi} - \log \Delta_{\Psi|\tilde\Phi} ,
\end{align}
where the second equality is a standard identity that follows from taking a derivative of Connes' cocycle flow. The entropy of the state $\ket{\Phi}$ on the algebra $\widetilde{\mathcal{A}}$ is therefore
\begin{align} \label{eq:inflatonentropy}
S(\Phi) = - \braket{\Phi|\log \rho_{\Phi}|\Phi} = \braket{\Phi|\log \Delta_{\Psi|\tilde\Phi}|\Phi},
\end{align}
where the expectation of the first term from \eqref{eq:inflatonlogrho} vanishes because $\log \Delta_{\tilde\Phi}$ is boost invariant and $\braket{\tilde\Phi| \log \Delta_{\tilde\Phi} |\tilde\Phi}$ vanishes by definition.

Since $\log \Delta_{\Psi|\tilde\Phi}$ is boost invariant, the formula \eqref{eq:inflatonentropy} can be formally interpreted as the relative entropy of $\ket{\tilde\Phi}$ with respect to $\ket{\Psi_{BD}}$, divided by a divergent factor equal to the squared norm of $\ket{\tilde\Phi}$. There is a general relationship between relative entropy and generalised entropy that was first developed by Casini \cite{Casini_2008} and then later Wall \cite{Wall_2012}. We can formally write 
\begin{align}
\log \Delta_{\Psi|\tilde\Phi} =\log \boldsymbol{\rho}_\Psi - \log \boldsymbol{\rho}_{\tilde\Phi} = 2 \pi \boldsymbol{h}- \log \boldsymbol{\rho}_{\tilde\Phi}' +{\rm const},
\end{align}
where $\boldsymbol{\rho}_\Psi$ and $\boldsymbol{\rho}_{\tilde\Phi}'$ after formal density matrices for $\ket{\Psi_{BD}}$ and $\ket{\tilde\Phi}$ on $\mathcal{A}$ and $\mathcal{A}'$ respectively and the constant is again divergent. It follows that
\begin{align} \label{eq:formalgenentinflaton}
    \braket{\Phi| \log \Delta_{\Psi|\tilde\Phi}|\Phi} = 2 \pi \braket{\Phi|\boldsymbol{h} |\Phi} - \braket{\Phi|\log \boldsymbol{\rho}_{\tilde\Phi}'|\Phi} + {\rm const},
\end{align}
Up to a divergent additive constant, the second term in \eqref{eq:formalgenentinflaton} is the entropy of $\ket{\tilde\Phi}$ divided by its divergent normalisation. Meanwhile, comparing the surface integral and volume formulas for the Komar mass of the static patch, we have
\begin{align}
\boldsymbol{\hat A}_{\rm hor} - 4 \pi = 8 \pi G \boldsymbol{h}.
\end{align}
The first term in \eqref{eq:formalgenentinflaton} is therefore equal to $\braket{\Phi|\boldsymbol{\hat A}_{\rm hor}/4G|\Phi}$ minus another divergent additive constant. We conclude that the Type II$_\infty$ entropy $S(\Phi)$ is equal to the generalised entropy for the boost-invariant weight $\ket{\tilde\Phi}$ up to the divergent constant needed to render Type II entropies finite.

 \section{Black holes}\label{sec:blackholes}

 \subsection{Black holes in asymptotically flat spacetimes} \label{sec:bhflat}

 We now turn from slow-rolling inflaton to our second example of a cosmological clock -- an evaporating black hole. As a warm up, we first consider black holes in asymptotically flat spacetimes. Our discussion will closely follow \cite{Witten_2022, CPW}, although some details are new. Our construction can be applied to any stationary two-sided black hole solution including black holes that break spherical symmetry (i.e. Kerr or Kerr-Newman metrics), which were first studied in this context in \cite{kudlerflam2024generalized}.

 As in Section \ref{sec:inflaton}, we take a strict $G \to 0$ limit where fluctuations of free quantum fields, including fluctuations in the semiclassical gravitational field $h_{\mu\nu} = \sqrt{G}g_{\mu \nu}$ are described in the $G \to 0$ limit by local quantum field theory on a fixed black hole background.

 Unlike in Section \ref{sec:inflaton}, we are now working with quantum field theory in a curved spacetime with noncompact Cauchy slices. As a result, we have to be somewhat more careful when defining our Hilbert space. Because the spacetime is asymptotically flat, there exists a particularly natural choice of Hilbert space that we will call $\mathcal{H}_{\rm QFT}$. Heuristically, this Hilbert space consists of states that look like the Minkowski vacuum both close to the black hole bifurcation surface and near to spatial infinity. Like the Hilbert space for an inflaton in Section \ref{sec:inflaton}, the Hilbert space $\mathcal{H}_{\rm QFT}$ does not contain any normalisable preferred states that are invariant under the black hole isometry group. The Hartle-Hawking state, for example, satisfies the first requirement but not the second, because it looks like a thermal state far from the black hole. The Boulwar\'{e} vacuum on the other hand satisfies the second requirement but not the first.
 Instead all states in $\mathcal{H}_{\rm QFT}$ will look like an evaporating black hole at sufficiently late times. At sufficiently early times, meanwhile, they look like a white hole that is steadily absorbing thermal radiation and thereby growing in size. Since this is the time reverse of an evaporating black hole, we will call it an evaporating white hole; it is evaporating in the direction of the thermodynamic arrow of time (i.e. towards the past).
 
\begin{figure}
\begin{subfigure}[t]{0.4 \textwidth}
    \begin{tikzpicture}[scale =1]
     \pgftext{ 
     \includegraphics[scale =.3]{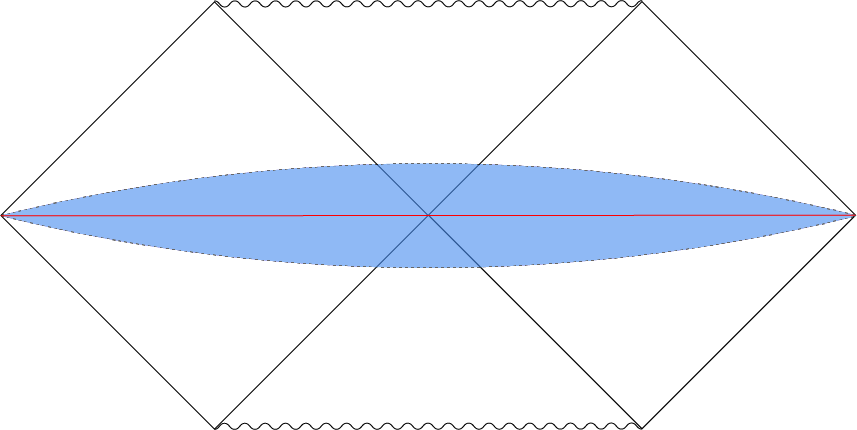} 
        \draw (-2,2.3) node {\color{red}  $\Sigma$};}
 \end{tikzpicture}
    \caption{}
\end{subfigure}
\hspace{4 cm}
\begin{subfigure}[t]{0.4\textwidth}
      \begin{tikzpicture}[scale =1]
     \pgftext{ 
     \includegraphics[scale =.33]{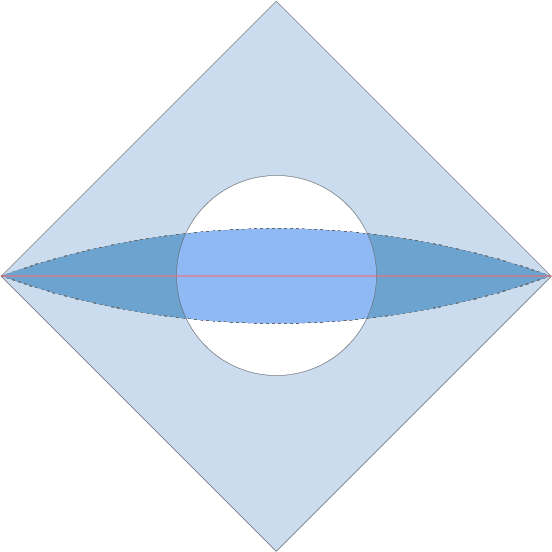} 
         \draw (-2,3.3) node {\color{red}  $\Sigma$};
         \draw (-4,5.3) node {\LARGE  $g_+$};
          \draw (-4,1.3) node {\LARGE  $g_-$};
}
 \end{tikzpicture}
 \caption{}
 \end{subfigure}
 \caption{\label{fig:preferredmodes} The metric near a Cauchy slice $\Sigma$ (the dark-blue region) of a Schwarzschild (or other asymptotically flat) black hole (left) can be smoothly connected to time-independent metrics on an asymptotically flat space (the light-blue region on the right). This allows us to construct an unambiguous Hilbert space $\mathcal{H}_{\rm QFT}$ for the black hole.}
 \end{figure}

 A somewhat more careful definition of the Hilbert space $\mathcal{H}_{\rm QFT}$, using the constructions of \cite{Wald:1995yp,witten2022does}, is as follows. We pick a Cauchy slice $\Sigma$ of the asymptotically flat black hole and deform the spacetime outside of a small neighbourhood of $\Sigma$ so that it is given by a time-independent metric $g_+$ sufficiently far in the future and by a potentially different metric $g_-$ sufficiently far in the past. See Figure \ref{fig:preferredmodes}. Specifically, we choose the metrics $g_-$ and $g_+$ to agree with the time-independent metrics of the left and right black hole outside of some finite radius, while connecting those regions by some smooth geometry so that there exists a globally timelike Killing vector.

 Since the metrics $g_-$ and $g_+$ are time independent, we can separate solutions to the equations of motion in those regions into positive and negative frequencies. This allows us to define Hilbert spaces $\mathcal{H}_{g_-}$ and $\mathcal{H}_{g_+}$ as Fock spaces. Moreover, the transition amplitudes $\mathcal{H}_{g_-} \to \mathcal{H}_{g_+}$ are well defined and unitary. This is because the full deformed spacetime is smooth and time independent outside of a finite region. The former property ensures that the transition amplitude is not UV divergent from particle production at asymptotically high energies, while the latter means that there are no IR issues from particle production at asymptotic infinity. As a result, one obtains the same Hilbert space $\mathcal{H}_{\rm QFT}$ in a neighbourhood of $\Sigma$ by either evolving states forwards from $\mathcal{H}_{g_-}$ or backwards from $\mathcal{H}_{g_+}$. Since $g_+$ and $g_-$ can be independently varied, it follows that the Hilbert space constructed in this way is independent of the choice of $g_\pm$.

To be clear, the Hilbert space constructed in this way is not the only Hilbert space that one might reasonably consider. For Schwarzschild black holes, another obvious choice is to construct a GNS Hilbert space using the Hartle-Hawking state. This does not contain states that look like the Minkowski vacuum far from the black hole. Instead all states will asymptote to a thermal state at the Hawking temperature, purified by thermal modes in the other asymptotic region. In other words, far from the black hole the Hartle-Hawking GNS Hilbert space becomes the thermofield double Hilbert space described in Section 3 of \cite{witten2022does}. In quantum gravity, thermal states with infinite extent cannot exist in an asymptotically flat spacetime at finite $G$ because of backreaction issues: if the radius $r_{max}$ of the thermal bath becomes sufficiently large, the system will collapse to form a larger black hole. To construct a semiclassical limit of quantum gravity that reproduces the Hartle-Hawking GNS Hilbert space, one would therefore need to take a double scaling limit where at finite $G$ states only look like the Hartle-Hawking state at radii $r_{BH} \ll r \ll r_{max}$. After first taking $G \to 0$ one could then safely take $r_{max}\to \infty$.

A related issue is that even after taking $G \to 0$, both the expectation value and fluctuations of the ADM mass of a spacetime with quantum fields in the Hartle-Hawking state receive divergent contributions from thermal modes far from the black hole. To obtain a Type II von Neumann algebra, we need the fluctuations in the ADM mass to be finite. But, if the quantum fields are in the Hartle-Hawking state, this can only happen if the divergent fluctuations in the energy of thermal modes far from the black hole are cancelled by divergent fluctuations in the quasilocal mass of the black hole at finite distances from the horizon. This is not only  somewhat physically unnatural condition but also means that means that the ADM mass cannot be measured (even approximately) by an observer at finite radius.

For Kerr black holes, the analogue of the Hartle-Hawking state, sometimes called the Frolov-Thorne state, cannot exist at all as a normalisable state. This is because the near-horizon boost symmetry generator becomes spacelike sufficiently far from the black hole. On the other hand, the Unruh state, where ingoing modes from infinity are in the Boulwar\`{e} vacuum while outgoing near horizon modes are in the Hartle-Hawking state, does exist and one can construct a GNS Hilbert space from it. This Hilbert space was used in \cite{kudlerflam2024generalized} to construct gravitational algebras for Kerr black holes. 

Unlike perturbations of the Hartle-Hawking state, perturbations of the Unruh state do not run into gravitational backreaction issues at large radius, because only a small number of low angular momentum Hawking modes escape the near-horizon region and radiate to infinity. However, the Unruh state still has the divergent contributions to the ADM mass from modes asymptotically far from the black hole. Additionally, the Unruh state, and perturbations thereof, are singular at the white hole horizon. As a result, the Unruh state is only expected to describe a late-time approximation to equilibrating states such as black holes formed from collapse; the \emph{global} Unruh state is not expected to exist as a well defined state in nonperturbative quantum gravity.
To obtain the Hilbert space described in \cite{kudlerflam2024generalized} in a $G \to 0$ limit of quantum gravity, one would need to e.g. a) fix a semclassical method of forming a black hole from collapse, b) take the coupling $G \to 0$, and then c) take the time at which the collapse occurred to the infinite past. This can certainly be done: indeed a similar limit was described explicitly in Section 4 of \cite{CPW} in the context of asymptotically-AdS black holes. However, it is a somewhat convoluted procedure.

In contrast to the issues above, we expect that any semiclassical state $\ket{\Phi}$ in $\mathcal{H}_{\rm QFT}$ should exist in nonperturbative quantum gravity at any sufficiently small but finite $G$. By this, we mean that there should exist a state $\ket{\tilde\Phi}$ in the nonperturbative quantum gravity Hilbert space that is defined uniquely up to perturbatively small corrections and  that has the same expectation values $\ket{\Phi}$ for all semiclassical observables that do not probe either a) the singularities or b) parametrically early or late times where backreaction from the evaporation becomes large. In particular, there is no need to take any double scaling limit like the ones above. Finally, for states in $\mathcal{H}_{\rm QFT}$, the total energy of modes far from the black hole will be finite. As a result the ADM mass of the spacetime can be obtained as a limit (in the strong operator topology) of quasilocal observables.

The Hilbert space $\mathcal{H}_{\rm QFT}$ describes a perfectly good semiclassical limit of quantum gravity where all metric fluctuations are $O(\sqrt{G})$ and hence vanish as $G \to 0$. However, there is a particular mode of the metric, called the timeshift mode $T$, that describes the asymptotic right boundary time reached by starting at $t=0$ on the left boundary and travelling along a spacelike geodesic orthogonal to the time-translation Killing vector through the Einstein-Rosen bridge.\footnote{For Kerr (or Kerr-Newman) black holes, the time translation here should be replace by the isometry of the Kerr black hole that looks like a boost generator near the horizon. As described above, at asymptotic infinity this is actually becomes a combination of a time translation and a rotation and so is actually spacelike.} The operator $G^{-1/2} T$ has a finite $G \to 0$ limit acting on the Hilbert space $\mathcal{H}_{\rm QFT}$. We can therefore write 
\begin{align}\label{eq:HQFT}
    \mathcal{H}_{\rm QFT} \cong \mathcal{H}_0 \otimes L^2(\R),
\end{align}  
where $L^2(\R)$ describes the timeshift (with $G^{-1/2}T$ acting as the position operator), while ${\mathcal{H}}_0$ describes the Hilbert space of quantum fluctuations of matter and graviton fields with the timeshift frozen at zero.\footnote{There are also other similar modes associated to the rotational symmetry of the black hole solution. Those modes turn out to be comparatively unimportant. In particular their treatment does not affect the type classification of the algebra of observables; see \cite{Witten_2022,CPW} for a detailed discussion. For simplicity of presentation, we will assume that, like the timeshift mode, the rotational modes are frozen when defining the Hilbert space $\mathcal{H}_0$ and we will not include operators that excite them in our algebra. (Since the rotational modes are included in $\mathcal{H}_{\rm QFT}$, there should therefore really be an additional factor of $L^2(\R^k)$ (with $k$ the dimension of the rotation group) on the right-hand side of \eqref{eq:HQFT}.)} The timeshift mode is locally pure gauge; it only affects the relationship between times at the left and right boundaries. As a result, we can consider a Hilbert space where the timeshift mode has $O(1)$ fluctuations without having to running into issues from nonperturbative backreaction on the spacetime.
 
The timeshift mode is canonically conjugate to both the left and right ADM masses $H_{L/R}$.\footnote{Again, for Kerr black holes, the ADM masses here should be replaced by the asymptotic charges associated to the boost isometry.} At leading order in $G$, the left and right ADM mass are equal to a constant value $E_0 = O(1/G)$ determined by the mass of the black hole background. Generically, $O(G^{1/2})$ metric fluctuations lead to $O(G^{-1/2})$ fluctuations in the ADM mass. Since backreaction from matter fields and graviton excitations only changes the ADM mass at $O(1)$, this leads to an operator $X = \sqrt{G}(H_L -E_0) = \sqrt{G}(H_R -E_0)$, defined by rescaling and shifting either the left or right ADM mass, that is finite in the $G \to 0$ limit and observable at both the left and right boundaries. In fact, it is in the joint center of the left and right boundary algebras acting on the Hilbert space $\mathcal{H}$.

However, if fluctuations $\Delta T$ in the timeshift $T$ are $O(1)$, the fluctuations $\scalebox{.8}{$\Delta$} H_{L/R}$ can also be $O(1)$ without violating the uncertainty principle $\scalebox{.8}{$\Delta$} H_{L/R}\, \Delta T \geq 1/2$. We can then define renormalised ADM masses $h_{L/R} = H_{L/R} - E_0$ that have a finite $G \to 0$ limit. These act on a (different!) semiclassical Hilbert space $\mathcal{H}$ that can again be written as
\begin{align}\label{eq:hsemi}
\mathcal{H} \cong \mathcal{H}_0 \otimes L^2(\R)
\end{align}
but where the position operator on $L^2(\R)$ is now the timeshift $T$ without any rescaling. It is convenient to choose a gauge where bulk coordinates, and in particular the location of bulk QFT operators acting on $\mathcal{H}_0$, are defined relative to right boundary coordinates. This is commonly described as operators on $\mathcal{H}_0$ being ``dressed'' to the right boundary. In this gauge, the timeshift $T$ is defined as the bulk time at which the left boundary time is zero. The (renormalised) left ADM mass $h_R$ then acts only on $L^2(\R)$ as the momentum operator $-i\partial_T.$ 

In contrast, the right ADM mass acts nontrivially on the Hilbert space $\mathcal{H}_0$. The difference between the two ADM masses can be written as
\begin{align}
H_R -H_L = h_R - h_L = H
\end{align}
where $H$ acts on the quantum fields $\mathcal{H}_0$ as the boost generator (i.e. as the generator of Schwarzschild time translations in the case of a Schwarzschild black holes). For AdS-Schwarzschild black holes, where the Hartle-Hawking state is contained in the natural QFT Hilbert space and exists in quantum gravity at small but finite $G$, we have
\begin{align} \label{eq:Hboostsplit}
\beta H = -\log \Delta_{\Psi}
\end{align}
where $\Delta_{\Psi}$ is the modular operator for the Hartle-Hawking state on the right exterior QFT algebra and $\beta$ is the inverse temperature of the black hole.

For an asymptotically flat black hole, $H$ is \emph{not} directly related to $\log \Delta_{\Phi}$ for any normalisable state in $\ket{\Psi} \in \mathcal{H}$. However, given any state $\ket{\Phi} \in \mathcal{H}$ that is cyclic and separating on the right exterior, we \emph{can} write
\begin{align}\label{eq:tildesplitting}
\beta H = - \log \Delta_{\Phi} + \beta \scalebox{.8}{$\Delta$} H_{\Phi,r} - \beta \scalebox{.8}{$\Delta$} H_{\Phi,\ell},
\end{align}
where $\scalebox{.8}{$\Delta$} H_{\Phi,r}$ and $\scalebox{.8}{$\Delta$} H_{\Phi,\ell}$ are densely defined operators (and not just sesquilinear forms!) on $\mathcal{H}$ localised in the left and right exteriors respectively. The operators $\scalebox{.8}{$\Delta$} H_{\Phi,r}$ and $\scalebox{.8}{$\Delta$} H_{\Phi,\ell}$ are unique up to the addition of a common $c$-number. The splitting \eqref{eq:tildesplitting} is possible because the obstruction to writing $H$ directly as a sum of operators in the left and right exteriors comes from local physics near the horizon. And near the horizon all states $\ket{\Phi} \in \mathcal{H}$ look like the Hartle-Hawking state, so that $\log \Delta_\Phi$ and $\beta H$ act in the same way. Far from the horizon, $\beta H$ and $\log \Delta_\Phi$ of course act very differently; these differences are captured by the operators $\scalebox{.8}{$\Delta$} H_{\Phi,r}$ and $\scalebox{.8}{$\Delta$} H_{\Phi,\ell}$.

The operators appearing in \eqref{eq:tildesplitting} are unbounded and hence are only defined on a dense subset of Hilbert space. To make the above discussion completely rigorous we would need to specify those domains explicitly. Instead, however, we can rephrase things entirely in terms of bounded operators as follows. As in Section \ref{sec:inflaton}, the Hilbert space $\mathcal{H}_0$ is acted on by two Type III von Neumann factors $\mathcal{A}_{\ell}$ and its commutant $\mathcal{A}_r = \mathcal{A}_{\ell}'$ that describe bounded operators localised in the left and right exteriors respectively.\footnote{If we had not frozen the modes associated to relative rotations of the left and right boundaries when defining $\mathcal{H}_0$, these would form a common center to left and right exterior algebras. The center would disappear if we took a limit where the fluctuations in the relative rotations were $O(1)$, which leads to algebras that are the crossed product of the factors  $\mathcal{A}_{\ell}$ and $\mathcal{A}_r$ by the rotation group \cite{Witten_2022, CPW}. One can also obtain nontrivial center if the QFT Hilbert space contains multiple superselection sectors.} We claim that one can write
\begin{align} \label{eq:cclikeflow}
    e^{-i Ht}\Delta_\Phi^{-i t/\beta} = U_{\Phi,\ell}(t)\,U_{\Phi,r}(t)\,
\end{align}
where $U_{\Phi,\ell}(t) \in \mathcal{A}_{\ell}$ and $U_{\Phi,r}(t) \in \mathcal{A}_r$ are unitary operators in the left and right exteriors respectively. Because the intersection $\mathcal{A}_{\ell} \cap \mathcal{A}_r$ is trivial, the unitaries $ U_{\Phi,\ell}(t)$ and $U_{\Phi,r}(t)$ (if they exist) must be unique up to a shift of a phase between the two.  

If the Hartle-Hawking state $\ket{\Psi}$ (i.e. a state satisfying \eqref{eq:Hboostsplit}) existed as a normalisable state in the Hilbert space $\mathcal{H}$, the operators 
\begin{align}\label{eq:Uasccflows}
    U_{\Phi,\ell}(t) = \Delta_{\Psi}^{it/\beta} \Delta_{\Psi|\Phi}^{-it/\beta} = \Delta_{\Phi|\Psi}^{it/\beta} \Delta_{\Phi}^{-it/\beta}
~~~~~~\text{and}~~~~~~
U_{\Phi,r}(t) = \Delta_{\Psi}^{it/\beta} \Delta_{\Phi|\Psi}^{-it/\beta} = \Delta_{\Psi|\Phi}^{it/\beta} \Delta_{\Phi}^{-it/\beta}
\end{align}
would be Connes' cocycle flows for $\ket{\Phi}$ relative to $\ket{\Psi}$. In that case it is a standard result of Tomita-Takesaki theory that $U_{\Phi,\ell}(t) \in \mathcal{A}_{\ell}$ and $ U_{\Phi,r}(t) \in \mathcal{A}_r$ and that all the definitions given in \eqref{eq:Uasccflows} agree. Since the Hartle-Hawking state is not contained in $\mathcal{H}_0$, we cannot directly use those results. However, the reason the Hartle-Hawking state fails to be contained in $\mathcal{H}_0$ only involves its properties at asymptotic infinity, and not the behaviour of $\ket{\Psi}$ (or of the boost Hamiltonian $H$) in the near-horizon region. The latter is all we need to determine whether $e^{-i Ht} \Delta_\Phi^{-i t/\beta}$ splits into a product of operators in the left and right algebras. So we can still write \eqref{eq:cclikeflow} even though $\ket{\Psi} \not\in \mathcal{H}_0$. Taking the derivative of \eqref{eq:cclikeflow} at $t=0$ leads to our original claim \eqref{eq:tildesplitting}, with
\begin{align}
    \scalebox{.8}{$\Delta$} H_{\Phi,\ell} = \lim_{t\to 0} \frac{ U_{\Phi,\ell}(t) - 1}{i t} ~~~\text{and}~~~ \scalebox{.8}{$\Delta$} H_{\Phi,r} = \lim_{t\to 0} \frac{U_{\Phi,r}(t) - 1}{ -i t}
\end{align}
which are unbounded operators affiliated to $\mathcal{A}_{\ell}$ and $\mathcal{A}_r$ respectively. One can then easily check that
\begin{align}\label{eq:tildeUell}
    U_{\Phi,\ell}(t) &= e^{-iHt} e^{i \Lambda_{\Phi,\ell t}} = e^{-i\Lambda_{\Phi,r} t} \Delta_\Phi^{-it/\beta}\\ \label{eq:tildeUr}
U_{\Phi,r}(t) &= e^{-iHt} e^{i\Lambda_{\Phi,r} t)} = e^{-i\Lambda_{\Phi,\ell} t} \Delta_\Phi^{-it/\beta},
\end{align}
where $\Lambda_{\Phi,\ell} = H + \scalebox{.8}{$\Delta$} H_{\Phi,\ell}$ generates modular flows of $\mathcal{A}_\ell$ and boosts of $\mathcal{A}_r$, while $\Lambda_{\Phi,r} = H - \scalebox{.8}{$\Delta$} H_{\Phi,r}$ generates boosts of $\mathcal{A}_\ell$ and modular flows of $\mathcal{A}_r$. To see this, note that both formulas for $U_{\Phi,\ell}(t)$ given in \eqref{eq:tildeUell} commute with $\mathcal{A}_r$ (and hence are contained in $\mathcal{A}_\ell$) and generate the same automorphism for $\mathcal{A}_\ell$ as \eqref{eq:cclikeflow}. Meanwhile the formulas for $U_{\Phi,r}(t)$ given in \eqref{eq:tildeUell} commute with $\mathcal{A}_\ell$ and generate the correct automorphism for $\mathcal{A}_r$.

In our later discussion of Schwarzschild-de Sitter black holes, we will often, out of practical convenience, manipulate unbounded densely defined operators without being careful to define their domains. However, those manipulations should always be possible to make rigorous by similar methods to those we have just used. 

The full gravitational algebra $\mathcal{A}_R$ of right exterior observables is generated by $\mathcal{A}_r$ along with (bounded functions of) $h_R$. But, since we have already established \eqref{eq:cclikeflow}, this is the same as the crossed product algebra $\mathcal{A}_R$ generated by $\mathcal{A}_r$ along with (bounded functions of) $\Delta_\Phi$.  

An alternative way to understand this result is the following. Classically, the horizon area $A_{\rm hor}$ generates timeshifts between the left and right boundaries via the Hamiltonian flow $\exp(\{A_{\rm hor}, \cdot\} t/\beta)$ but preserves the left and right exteriors separately. Quantum mechanically, the operator $\boldsymbol{\hat A}_{\rm hor}/4G$ is UV-divergent. However the operator $ \boldsymbol{\hat A}_{\rm hor}/4G - \log \boldsymbol{\rho}_{\Phi,r}$,
with $\boldsymbol{\rho}_{\Phi,r}$ a (formal) QFT density matrix on the right exterior, is UV finite. In the $G \to 0$ limit we need to subtract a divergent constant to obtain a finite operator
\begin{align} \label{eq:hphiR}
    \beta X_{\Phi,R} = \frac{\boldsymbol{\hat A}_{\rm hor}}{4G} - \log \boldsymbol{\rho}_{\Phi,r} - \mathrm{const}. 
\end{align}
We only need to add this operator for one state $\ket{\Phi}$ because, for any pair of states $\ket{\Phi_1}$, $\ket{\Phi_2}$, the difference $(X_{\Phi_1,R} - X_{\Phi_2,R})$ is already contained in the QFT algebra $\mathcal{A}_r$. We therefore obtain the modular crossed product algebra $\mathcal{A}_R$ that we just described. In fact the operator $X_{\Phi,R}$ can be written (up to an arbitrary constant) as
\begin{align}\label{eq:hrmeaning}
    X_{\Phi,R} = h_R - \scalebox{.8}{$\Delta$} H_{\Phi,r}.
\end{align}
Interesting, the ADM mass itself did not show up anywhere in this version of the derivation. Instead, we just used the fact that \eqref{eq:hphiR} is a UV-finite operator that is localised in the right exterior. An advantage of this sort of approach is therefore that it can be applied in cosmological spacetimes where there are no asymptotic charges. 

Since the crossed product algebra $\mathcal{A}_{R}$ is a Type II$_\infty$ von Neumann factor, it has a unique semifinite trace. One way to describe this trace is as follows. We have
\begin{align}
     U_{\Phi,\ell}(T)^\dagger X_{\Phi,R} U_{\Phi,\ell}(T) =  U_{\Phi,\ell}(T)^\dagger \left( h_L +  \Lambda_{\Phi,r}\right) U_{\Phi,\ell}(T) = \Delta_\Phi^{iT/\beta} h_L \Delta_{\Phi}^{-iT/\beta} = h_L - \frac{1}{\beta} \log \Delta_\Phi.
\end{align}
We also have $U_{\Phi,\ell}(T)^\dagger \mathcal{A}_r U_{\Phi,\ell}(T) = \mathcal{A}_r$. The algebra $U_{\Phi,\ell}(T)^\dagger \mathcal{A}_R U_{\Phi,\ell}(T)$ is therefore generated by $\mathcal{A}_r$ and $\beta h_L - \log \Delta_\Phi$. Since $h_L$ acts only on $L^2(\R)$, this is a canonical description of a modular crossed product algebra. A standard formula then says that
\begin{align}\label{eq:traceflatbh1}    
\Tr(a) = \bra{0}_Te^{\beta h_L/2}\braket{\Phi|\,U_{\Phi,\ell}(T)^\dagger \,a\, U_{\Phi,\ell}(T) \,|\Phi}e^{\beta h_L/2}\ket{0}_T,
\end{align}
where the states $\ket{T_0}_T$ are delta-function normalisable position eigenstates for $ L^2(\R)$ satisfying $\braket{T_0|T_1}_T = \delta(T_0 - T_1)$. In terms of the conjugate momentum $h_L$, the state $\ket{0}_T$ is the constant wavefunction.

It turns out that there is a much more natural expression that we can instead use. There is a converse theorem \cite{connes1973classification, takesaki2003theory, Jensen_2023} to the statement that Connes' cocycle flow for a von Neumann algebra is contained in that algebra: given any operator $\beta H$ generating automorphisms of $\mathcal{A}_r$ for which it is possible to write \eqref{eq:cclikeflow} there must exist a faithful normal semifinite weight $\ket{\Psi_0}$ for $\mathcal{A}_r$ such that 
\begin{align}
    \beta H = - \log \Delta_{\Psi_0}.
\end{align} In other words, while the normalisable Hartle-Hawking state $\ket{\Psi}$ does not exist as a normalisable state in $\mathcal{H}_0$, there does exist an unnormalisable weight $\ket{\Psi_0}$ that generates boosts as modular flows. All of our work above was therefore really just reproducing the standard result that a modular crossed product algebra, despite appearances, is independent of the choice of normalisable state, or unnormalisable semifinite weight, used to define it \cite{Witten_2022}.

The fact that $\ket{\Psi_0}$ exists as a weight on $\mathcal{A}_r$ may be unfamiliar to most readers (it certainly was to the authors of this paper!) but it should not in fact be surprising. The reason that the Hartle-Hawking state cannot exist as a normalisable state on $\mathcal{H}_0$ is the same reason that the thermofield double state cannot exist as a normalisable state on two copies of the (vacuum sector) Minkowski Hilbert space $\mathcal{H}_{\rm Mink}$, namely that the Minkowski partition function $Z(\beta) = \Tr(e^{-\beta H})$ (with $H$ the Minkowski Hamiltonian) is infinite. However the linear functional $\Tr(e^{-\beta H}[\cdot])$ (just like the trace $\Tr$ itself) is still a semifinite weight on $\mathcal{H}_{\rm Mink}$. Indeed, because $H \geq 0$, any trace-class operator $a$ has finite expectation value $\Tr(e^{-\beta H}a)$. Furthermore, the linear functional $\Tr(e^{-P_i} [\cdot])$ is also semifinite, even though the momentum $P_i$ is not bounded from below, because trace-class operators that vanish exponentially fast as $P_i \to -\infty$ above some minimal value are also s.o.t. dense in $\mathcal{B}(\mathcal{H}_{\rm Mink})$. So the existence of $\ket{\Psi_0}$ as a semifinite weight on $\mathcal{A}_r$ for Kerr black holes with a modular Hamiltonian that generates boosts at the horizon (and hence generates a spacelike isometry at infinity) should also not be surprising.

It is important not to confuse the Hartle-Hawking weight $\ket{\Psi_0}$ on the algebra $\mathcal{A}_r$ with the normalised Hartle-Hawking state $\ket{\Psi}$ viewed as an element of its own GNS Hilbert space $\mathcal{H}_\Psi$. The algebra $\mathcal{A}_{\Psi,r}$ of right exterior observables on $\mathcal{H}_\Psi$ is fundamentally different from $\mathcal{A}_r$, just like the Type III algebra $\mathcal{A}_{TFD}$ of operators on one copy of Minkowski space in the thermofield double Hilbert space $\mathcal{H}_{\rm TFD}$ is fundamentally different from the Type I algebra of operators $\mathcal{B}(\mathcal{H}_{\rm Mink})$. Even when an operator like the identity is apparently in both algebras, we have e.g. $\braket{\Psi_0|\mathds{1}|\Psi_0} = \infty$, while $\braket{\Psi|\mathds{1}|\Psi} = 1$. In fact, we do not expect that any operators $a \in \mathcal{A}_r$ with $\braket{\Psi_0|a|\Psi_0}$ finite make sense as operators on $\mathcal{H}_{\Psi}$.

Given the existence of the semifinite weight $\ket{\Psi_0}$, we can write the trace on $\mathcal{A}_R$ is a more natural form as
\begin{align}\label{eq:traceflatbh2}
    \Tr(a) = \int dh_L e^{\beta h_L} \braket{\Psi_0|a|\Psi_0},
\end{align}
where $\braket{\Psi|a|\Psi}$ is interpreted as a function of $h_L$. Since the trace on a Type II$_\infty$ von Neumann factor is unique up to rescaling, \eqref{eq:traceflatbh2} must be proportional to \eqref{eq:traceflatbh1}. In fact, one can show that they are the same \cite{CPW}. 

Given that the final conclusion of this section was that the only difference between asymptotically flat and asymptotically AdS black holes is whether the Hartle-Hawking state is a normalisable state or merely a semifinite weight, it might seem peculiar that we included such a long discussion of it at all. The reason is that our strategy above will play a crucial role in understanding the gravitational algebras for black holes in de Sitter space, to which we now turn. And for black holes in de Sitter space, there does not exist even a semifinite weight whose modular Hamiltonian generates boosts, because the black hole and cosmological horizons have different temperatures.

 \subsection{Schwarzschild-de Sitter black holes} \label{sec:sds}
 Roughly speaking, black holes in de Sitter space combine all the ingredients from our studies of both slow-rolling inflatons in Section \ref{sec:inflaton} and asymptotically flat black holes in Section \ref{sec:bhflat}. They feature a boost symmetry that needs to be imposed as a gauge constraint that remains out-of-equilbrium for parametrically long times. But they also feature a timeshift mode that can have $O(1)$ fluctuations which are not described by quantum field theory in a fixed spacetime background. As a result, their analysis is considerably more technical than either; the rest of this section should probably not be read without first reading those. Black holes in de Sitter space were previously considered in \cite{kudlerflam2024generalized}, but with significant differences that we discuss further below. In particular, the analysis in \cite{kudlerflam2024generalized} was based on the GNS Hilbert space for the Unruh state, and made use of both asymptotic gravitational charges and of an observer with a clock.
 
\begin{figure}
  \centering
  \begin{subfigure}[t]{0.4 \textwidth}
      \includegraphics[width=8cm]{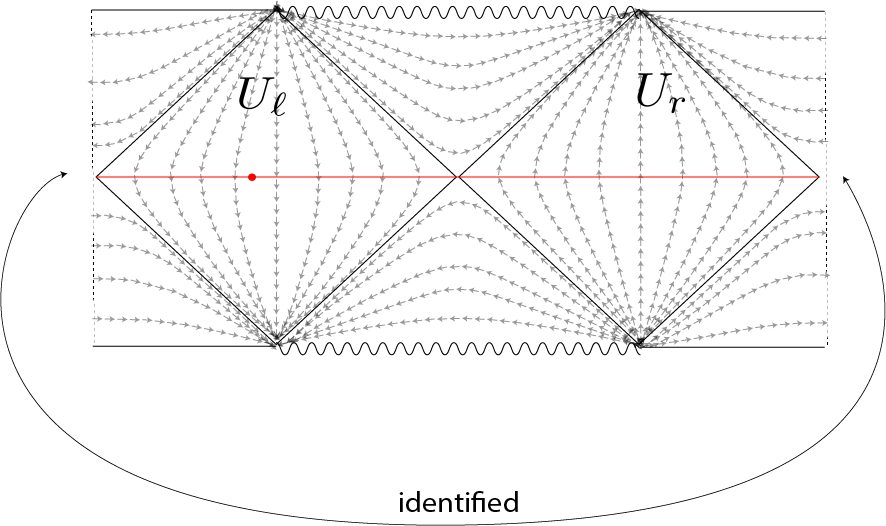} 
      \caption{}
  \end{subfigure}
  \hfill
  \begin{subfigure}[t]{0.4\textwidth}
      \includegraphics[width=8 cm]{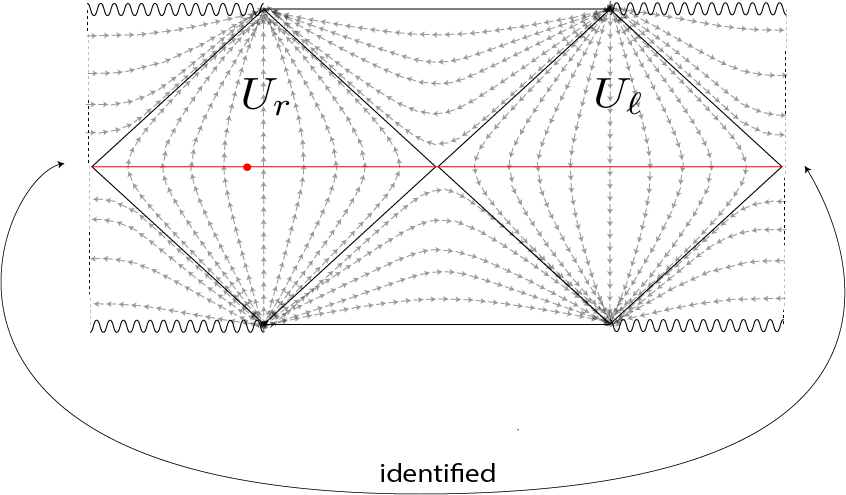} 
      \caption{}
  \end{subfigure}
  \hfill
 \caption{\label{fig:sds}Penrose diagrams for Schwarzschild-de Sitter black hole, with the red slice being a spacelike geodesics that winds around the closed universe. The left and right edges are periodically identified as shown. The dashed lines indicate the direction of the boosts. On the left, the black hole bifurcation surface is placed in the middle, while on the right the cosmological bifurcation surface is placed there.}
 \end{figure}

 We focus for simplicity on the case of Schwarzschild-de Sitter (SdS) black holes  but the analysis can be easily extended to e.g. Kerr-Newman-de Sitter black holes. Specifically, we focus on a spatially compact, two-sided Schwarzschild-de Sitter solution where the left and right black hole exteriors $U_\ell$ and $U_r$ are glued together at the cosmological horizon. This is shown in Figure \ref{fig:sds}. The metric of the SdS spacetime in Schwarzschild coordinates is
 \begin{align}
     ds^2 = - f(r)dt^2 + f(r)^{-1} dr^2 + r^2 d\Omega_2^2
 \end{align}
 where $f(r) = 1 - r_s/r -  r^2/ \ell_{dS}^2$. These coordinates are singular at the black hole and cosmological horizons.

\begin{figure}
  \centering
  \begin{subfigure}[t]{0.4 \textwidth}
      \includegraphics[width=8cm]{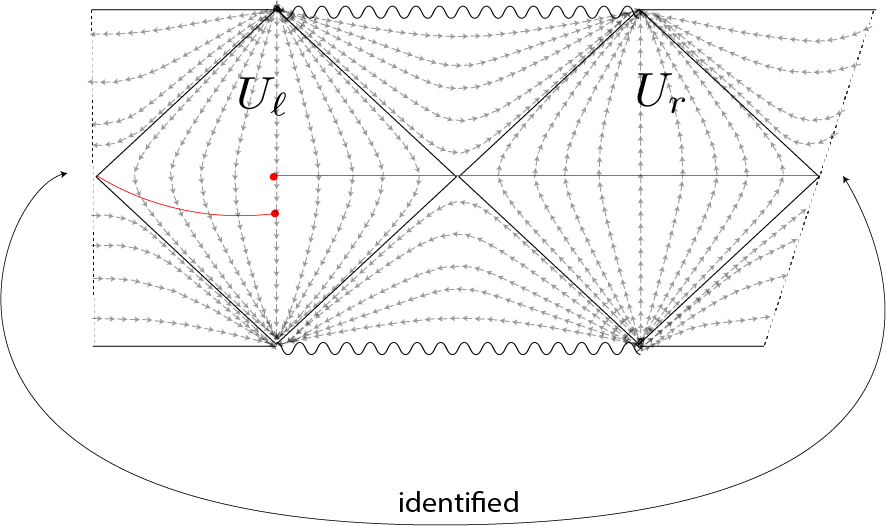} 
      \caption{}
  \end{subfigure}
  \hfill
  \begin{subfigure}[t]{0.4\textwidth}
      \includegraphics[width=8 cm]{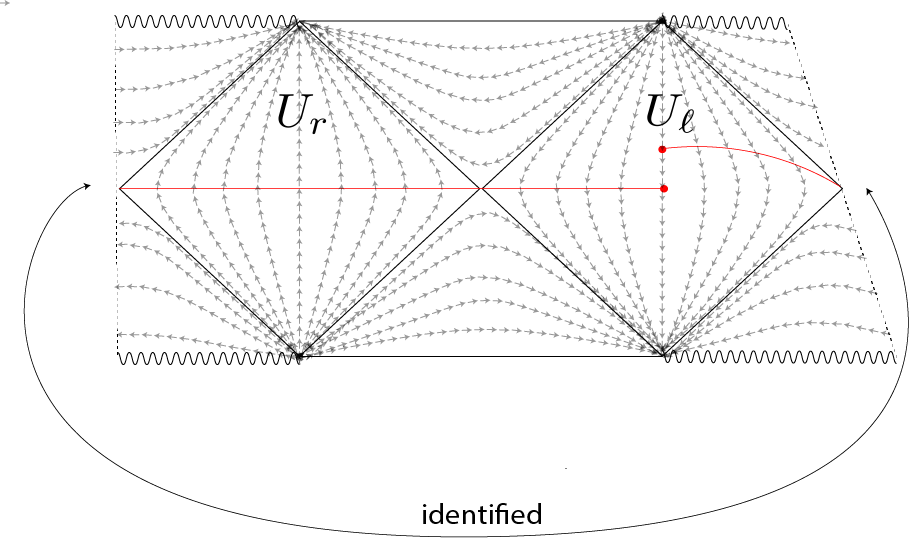} 
      \caption{}
  \end{subfigure}
  \hfill
 \caption{\label{fig:sdst}Penrose diagrams for time-shifted Schwarzschild-de Sitter black hole. This time, the left and right edges are periodically identified up to a time-shift. The effect is that a geodesic, such as the red slice, gets boosted at the bifurcation surface at the edges, {\it i.e.,} picks up a non-zero boost angle.}
 \end{figure}

 As usual, there is a $G \to 0$ limit of an SdS black hole where the quantum gravity Hilbert space becomes, up to gauge constraints, the QFT Hilbert space describing perturbations of matter fields and gravitons on a fixed classical SdS spacetime background. This Hilbert space is unambiguous because the SdS background has compact Cauchy slices. However, as in Section \ref{sec:bhflat}, we can consider an alternative limit, where a locally pure-gauge metric mode has $O(1)$, rather than $O(\sqrt{G})$, fluctuations. In this case, that mode is the shift $T$ in boost time $t$ from travelling on a static cycle through the black hole and cosmological horizons; see Figure \ref{fig:sdst}.\footnote{As in Section \ref{sec:bhflat}, there are additional locally pure-gauge modes associated to rotations from travelling around the nontrivial cycle. The treatment of these modes is again inessential to our discussion, and we will assume for simplicity that they are frozen out. In a Kerr-de Sitter black hole, however, the rotational mode would need to be treated more carefully, however, because the boost generators at the black hole and cosmological horizons differ by a rotation generator.}

 If we allow $O(1)$ fluctuations in the timeshift, the  Hilbert space becomes the direct integral
 \begin{align} \label{eq:directintegral}
     \mathcal{H} = \int^{\oplus}_{\R} dT\, \mathcal{H}_T
 \end{align}
 where $\mathcal{H}_T$ is the QFT Hilbert space for an SdS black hole background with fixed timeshift $T$. Unlike for asymptotically flat black holes, there is no (global) isometry between SdS black holes with different timeshifts that can be used to (almost) canonically identify the Hilbert spaces $\mathcal{H}_T$.\footnote{There were two (or perhaps three) natural identifications we could have made: fixing the bulk coordinates to match the right boundary coordinates, the left boundary coordinates or possibly an average of the two. All three choices of identification, in slightly different language, were discussed in \cite{Witten_2022}.} Of course, since they are all separable infinite-dimensional Hilbert spaces, the Hilbert spaces $\mathcal{H}_T$ are in principle isomorphic: we simply choose a countable basis for each and define a unitary to map one basis to the other. For practical purposes, it is convenient to pick a (hopefully somewhat less arbitrary) choice of identification $\mathcal{H}_T \cong \mathcal{H}_0$ so that we can write \begin{align}
     \mathcal{H} \cong \mathcal{H}_0 \otimes L^2(\R),
 \end{align}
 with the timeshift $T$ again acting as the position operator on $L^2(\R)$.
 
 In classical general relativity, a natural choice would be to identify the (perturbative) phase spaces for different timeshifts $T$ as follows.  First define a Cauchy slice $\Sigma_0$ for the zero-timeshift spacetime that is defined by $t=0$ for some choice of boost time $t$ on each exterior and that is smooth at the bifurcation surfaces. Then for $T \neq 0$, we pick a slice $\Sigma_T$ that is locally diffeomorphic to $\Sigma_0$ everywhere except at the cosmological horizon, where there will necessarily be a kink with boost angle $T/\beta_{CH}$ (with $\beta_{CH}$ the inverse temperature of cosmological horizon). (We will similarly denote the inverse temperature of the black hole horizon by $\beta_{BH}$.) We then identify classical states of the matter (and graviton) fields for different timeshifts $T$ and $T'$ if their field configurations on the respective Cauchy slices $\Sigma_T$ and $\Sigma_{T'}$ are the same. With this identification, the area $A_{CH}$ of the cosmological horizon generates a Hamiltonian flow $\exp(\{A_{CH}/4G, \cdot \}\Delta T/\beta_{CH})$ on the phase space of general relativity that increases the timeshift by $\Delta T$ while leaving the matter fields unchanged.

 In quantum field theory, however, a transformation that creates a kink at the cosmological bifurcation surface while leaving the state of the quantum fields unchanged is singular because it locally acts like a one-sided boost. As we discussed in Section \ref{sec:bhflat}, the singular behaviour of a one-sided boost is closely related to the fact that the quantum operator $\boldsymbol{\hat A}_{CH}/4G$ is UV-divergent. To make a nonsingular operator that we can use to identify the Hilbert spaces $\mathcal{H}_t$ we need to add $-\log \boldsymbol{\rho}$ with $\boldsymbol{\rho}$ a formal density matrix on an appropriate algebra.

 But what is the appropriate algebra? A naive choice would be e.g. the algebra $\mathcal{A}_r$ of operators on the right exterior $U_r$. But density matrices on this algebra have divergences both from modes near the cosmological horizon and from modes near the black hole horizon. The latter cannot be cancelled solely by adding $\boldsymbol{\hat A}_{CH}/4G$. So $\mathcal{A}_r$ (or the algebra $\mathcal{A}_\ell$ of operators in the left exterior) won't work.

\begin{figure}
  \centering
  \begin{subfigure}[t]{0.25\textwidth}
      \begin{tikzpicture}[scale =1]
     \pgftext{\includegraphics[scale =.25]{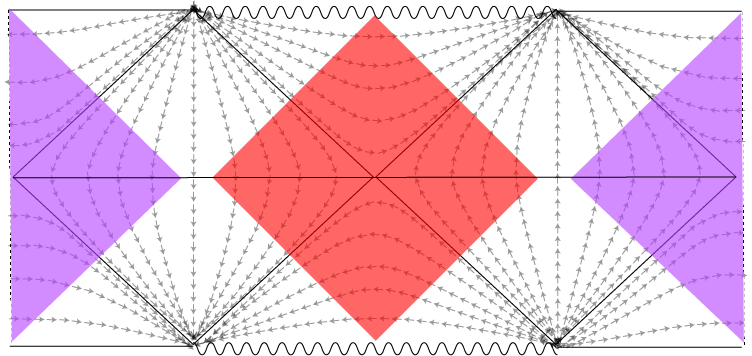} 
     \draw (-3.9,2.2) node {$U_{BH}$};
     \draw (-1.1,2.2) node {$U_{CH}$};}
     \end{tikzpicture}
     \caption{}
  \end{subfigure}
  \hspace{5 cm}
\begin{subfigure}[t]{0.25\textwidth}
      \begin{tikzpicture}[scale =1]
     \pgftext{\includegraphics[scale =.27]{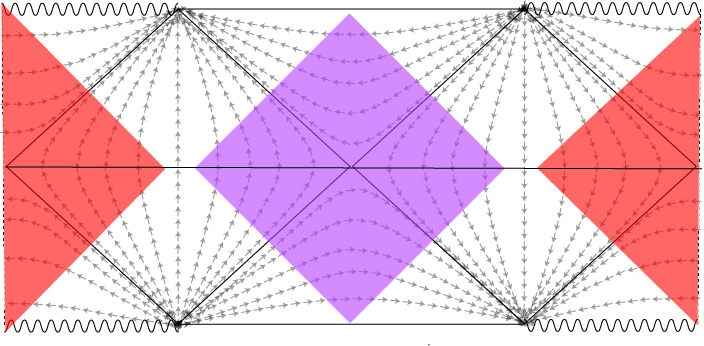} 
      \draw (-4.,2.2) node {$U_{CH}$};
     \draw (-1.2,2.2) node {$U_{BH}$};
}
 \end{tikzpicture}
 \caption{}
\end{subfigure}
 \caption{\label{fig:wedgenot} The black hole and cosmological wedges with zero timeshift. The black hole wedge $U_{BH}$ is a causal diamond with edges in the Cauchy slice $\Sigma_0$. The cosmological wedge $U_{CH}$ is spacelike separated from $U_{BH}$ but contains almost its entire causal complement. We can factorise the Hilbert space $\mathcal{H}_0 \cong \mathcal{H}_{BH} \otimes \mathcal{H}_{CH}$ so that operators in $U_{BH}$ act only on $\mathcal{H}_{BH}$ while operators in $U_{CH}$ act only on $\mathcal{H}_{CH}$.}
 \end{figure}

Instead, we first use the split property of quantum field theory \cite{split1,split2,split3,split4,split5} to write the zero-timeshift Hilbert space as 
\begin{align}\label{eq:zerotimeshiftHdecomp}
\mathcal{H}_0 \cong \mathcal{H}_{BH} \otimes \mathcal{H}_{CH}
\end{align}
such that operators within a causal diamond $U_{BH}$ containing the black hole bifurcation surface act only on $\mathcal{H}_{BH}$ while operators acting within an almost complementary causal diamond $U_{CH}$ containing the cosmological bifurcation surface act only on $\mathcal{H}_{CH}$; see Figure \ref{fig:wedgenot}. We call $U_{BH}$ and $U_{CH}$ the black hole and cosmological wedges respectively. A decomposition \eqref{eq:zerotimeshiftHdecomp} can be found in reasonable quantum field theories so long as $U_{BH}$ and $U_{CH}$ are spacelike separated, even if there is an arbitrarily small gap between them. For convenience, we can choose this splitting to be invariant under rotations so that we can write the total angular momentum $L_i$ of the quantum fields as 
\begin{align}
    L_i = L_{BH,i} + L_{CH,i}
\end{align} 
with $L_{BH,i} \in \mathcal{B}(\mathcal{H}_{BH})$ and $L_{CH,i} \in \mathcal{B}(\mathcal{H}_{CH})$.
We can also choose it to be invariant under the discrete antiunitary symmetry 
\begin{align}\label{eq:Jdef}
    J = J_{BH} + J_{CH}
\end{align}
that exchanges the left and right exteriors $U_\ell$ and $U_r$ while also reflecting time about $t = 0$. (This symmetry is the de Sitter version of the CRT symmetry of Minkowski space.) It cannot however be made boost invariant, because the only boost-invariant causal diamonds in the SdS spacetime are $U_\ell$ and $U_r$.

 There are Type I von Neumann factors $\mathcal{B}(\mathcal{H}_{BH})$ and $\mathcal{B}(\mathcal{H}_{CH})$ that contain all bounded operators acting on only $\mathcal{H}_{BH}$ or $\mathcal{H}_{CH}$ respectively. Each of these algebras contains Type III subfactors $\mathcal{A}_{BH,r}$ and $\mathcal{A}_{CH,r}$ of operators localised within the right black hole exterior $U_r$. The commutants of these algebras, acting respectively on $\mathcal{H}_{BH}$ and $\mathcal{H}_{CH}$, are the algebras $\mathcal{A}_{BH,\ell}$ and $\mathcal{A}_{CH, \ell}$ consist of all operators on their respect Hilbert spaces that are localised in the left black hole exterior $U_\ell$. The full algebras $\mathcal{A}_\ell$ and $\mathcal{A}_r$ associated to the left and right black hole exteriors $U_\ell$ and $U_r$ can then be written as 
 \begin{align}
     \mathcal{A}_{\ell} \cong \mathcal{A}_{BH,\ell} \otimes \mathcal{A}_{CH,\ell} ~~~\text{and} ~~~\mathcal{A}_r \cong \mathcal{A}_{BH,r} \otimes \mathcal{A}_{CH,r}.
 \end{align} 
 Unlike density matrices on $\mathcal{A}_\ell$ and $\mathcal{A}_r$, density matrices on $\mathcal{A}_{CH,\ell}$ and $\mathcal{A}_{CH,r}$ only feature divergences from modes near the cosmological bifurcation surface. They are therefore much more hopeful candidates to regularise the area operator.
 
 Importantly, because the left and the right wedges are boost-invariant, while the black hole and the cosmological wedges $U_{BH}$ and $U_{CH}$ are not, the boost Hamiltonian $H_0$ generates automorphisms of $\mathcal{A}_{\ell}$ and $\mathcal{A}_r$ but couples $\mathcal{H}_{BH}$ and $\mathcal{H}_{CH}$ so that e.g. $\mathcal{A}_{BH,r}$ and $\mathcal{A}_{CH,r}$ mix. We denote the image of those algebras under a boost by
 \begin{align}
     \mathcal{A}_{BH,r}^{(t)} \cong e^{iH_0 t} \mathcal{A}_{BH,r} e^{-iH_0 t}~~~\text{ and } ~~~\mathcal{A}_{CH,r}^{(t)} \cong e^{iH_0 t} \mathcal{A}_{CH,r} e^{-iH_0 t}
 \end{align}
 respectively.\footnote{Note that we have labelled the boost Hamiltonian $H_0$ rather than $H$ as in Section \ref{sec:bhflat}. This is because, once we identify $\mathcal{H}_T$ with $\mathcal{H}_0$, the operator $H_0$ will generate boosts on the zero-timeshift Hilbert space $\mathcal{H}_0$ but not on the Hilbert spaces $\mathcal{H}_T$ for nonzero $T$.}

\begin{figure}
  \centering
\begin{tikzpicture}[scale =1]
     \pgftext{\includegraphics[scale =.25]{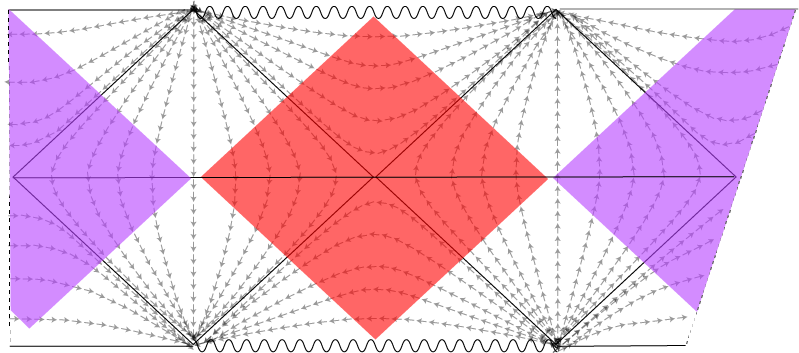} 
     \draw (-4.4,2.2) node {$U_{BH}$};
     \draw (-1.6,2.2) node {$U_{CH,T}$};
     \draw (-1.1,1.5) node {$\Sigma_T$};
     
     \hspace{2 cm}

     \includegraphics[scale =.27]{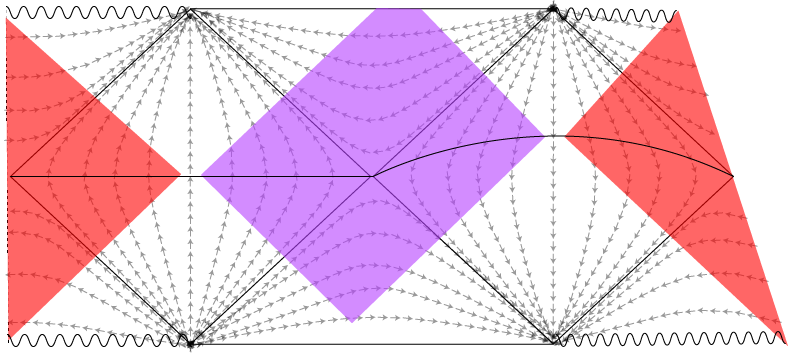} 
      \draw (-4.7,2.2) node {$U_{CH,T}$};
     \draw (-2,2.2) node {$U_{BH}$};
          \draw (-1.1,1.7) node {$\Sigma_T$};
}
 \end{tikzpicture}
 \caption{\label{fig:wedge} The black hole and cosmological wedges in a spacetime with non-zero timeshift. We define the black hole wedge $U_{BH}$ to be locally indistinguishable from the black hole wedge with zero-timeshift. Consequently, the cosmological wedge $U_{CH}(t)$ needs to locally look like its right edge has been boosted forwards in time by the timeshift $T$ relative to the zero-timeshift cosmological wedge $U_{CH}$ so that its edges are both spacelike separated from $U_{BH}$.}
 \end{figure}
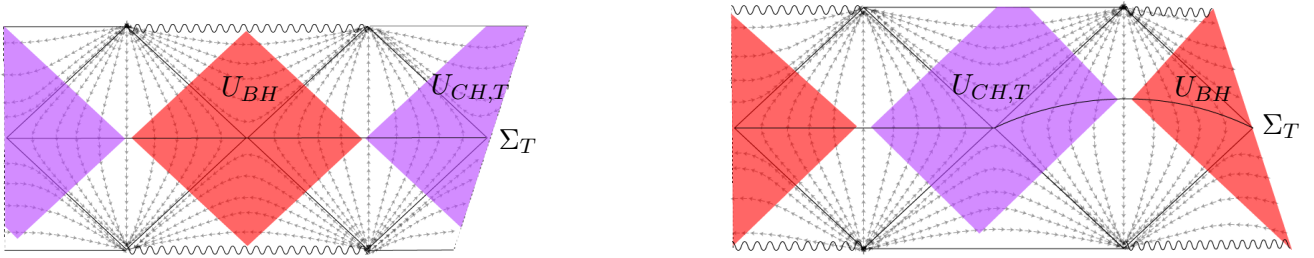

What about the Hilbert spaces $\mathcal{H}_T$ for timeshifts $T \neq 0$? Locally, the time shift is not detectible. So identifying Cauchy slices $\Sigma_T$ for different $T$ allows us identify the causal complement of the cosmological bifurcation surface, and in particular to identify diffeomorphic wedges $U_{BH}$ for SdS spacetimes with different timeshifts $T$. In turn, this leads to an action of the Type I von Neumann factor $\mathcal{B}(\mathcal{H}_{BH})$ on the Hilbert space $\mathcal{H}_T$ for any $T$.  However $\mathcal{H}_T \not\cong \mathcal{H}_{BH} \otimes \mathcal{H}_{CH}$ because any causal diamond in the timeshifted SdS spacetime that we could locally identify with $U_{CH}$ will end up partially timelike-separated from $U_{BH}$. As a result, operators in  $\mathcal{B}(\mathcal{H}_{BH})$ would not commute with operators in $\mathcal{B}(\mathcal{H}_{CH})$. Suppose that we instead define a region $U_{CH,T}$ in the timeshifted spacetime by fixing the location of its edges in $\Sigma_T$, as shown in Figure \ref{fig:wedge}, so that by construction it is spacelike separated from $U_{BH}$. Then locally $U_{CH,T}$ will look like we have taken the right edge of $U_{CH}$ and boosted forwards in time by $T$ relative to the left edge. We therefore have a natural identification 
\begin{align}
\mathcal{H}_T \cong \mathcal{H}_{BH} \otimes \mathcal{H}_{CH,T}
\end{align}
where $\mathcal{B}(\mathcal{H}_{CH,T})$ can be identified with the algebra of operators generated by $\mathcal{A}_{CH,\ell}$ and $\mathcal{A}_{CH,r}^{(T)}$.\footnote{The algebra generated by $\mathcal{A}_{CH,\ell}$ and $\mathcal{A}_{CH,r}^{(T)}$ must be a Type I factor because we can identify it as the commutant of the Type I factor $\mathcal{B}(\mathcal{H}_{BH})$ on the Hilbert space $\mathcal{H}_T$.} 

To construct an isomorphism $\mathcal{H}_T \cong \mathcal{H}_0$ it therefore remains to construct an isomorphism $\mathcal{H}_{CH,T} \cong \mathcal{H}_{CH}$.
Suppose there existed a splitting of the boost Hamiltonian 
\begin{align} \label{eq:formalHamsplit}
    H_0 = \boldsymbol{h}_{r,0} - \boldsymbol{h}_{\ell,0}
\end{align}
into one-sided boost generators $\boldsymbol{h}_{r,0}$ and $\boldsymbol{h}_{\ell,0}$ that are localised to the right and left exteriors respectively. Then, since 
\begin{align}
    \mathcal{A}_{CH,r}^{(T)} = e^{\i H_0 T}\mathcal{A}_{CH,r}e^{-\i H_0 T} = e^{\i \boldsymbol{h}_{r,0} T}\mathcal{A}_{CH,r}e^{-\i \boldsymbol{h}_{r,0} T},
\end{align} 
we would have 
\begin{align}\label{eq:naiveHCHTid}
    \mathcal{B}(\mathcal{H}_{CH,T}) \cong e^{\i \boldsymbol{h}_{r,0} T} \mathcal{B}(\mathcal{H}_{CH}) e^{-\i \boldsymbol{h}_{r,0} T}.
\end{align} 
However, as discussed around \eqref{eq:Sec2formalHamsplit}, $\boldsymbol{h}_{r,0}$ and $ \boldsymbol{h}_{\ell,0}$ only exist as sesquilinear forms mapping pairs of bras and kets to expectation values and not as densely defined operators. They therefore cannot be exponentiated. Indeed, attempting to identify $\mathcal{H}_{CH,T} \cong \mathcal{H}_{CH}$ via \eqref{eq:naiveHCHTid} is exactly the same strategy that we already tried, and failed, to use to identify $\mathcal{H}_{T} \cong \mathcal{H}_{0}$ -- namely localising the timeshift at the cosmological horizon while leaving the QFT state otherwise unchanged.

What we need is a nonsingular unitary $U(T)$ such that 
\begin{align}
    U(T)^\dagger \mathcal{A}_{CH,r} U(T) = \mathcal{A}_{CH,r}^{(T)}
\end{align}
while
\begin{align}\label{eq:preserveAchl}
    U(T)^\dagger \mathcal{A}_{CH,\ell}U(T) = \mathcal{A}_{CH,\ell}.
\end{align}
The choice $U(T) = e^{-\i \boldsymbol{h}_{r,0} T}$ didn't work because it was singular. Meanwhile the choice $U(T) = e^{-\i H_0 T}$ doesn't work because it doesn't satisfy \eqref{eq:preserveAchl}. Suppose we fix some choice of states $\ket{\Phi_{BH}} \in \mathcal{H}_{BH}$ and $\ket{\Phi_{CH}} \in \mathcal{H}_{CH}$. Let $\Delta_{\Phi_{BH}}$ and $\Delta_{\Phi_{CH}}$ be the corresponding modular operators on $\mathcal{A}_{BH,r}$ and $\mathcal{A}_{CH,r}$ respectively. As in our choice of $\mathcal{H}_{BH}$ and $\mathcal{H}_{CH}$, it is convenient to choose $\ket{\Phi_{BH}}$ and $\ket{\Phi_{CH}}$ to be rotation invariant, so that
\begin{align}
    L_{BH,i} \ket{\Phi_{BH}} = L_{CH,i}\ket{\Phi_{CH}} = 0
\end{align}
and to be canonically purifications with respect to the symmetry $J$ so that their respective modular conjugation operators are $J_{\Phi_{BH}} = J_{BH}$ and $J_{\Phi_{CH}} = J_{CH}$.  

While the splitting \eqref{eq:naiveHCHTid} does not exist, we can, in close analogy with \eqref{eq:tildesplitting}, write
\begin{align}\label{eq:hamsplitting}
H_0 = -\frac{1}{\beta_{BH}} \log \Delta_{\Phi_{BH}} -\frac{1}{\beta_{CH}} \log \Delta_{\Phi_{CH}} + \scalebox{.8}{$\Delta$} H_{\Phi,r} - \scalebox{.8}{$\Delta$} H_{\Phi,\ell}
\end{align}
with $\scalebox{.8}{$\Delta$} H_{\Phi,\ell}$ and $\scalebox{.8}{$\Delta$} H_{\Phi,r}$ densely defined unbounded operators affiliated to $\mathcal{A}_\ell$ and $ \mathcal{A}_r$ respectively. As in \eqref{eq:tildesplitting}, $\scalebox{.8}{$\Delta$} H_{\Phi,\ell}$ and $\scalebox{.8}{$\Delta$} H_{\Phi,r}$ are uniquely defined up to the addition of a common $c$-number. Equivalently, there exist densely defined operators
\begin{align} \label{eq:firstlambdadef}
    \Lambda_{\Phi,\ell} = H_0 + \scalebox{.8}{$\Delta$} H_{\Phi,\ell} ~~~\text{and}~~~ \Lambda_{\Phi,r} = H_0 - \scalebox{.8}{$\Delta$} H_{\Phi,r}
\end{align}
that generate boosts of $\mathcal{A}_r$ (respectively $\mathcal{A}_\ell$) and modular flows of $\mathcal{A}_{BH,\ell}$ and $\mathcal{A}_{CH,\ell}$ (respectively $\mathcal{A}_{BH,r}$ and $\mathcal{A}_{CH,r}$).

As we did for the boost Hamiltonian $H_0$ in \eqref{eq:formalHamsplit}, we can write
\begin{align} \label{eq:modopsplit}
 -\log \Delta_{\Phi_{CH}} = \beta_{CH} \boldsymbol{h}_{\Phi_{CH},r} - \beta_{CH} \boldsymbol{h}_{\Phi_{CH},\ell}
\end{align}
with $\boldsymbol{h}_{\Phi_{CH},\ell}$ and $\boldsymbol{h}_{\Phi_{CH},r}$ sesquilinear forms on $\mathcal{H}_{CH}$ that are localised in the left and right exteriors respectively.\footnote{One way to define $\boldsymbol{h}_{\Phi_{CH},\ell}$ and $\boldsymbol{h}_{\Phi_{CH},r}$ is by 
\begin{align}
    \beta_{CH}\braket{\Phi'|  \boldsymbol{h}_{\Phi_{CH},r}|\Phi'} = -\braket{\Phi'| \log \Delta_{\Phi_{CH}|\Phi'}|\Phi'}  ~~~\text{and}~~~
    \beta_{CH}\braket{\Phi'|  \boldsymbol{h}_{\Phi_{CH},\ell}|\Phi'} = \braket{\Phi'| \log \Delta_{\Phi'|\Phi_{CH}}|\Phi'}.
\end{align}
Since taking a derivative of Connes' cocycle flow leads to 
\begin{align}
\log \Delta_{\Phi_{CH}|\Phi'} + \log \Delta_{\Phi'|\Phi_{CH}} = \log \Delta_{\Phi_{CH}} + \log \Delta_{\Phi'},
\end{align} 
we can then obtain the desired relation \eqref{eq:modopsplit} via the standard identity $\braket{\Phi'|\log \Delta_{\Phi'} |\Phi'} = 0$.} Similarly, we can write
\begin{align}\label{eq:deltaphiBHsplit}
     -\log \Delta_{\Phi_{BH}} = \beta_{BH} \boldsymbol{h}_{\Phi_{BH},r} - \beta_{BH} \boldsymbol{h}_{\Phi_{BH},\ell}.
\end{align}
We can then write e.g.
\begin{align} \label{eq:lambda}
   \Lambda_{\Phi,\ell} = \boldsymbol{h}_{r,0} - \boldsymbol{h}_{\Phi_{BH},\ell} - \boldsymbol{h}_{\Phi_{CH},\ell}.
\end{align}
Naively one might think that $\Lambda_{\Phi,\ell}$ is also only a sesquilinear form. However, in a small neighbourhood of the black hole bifurcation surface, with size $L \ll r_s$, there locally exists a Minkowski vacuum state $\ket{\Psi_{BH}}$ with modular operator $\Delta_{\Psi_{BH}} = e^{-\beta_{BH} H_0}$. Near this horizon, the operator $\beta_{BH} \Lambda_{\Phi,\ell}$ acts like the relative modular Hamiltonian $-\log \Delta_{\Psi_{BH}|\Phi_{BH}}$ which is a densely defined operator. Similarly, in a small neighbourhood of the cosmological horizon with size $L \ll \ell_{dS}$, there locally exists a Minkowski vacuum $\ket{\Psi_{CH}}$ with $\Delta_{\Psi_{CH}} = e^{-\beta_{CH} H_0}$ and the operator $\beta_{CH} \Lambda_{\Phi,\ell}$ acts like the relative modular Hamiltonian $-\log \Delta_{\Psi_{CH}|\Phi_{CH}}$. In other words, the singular behaviour of the one-sided boost of $\mathcal{A}_{r}$ is cancelled out by the singular one-sided modular flows on $\mathcal{A}_{BH,\ell}$ and $\mathcal{A}_{CH,\ell}$. Since the two horizons are the only locations where potential divergences in \eqref{eq:lambda} could occur, it follows that $\Lambda_{\Phi,\ell}$, and hence also $\scalebox{.8}{$\Delta$} H_{\Phi,\ell}$ via \eqref{eq:firstlambdadef}, exist as densely defined operators. Similar arguments apply to $\Lambda_{\Phi,r}$ and $\scalebox{.8}{$\Delta$} H_{\Phi,r}$.

Since the modular flow $\Delta_{\Phi_{CH}}^{iT/\beta_{CH}}$ preserves $\mathcal{A}_{CH,\ell}$, we have
\begin{align}\label{eq:H_CHtId}
    \mathcal{B}(\mathcal{H}_{CH,T}) \cong e^{i \Lambda_{\Phi,\ell} T} \mathcal{B}(\mathcal{H}_{CH}) e^{-i\Lambda_{\Phi,\ell} T}.
\end{align}
This gives our desired identification of $\mathcal{H}_{CH,T}$ with $\mathcal{H}_{CH}$ and hence trivialisation of the direct integral \eqref{eq:directintegral} as
\begin{align}
    \mathcal{H} \cong \mathcal{H}_0 \otimes L^2(\R),
\end{align} 
with the timeshift $T$ acting as the position operator on $L^2(\R)$. Note that this trivialisation only depends on the choice of state $\ket{\Phi_{CH}} \in \mathcal{H}_{CH}$ and not on $\ket{\Phi_{BH}} \in \mathcal{H}_{BH}$. In effect, the identification \eqref{eq:H_CHtId} cancels the singular behaviour of \eqref{eq:naiveHCHTid} with an equally one-sided modular flow of $\mathcal{A}_{CH,\ell}$.

We already saw that the classical area of the cosmological bifurcation surface generates a timeshift localised at that surface via Hamiltonian flow. Consequently, if we had used the singular identification $\mathcal{H}_T \cong \mathcal{H}_0$ induced by \eqref{eq:naiveHCHTid}, the quantum operator $\boldsymbol{\hat A}_{CH}$ describing that area would generate timeshifts while leaving the state of the quantum fields unchanged. More precisely, \eqref{eq:naiveHCHTid} would lead to an identification of $X= - i\partial_T \in \mathcal{B}(L^2(\R))$ with $(\boldsymbol{\hat A}_{CH} - A_{CH,0})/4G \beta_{CH}$, where we have subtracted the area $A_{CH,0}$  of the cosmological horizon in the classical background to remove the divergence as $G \to 0$. Because we instead used the identification \eqref{eq:H_CHtId}, which featured an additional one-sided modular flow of $\mathcal{A}_{CH,\ell}$ to make the isomorphism nonsingular, we actually have 
\begin{align} \label{eq:ptinterpretation}
\beta_{CH}\, X_\Phi= \frac{\boldsymbol{\hat A}_{CH} - A_{CH,0}}{4G} + \beta_{CH} \boldsymbol{h}_{\Phi_{CH},\ell}  = \frac{\boldsymbol{\hat A}_{CH}}{4G} - \log \boldsymbol{\rho}_{\Phi_{CH},\ell} -\mathrm{const}.
\end{align}
We have added a subscript to $X_\Phi= - i\partial_T \in \mathcal{B}(L^2(\R))$ to indicate that the identification \eqref{eq:H_CHtId}, which was crucial to its physical interpretation, depended on the arbitrary choice of state $\ket{\Phi_{CH}}$. In the last equality of \eqref{eq:ptinterpretation}, we used the fact that the formal density matrix $\boldsymbol{\rho}_{\Phi_{CH},\ell}$ satisfies  
\begin{align}
    -\log \boldsymbol{\rho}_{\Phi_{CH},\ell} = \beta_{CH}\boldsymbol{h}_{\Phi_{CH},\ell} + S({\Phi_{CH}}),
\end{align} 
with $S(\Phi_{CH})$ the divergent entropy of $\ket{\Phi_{CH}}$ on $\mathcal{A}_{CH,\ell}$, in order to write \eqref{eq:ptinterpretation} in a form analogous to \eqref{eq:hphiR}.

 \subsection{The algebras of observables}
 
We now turn to describe the algebra $\widetilde{\mathcal{A}}_{R}$ of quantum gravity operators localised in the right black hole exterior $U_r$. In order to be gauge-invariant, operators in $\widetilde{\mathcal{A}}_{R}$ need to commute with the boosts and rotations of the SdS background. First, however, we describe the algebra $\mathcal{A}_{R}$ of operators acting on $\mathcal{H}$ that are localised within the right black hole exterior but that are not necessarily gauge-invariant. This algebra contains a subalgebra of timeshift-preserving right exterior operators that, for each possible timeshift $T$, is isomorphic to the algebra $\mathcal{A}_r$ of right-exterior operators on $\mathcal{H}_0$. In fact, the isomorphism $\mathcal{H}_T \cong \mathcal{H}_0$ described around \eqref{eq:H_CHtId} allows us to identify this algebra directly with $\mathcal{A}_r$.
 
 However there also exist operators that are localised in the right exterior but that do not commute with the timeshift $T$. These include, in particular, the operator
 \begin{align} \label{eq:hCH,R}
     \beta_{CH} \,X_{\Phi , R} &=   \frac{\boldsymbol{\hat A}_{CH} - A_{CH,0}}{4G} + \beta_{CH} \,\boldsymbol{h}_{\Phi_{CH},r} \\&= \beta_{CH}\, X_\Phi- \log \Delta_{\Phi_{CH}} \label{eq:hCH,Rnotsing}
 \end{align}
In fact, the algebra $\mathcal{A}_r$ and (bounded functions of) $X_{\Phi, BH, R}$ will turn out to generate the full algebra $\mathcal{A}_{R}$. We therefore have
 \begin{align}
     \mathcal{A}_{R} \cong \mathcal{A}_{BH,r} \otimes \left(\mathcal{A}_{CH,r} \rtimes \R\right)
 \end{align}
 where $\left(\mathcal{A}_{CH,r} \rtimes \R\right)$ is the modular crossed product of $\mathcal{A}_{BH,r}$. Since $\mathcal{A}_{BH,r}$ is a Type III$_1$ factor while  $\left(\mathcal{A}_{CH,r} \rtimes \R\right)$ is a Type II$_\infty$ factor, the algebra $\mathcal{A}_{R}$ is a Type III$_1$ factor.
 
 One might think that additionally we have e.g. the operator
 \begin{align} \label{eq:hBH,R}
     \beta_{BH} \,X_{\Phi, BH, R} &=   \frac{\boldsymbol{\hat A}_{BH} - A_{BH,0}}{4G} + \beta_{BH} \,\boldsymbol{h}_{\Phi_{BH},r}
 \end{align}
 However it follows from the Einstein equations -- specifically the conservation of the Komar mass associated to boost time translations -- that
 \begin{align} \label{eq:nomissingoperators}
    0 &= \frac{1}{\beta_{BH}} \frac{\boldsymbol{\hat A}_{BH} - A_{BH,0}}{4G}   + \frac{1}{\beta_{CH}}\frac{\boldsymbol{\hat A}_{CH} -  A_{CH,0}}{4G} + \boldsymbol{h}_r 
    \\&= X_{\Phi, BH, R} + X_{\Phi , R} + \scalebox{.8}{$\Delta$} H_{\Phi,r}.
 \end{align}
 Since $\scalebox{.8}{$\Delta$} H_{\Phi,r}$ is affiliated to $\mathcal{A}_r$, \eqref{eq:nomissingoperators} tells us that the operator $X_{\Phi, BH, R}$ is \emph{already} affiliated to $\mathcal{A}_{R}$ and does not need to be added separately.\footnote{There is a subtlety here that should be briefly commented on. Conservation of the Komar mass also tells us that 
 \begin{align}\label{eq:leftkomarcons}(\boldsymbol{\hat A}_{BH} - A_{BH,0})/4G\beta_{BH}   + (\boldsymbol{\hat A}_{CH} -  A_{CH,0})/4G\beta_{CH} + \boldsymbol{h}_l = 0.\end{align} After imposing the gauge constraint $H = \boldsymbol{h}_r - \boldsymbol{h}_\ell =0$, \eqref{eq:nomissingoperators} and \eqref{eq:leftkomarcons} are equivalent. But before doing so they lead to different definitions of $X_{\Phi, BH, R}$ as an operator on $\mathcal{H}$. Adding $X_{\Phi, BH, R}$ as defined via \eqref{eq:leftkomarcons} to $\mathcal{A}_R$ is therefore equivalent to adding functions of the boost generator $H$. Doing so would not change the final gauge-invariant algebra $\widetilde{\mathcal{A}}_R$.} Similarly, operators analogous to \eqref{eq:hCH,R} but with the state $\ket{\Phi_{CH}}$ replaced by a different state $\ket{\Phi'_{CH}}$ are already affiliated to $\mathcal{A}_{R,0}$ because Connes cocycle flow on $\mathcal{A}_{CH,r}$ is an inner automorphism.

 What about operators in the left black hole exterior? The identification $\mathcal{H}_T \cong \mathcal{H}_0$ not only created a kink at the cosmological bifurcation surface, when identifying $\Sigma T$ with $\Sigma_0$, but also conjugated $\mathcal{A}_{CH,\ell}$ by $e^{i\Lambda_{\Phi,\ell}t}$, or equivalently by $\Delta_{\Phi_{CH}}^{-iT/\beta_{CH}}$. As a result, the algebra of timeshift-preserving left exterior operators is
\begin{align}\label{eq:Tconstleftext}
   \mathcal{A}_{BH,\ell} \otimes \Delta_{\Phi_{CH}}^{iT/\beta_{CH}} \,\mathcal{A}_{CH,\ell}\, \Delta_{\Phi_{CH}}^{-iT/\beta_{CH}} \cong \Delta_{\Phi_{CH}}^{iT/\beta_{CH}} \,\mathcal{A}_{\ell}\, \Delta_{\Phi_{CH}}^{-iT/\beta_{CH}}.
 \end{align}
 A useful check of \eqref{eq:Tconstleftext} is that any operator $a'$ in \eqref{eq:Tconstleftext} satisfies
 \begin{align}
     \beta_{CH}[ X_{\Phi,R}, a'] &= \beta_{CH}[X_\Phi, a'] - [\log \Delta_{\Phi_{CH}}, a'] = 0
 \end{align}
 and so commutes with the right exterior operator $X_{\Phi,R}$. In addition, we know from \eqref{eq:ptinterpretation} that the left exterior algebra should also bounded functions of $X_\Phi= -i \partial_T$. Together these generate the commutant algebra
 \begin{align}\label{eq:ALdef}
     \mathcal{A}_{L} \cong \mathcal{A}_{R}' \cong \mathcal{A}_{BH,\ell} \otimes \left(\mathcal{A}_{CH,r} \rtimes \R\right)'
 \end{align}
 where $\left(\mathcal{A}_{CH,r} \rtimes \R\right)'$, which is generated by $X_\Phi$ and $\Delta_{\Phi_{CH}}^{iT/\beta_{CH}} \mathcal{A}_{CH,\ell} \Delta_{\Phi_{CH}}^{-iT/\beta_{CH}}$, is the commutant of  $\left(\mathcal{A}_{CH,r} \rtimes \R\right)$ on $\mathcal{H}_{CH}$ and is also a modular crossed product algebra. 
 
 As one might hope, the algebras $\mathcal{A}_L$ and $\mathcal{A}_R$ are isomorphic, as can be easily verified by conjugating by $\Delta_{\Phi_{CH}}^{iT/\beta_{CH}}$. In effect, this conjugation switches from the identification \eqref{eq:H_CHtId} to an identification $\mathcal{B}(\mathcal{H}_{CH,T} \cong e^{-i \Lambda_{\Phi,r} t} \mathcal{B}(\mathcal{H}_{CH}) e^{i\Lambda_{\Phi,r}t}$. They are also independent of the choice of state $\ket{\Phi_{CH}}$. For example, conjugation by the Connes' cocycle flow $\Delta^{iT/\beta_{CH}}_{\Phi_{CH}}\Delta^{-iT/\beta_{CH}}_{\Phi|\Phi'_{CH}}$ maps the algebras $\mathcal{A}_L$ and $\mathcal{A}_R$ to the corresponding algebras with $\ket{\Phi_{CH}}$ replaced by $\ket{\Phi'_{CH}}$. Again, this can be physically interpreted as a replacement of $\ket{\Phi_{CH}}$ by $\ket{\Phi_{CH}'}$ in the identification \eqref{eq:H_CHtId}.

\subsection{The gauge constraints}
We now turn to the question of how to impose the gauge constraints associated to boosts and rotations of the SdS background. On the Hilbert space $\mathcal{H}_0$ boosts are generated by the operator $H_0$ while rotations are generated by $L_i$. However, when acting on the timeshifted SdS Hilbert space $\mathcal{H}_T$ (identified with $\mathcal{H}_0$ via \eqref{eq:H_CHtId}) we needed to conjugate operators in $\mathcal{A}_{CH,\ell}$ -- but not operators in $\mathcal{A}_{CH,r}$ -- by $\Delta_{\Phi_{CH}}^{-iT/\beta_{CH}}$. Formally, we could describe this by saying that all operators acting on $\mathcal{H}_T$ need to be conjuated by $\exp(-i \boldsymbol{h}_{\Phi_{CH},\ell}T)$. In general, it does not make sense to conjugate an operator in $\mathcal{B}(\mathcal{H}_0)$ (that isn't e.g. contained in $\mathcal{A}_\ell$ or $\mathcal{A}_r$) by a singular one-sided modular flow, which is why the algebras of operators $\mathcal{B}(\mathcal{H}_0)$ and $\mathcal{B}(\mathcal{H}_T)$ cannot be naturally identified. However, in this case, we know that boosts and rotations \emph{are} isometries of the timeshifted SdS background, and so the operators generating them need to exist. So somehow things have to work out. Because we chose the state $\ket{\Phi_{CH}}$ to be rotation invariant, we have 
\begin{align}
[\boldsymbol{h}_{\Phi_{CH},\ell}, L_i] = [h_{\Phi_{CH},\ell}, L_{CH,i}] = 0.
\end{align}
Rotations on $\mathcal{H}_T$, like rotations on $\mathcal{H}_0$ are therefore generated by $L_i$. 

On the other hand, the generator
\begin{align} \label{eq:Ht}
    H = e^{i \boldsymbol{h}_{\Phi_{CH},\ell}T} H_0 e^{-i \boldsymbol{h}_{\Phi_{CH},\ell}T}
\end{align}
of boosts on $\mathcal{H}_T$ is not equal to the generator $H_0$ of boosts on $\mathcal{H}_0$ because $\ket{\Phi_{CH}}\ket{\Phi_{BH}}$ and, for that matter, the Hilbert space $\mathcal{H}_{CH}$ itself are not boost invariant. However, it follows from \eqref{eq:lambda} that the operator $\Lambda_{\Phi,\ell}$ does commute with $h_{\Phi_{CH},\ell}$. Meanwhile
\begin{align}
      H_0 - \Lambda_{\Phi,\ell} = -\scalebox{.8}{$\Delta$} H_{\Phi,\ell}
\end{align}
is affiliated with $\mathcal{A}_\ell$ so that
\begin{align}
    e^{i h_{\Phi_{CH},\ell}T}\scalebox{.8}{$\Delta$} H_{\Phi,\ell}e^{-i h_{\Phi_{CH},\ell}T} = \Delta_{\Phi_{CH}}^{iT/\beta_{CH}}\scalebox{.8}{$\Delta$} H_{\Phi,\ell} \Delta_{\Phi_{CH}}^{-iT/\beta_{CH}}.
\end{align}
We therefore have
\begin{align}\label{eq:HdefSdS}
    H = \Lambda_{\Phi,\ell} - \Delta_{\Phi_{CH}}^{iT/\beta_{CH}}\scalebox{.8}{$\Delta$} H_{\Phi,\ell} \Delta_{\Phi_{CH}}^{-iT/\beta_{CH}},
\end{align}
which is manifestly densely defined.

It is important to check that $H$ generates automorphisms of the algebra $\mathcal{A}_{R}$. For any $a \in \mathcal{A}_r$, we have
\begin{align}
    [H,a] = [\Lambda_{\Phi,\ell},a] = [H_0, a].
\end{align}
So $H$, like $H_0$ generates an automorphism of $\mathcal{A}_r$. Meanwhile, taking a derivative of \eqref{eq:Ht} gives
\begin{align} \label{eq:dtXphi}
     [H, X_\Phi] = i \partial_T H =  -\frac{1}{\beta_{CH}}[\Delta_{\Phi_{CH}}^{iT/\beta_{CH}} \scalebox{.8}{$\Delta$} H_{\Phi,\ell} \Delta_{\Phi_{CH}}^{-iT/\beta_{CH}}, \log \Delta_{\Phi_{CH}}].
\end{align}
Hence, using \eqref{eq:hCH,Rnotsing}, we obtain
\begin{align}\label{eq:dthCHR}
    [H, X_{\Phi , R}]  &= [H, X_\Phi] - \frac{1}{\beta_{CH}} [H, \log \Delta_{\Phi_{CH}}]
    \\&=-\frac{1}{\beta_{CH}}[\Lambda_{\Phi,\ell}, \log \Delta_{\Phi_{CH}}]
    \\& = -\frac{1}{\beta_{CH}}[\scalebox{.8}{$\Delta$} H_{\Phi,r}, \log \Delta_{\Phi_{CH}}] \label{eq:thirdstep}
    \\&= [\Lambda_{\Phi,\ell}, X_{\Phi , R}].\label{eq:laststep}
 \end{align}
 \eqref{eq:thirdstep} is manifestly contained in $\mathcal{A}_r$ since $\log \Delta_{\Phi_{CH}}$ generates modular flows of $\mathcal{A}_{CH,r}$ and commutes with $\mathcal{A}_{BH,r}$. We conclude not only that 
 \begin{align}
     e^{iHt}\mathcal{A}_Re^{-iHt} = \mathcal{A}_R,
 \end{align}
 but that in fact the automorphism generated by $H$ is the same as the automorphism generated by $\Lambda_{\Phi,
 \ell}$.
 
An intuitive explanation for \eqref{eq:dthCHR} is that we know from \eqref{eq:hCH,R} $X_{\Phi , R}(t)$ can be written as a sum of the sesquilinear forms $\boldsymbol{h}_{\Phi , r}$ and $(\boldsymbol{\hat A}_{CH}-A_{CH,0})/4\beta_{CH} G$. Since the cosmological bifurcation surface is boost invariant, the latter is independent of boost time. We should therefore have
 \begin{align}\label{eq:intuitdHCR/dt}
     [H, X_{\Phi , R}] = [H, \boldsymbol{h}_{\Phi , r}] = [\Lambda_{\Phi,\ell}, \boldsymbol{h}_{\Phi , r}],
 \end{align}
 But, since  $[\Lambda_{\Phi,\ell},X_\Phi]= [\Lambda_{\Phi,\ell},\boldsymbol{h}_{\Phi , \ell}]= 0$, \eqref{eq:intuitdHCR/dt} is equivalent to \eqref{eq:laststep}.

  As in Section \ref{sec:inflaton}, it is not at all hard to find operators in the algebra $\mathcal{A}_{R}$ that are rotation invariant and hence commute with the operators $L_i$. There are already many such operators in the QFT algebra $\mathcal{A}_r$ and, in addition,
  \begin{align}
    [L_i, X_{\Phi , R}] = 0.
  \end{align}
  The challenge is to find a nontrivial algebra $\widetilde{\mathcal{A}}_R$ of operators in $\mathcal{A}_R$ that are also boost invariant. No such operators exist (except for $c$-numbers) in $\mathcal{A}_r$ because QFT observables acting on $\mathcal{H}_0$ equilibrate at late times to the Unruh state and at early times to the time-reflected Unruh state. It follows that, as in pure de Sitter space, integrating over boosts cannot lead to a finite operator \eqref{eq:introtildeadef}. 
  
  We will see that, like in Section \ref{sec:inflaton}, the existence of nontrivial boost invariant observables in the larger algebra $\mathcal{A}_R$ follows from the existence of a runaway instability in the boost time evolution of $\mathcal{H}$. In this case, that instability is the net energy flux that results from the black hole being at a higher temperature than the cosmological horizon. 

  In the Unruh state, outgoing Rindler modes from the white hole and past cosmological horizons will be in a thermal state at their respective horizon temperatures. For simplicity let us consider a single scalar matter field $\phi$ of mass $m \gg \ell_{dS}^{-1}$. Since the Rindler modes have a fixed, conserved boost energy $\omega$, they can be expanded as
  \begin{align}
      \phi_{\omega,l,m} = \frac{1}{r} Y_{\ell m}(\Omega) e^{-i\omega t} \psi_{\omega \ell}(r)
  \end{align}
  where $Y_{\ell m}$ are (two-dimensional) spherical harmonics. The classical equation of motion then becomes
  \begin{align}\label{eq:effectivescattering}
      - \frac{d^2}{dr_*^2} \psi_{\omega \ell} + V_{\rm eff}(r) \psi_{\omega \ell} = \omega^2 \psi_{\omega \ell},
  \end{align}
  where $r_*$ is a tortoise coordinate defined by $dr_* = f(r)^{-1} dr$ and the effective potential
  \begin{align}
      V_{\rm eff}(r) = f(r) \left[\frac{\ell(\ell+ 1)}{r^2}  + m^2 + \frac{f'(r)}{r}\right] 
  \end{align}
  that is everywhere positive but goes to zero at the black hole horizon $r_+ \to -\infty$ and the cosmological horizon $r_+ \to + \infty$. The equation \eqref{eq:effectivescattering} is therefore a one-dimensional scattering problem that unitarily maps the outgoing modes from the white hole and past cosmological horizons into a mixture of the Rindler modes with the same boost energy $\omega$ falling across the black hole and future cosmological horizons. Since the black hole temperature is always higher than the cosmological temperature, the Rindler modes crossing the black hole horizon will have less energy than the local Minkowski vacuum and hence carry a negative energy flux. Meanwhile the modes crossing the future cosmological horizon will carry a positive expected energy flux.
  
  It follows that at sufficiently late times or, by time-reflection symmetry, at sufficiently early times, the area of the black hole horizon will tend to linearly decrease, while the area of the cosmological horizon will linearly increase. The fluctuations in both horizon areas will also grow with time, but, like the inflaton fluctuations in Section \ref{sec:inflatonalgebra}, they only grow as the square root of time because the fluctuations in the energy of Hawking radiation passing from one horizon to the other are uncorrelated at long timescales and hence obey the central limit theorem. As a result, the probability of the black hole energy remaining above any fixed value decays exponentially at sufficiently early or late times.

 The late- and early-time growth of the cosmological horizon area is captured by the time evolution of the operator 
 \begin{align}
     X_{\Phi , R}(t) = e^{iHt}\, X_{\Phi , R} \,e^{-iHt}.
 \end{align} 
By \eqref{eq:intuitdHCR/dt}, this is equivalent to understanding the time evolution of the sesquilinear form $\boldsymbol{h}_{\Phi_{CH},r}$. Let $v$ be an affine coordinate on the future cosmological horizon (with the bifurcation surface at $v = 0$). Suppose $\ket{\Phi_{CH}}$ is very close to the time-reflected Unruh state everywhere on the future cosmological horizon prior to some affine time $v_0 > 0$. Since all states in $\mathcal{H}_0$ locally look like the Minkowski vacuum, this will always be true for sufficiently small $v_0$ with error that goes to zero as $v_0 \to 0$. Now we define the sesquilinear form 
\begin{align}
\boldsymbol{h}_{v_0,r} = \int d\Omega \int_0^\infty dv \,v f(v/v_0) T_{vv}
\end{align}
where we are integrating over the future cosmological horizon and the transverse sphere and $f(v/v_0)$ is a smooth cut-off function satisfying $f(v/v_0) = 1$ for $v_0 > v > 0$ and $f(v/v_0) = 0$ for large $v \gg v_0$. Heuristically, $\boldsymbol{h}_{v_0,r}$ measures the boost energy of modes crossing the future cosmological horizon before $v_0$. Also let
 \begin{align} \label{eq:Deltah}
{\Delta \boldsymbol{h}}_{\Phi_{CH},v_0} = \boldsymbol{h}_{\Phi_{CH},r} - \boldsymbol{h}_{v_0,r}.
 \end{align}
 On the future cosmological horizon at $v < v_0$, the two terms on the right hand side of \eqref{eq:Deltah} approximately cancel. Note that this does not mean ${\Delta \boldsymbol{h}}_{\Phi_{CH},v_0}$ is a densely defined operator. It is still only a sesquilinear form because $\boldsymbol{h}_{\Phi_{CH},r}$ also has divergent fluctuations from modes in $U_r$ that cross the past cosmological horizon very close to the bifurcation surface. Those modes are not acted on by $\boldsymbol{h}_{v_0,r}$.
 \begin{figure}
  \centering
\begin{tikzpicture}[scale =1]
     \pgftext{\includegraphics[scale =.45]{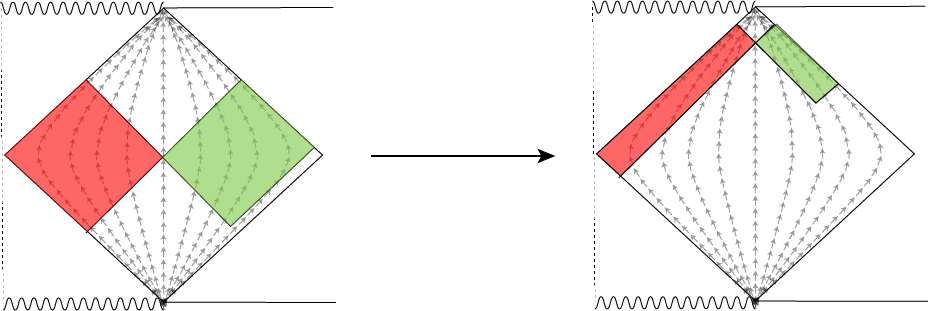} 
     \draw (-9.8 ,2.7) node { $v=v_0$};
     \draw (-11.6,2.9) node {\large $U_{v_0}$};
     \draw (-8.5,3) node {\LARGE $e^{-\i H_0 t}$};
     \draw (-1.1 ,3.8) node {\large $v=e^{2\pi t/\beta_{CH}}v_0$};
}
 \end{tikzpicture}
 \caption{\label{fig:boostwedge} The reduced state on the green wedge $U_{v_0}$ is well-approximated by the Unruh state at late boost times.}
 \end{figure}
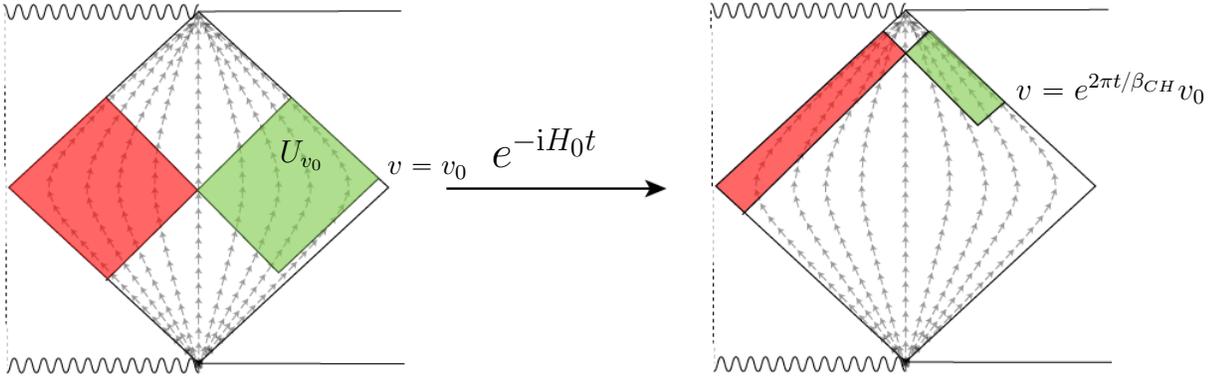

 However, as shown in Figure \ref{fig:boostwedge}, it does mean that we can approximate ${\Delta \boldsymbol{h}}_{\Phi_{CH},v_0}$, to arbitrarily high precision, by a sesquilinear form on a causal diamond $U_{v_0}$ bounded the $v = v_0$ on the future cosmological horizon and the right edge of $U_{BH}$. 
 Because $U_{v_0} \subseteq U_r$ does not touch either the white hole or past cosmological horizons, it is mapped under forwards translation by a sufficiently large boost time to a causal diamond $U_{v_0}(t)$ that is localised close the asymptotic future. As a result, in the limit $t\to\infty$, the expectation value of 
 \begin{align}
 {\Delta \boldsymbol{h}}_{\Phi_{CH},v_0}(t) = e^{ i H_0 t} {\Delta \boldsymbol{h}}_{\Phi_{CH},v_0} e^{- i H_0 t}
 \end{align}
 will converge to its expectation value of the Unruh state.
 
 On the other hand, the operator 
 \begin{align}
     \boldsymbol{h}_{v_0,r}(t) = e^{iH_0t} \boldsymbol{h}_{v_0,r} e^{-iH_0t} = \boldsymbol{h}_{ e^{2 \pi t/\beta} v_0,r}
 \end{align}
 describes the total energy flux across the future cosmological horizon prior to the affine time $e^{2 \pi t/\beta} v_0$. As we saw from \eqref{eq:effectivescattering}, this is positive and grows linearly with $t$. It follows that at late times $t$, $\boldsymbol{h}_{\Phi_{CH},r} (t)$ and hence $X_{\Phi , R}(t)$ also grow linearly. Similar arguments shows that at very early times $t \ll 0$ the dominant contribution to the expectation value of $X_{\Phi , R}(t)$ comes from Hawking modes in the time-reversed Unruh state crossing the past cosmological horizon. The net energy flux of these modes is again positive and grows linearly with $|t|$.

 Let $a(t) = e^{iHT} a e^{-iHT} \in \mathcal{A}_R$ be a rotation-invariant operator with matrix elements that decay faster than exponentially to zero as $X_{\Phi , R}(t) \to + \infty$, e.g. $a(t)$ could be $\exp(-X_{\Phi , R}(t)^2)$. We can then construct a boost-time invariant operator $\widetilde A \in \widetilde{\mathcal{A}}_R$ by
 \begin{align}
\widetilde{a} = \int dt \,a(t).
 \end{align}
 Since operators $a(t)$ that decay in this way are dense in the rotation-invariant subalgebra of $\mathcal{A}_R$, this leads to a large gauge-invariant algebra $\widetilde{\mathcal{A}}_R$.

 We can also ask about the algebra $\widetilde{\mathcal{A}}_L$ of gauge-invariant left exterior operators. It follows immediately from \eqref{eq:intuitdHCR/dt} that
 \begin{align} \label{eq:dHCL/dt}
     [H, X_\Phi] = [H, \boldsymbol{h}_{\Phi_{CH},\ell}].
 \end{align}
Hence, by arguments identical to those above except with right exterior sesquilinear forms replaced by the corresponding left exterior ones, we find that
 \begin{align}
 H_{\Phi_{CH},L}(t) = e^{iHt}X_\Phi e^{-iHT}
 \end{align}
 also grows linearly with $|t|$ at sufficiently early and late times. Given a rotation-invariant operator $a' \in \mathcal{A}_L$ that vanishes sufficiently fast as $H_{\Phi_{CH},L} = X_\Phi \to +\infty$, we can therefore construct a gauge-invariant operator $\widetilde{a}' \in \widetilde{\mathcal{A}}_L$ by
 \begin{align}
     \widetilde{a}' = \int dt a'(t).
 \end{align}
 
 On the Hilbert space $\mathcal{H}$, there exist operators that commute with both $\mathcal{A}_L$ and $\mathcal{A}_R$. The obvious examples are the gauge group generators $J_i$ and $H$. However, the Hilbert space $\mathcal{H}$ is not the physical Hilbert space $\widetilde{\mathcal{H}}$ of the quantum gravity theory because we have not yet imposed the gauge constraints. The rotation constraints can be imposed on $\mathcal{H}$ by restricting to the rotation invariant subspace. Because boosts generate a noncompact gauge group, the boost constraint must be imposed using the method of coinvariants. The algebras $\widetilde{\mathcal{A}}_L$ and $\widetilde{\mathcal{A}}_R$ have a natural action on the coinvariant Hilbert space $\widetilde{\mathcal{H}}$. As in Section \ref{sec:inflatonalgebra}, we suspect, but do not know how to prove, that in fact $\widetilde{\mathcal{A}}_L$ and $\widetilde{\mathcal{A}}_R$ act as commutants on $\widetilde{\mathcal{H}}$.

It is worth comparing the algebra we have defined here with the algebra for a Schwarzschild-de Sitter black hole that was previously defined in \cite{kudlerflam2024generalized}. An obvious distinction is that the algebra in \cite{kudlerflam2024generalized} involved an explicit clock Hilbert space similar to that used for de Sitter space in \cite{CLPW}. On the other hand, a crucial part of our story is that the dynamical evolution of the black hole itself plays the role of a clock, and hence leads to a nontrivial algebra. It turns out, however, that it is easy to obtain an algebra for Schwarzschild-de Sitter without an observer from the algebra described in \cite{kudlerflam2024generalized} by simply fixing the observer energy to zero.\footnote{During the preparation of this manuscript, this observation was independently made in \cite{faulkner2024gravitational}.}

The other major difference is that the algebra in \cite{kudlerflam2024generalized} was based on a Hilbert space constructed around the Unruh state. Working with the Unruh state leads to a number of technical simplifications, primarily because the boost Hamiltonian can be written as a sum of modular Hamiltonians on the black hole and cosmological horizons. However, as in Section \ref{sec:bhflat}, we believe that our approach has some important conceptual advantages that make up for this. In particular, as in flat space, the Unruh state is singular on both the white hole and past cosmological horizons and so presumably only exists in nonperturbative quantum gravity as a late-time approximation to the quantum state. Furthermore, in the absence of an explicit clock Hilbert space, the only finite physical observable that can be used as a clock for states in the Unruh GNS Hilbert space is the asymptotic gravitational mass of the black hole at past timelike infinity. At any finite time, quasilocal gravitational masses will have divergent fluctuations because of the infinite exchange of Hawking radiation between the two horizons that will have already occurred by that time. Both issues do not exist in our construction: all states are globally smooth and a dense set of operators in $\mathcal{A}_R$ (in the strong operator topology) can be measured quasilocally at finite times. Finally, by working in global de Sitter space, our construction makes clear the physical significance of the timeshift mode. 

 \subsection{The trace, density matrices and entropies}
 The algebra $\mathcal{A}_{R}$ is a Type III$_1$ factor (because of the presence of the Type III$_1$ subfactor $\mathcal{A}_{BH,r}$) and so does not have a trace. However, as in Section \ref{sec:inflaton}, there may nonetheless exist a trace for the subalgebra $\widetilde{\mathcal{A}}_R\subseteq \mathcal{A}_R$ of gauge-invariant observables. In fact, as in that section, we will find a trace for all boost-invariant operators (regardless of whether they are also rotation invariant). 
 
 Explicitly, if
 \begin{align}\label{eq:tildeAexp}
     \widetilde{a} = \int dt a(t) = \int dt e^{iHt} a e^{-iHt}
 \end{align}
 we define 
 \begin{align}\label{eq:bhtr}
     \Tr(\widetilde{a}) &= \int d X_{\Phi}\,e^{\scalebox{.6}{$\Delta$} \beta X_{\Phi}} \braket{\Phi|e^{-\beta_{BH} \Lambda_{\Phi,\ell}/2} a(t) e^{-\beta_{BH} \Lambda_{\Phi,\ell}/2}|\Phi}
     \\&= \bra{0}_T e^{\scalebox{.6}{$\Delta$} \beta X_{\Phi}/2} \braket{\Phi|e^{-\beta_{BH} \Lambda_{\Phi,\ell}/2} a(t) e^{-\beta_{BH} \Lambda_{\Phi,\ell}/2}|\Phi} e^{\scalebox{.6}{$\Delta$} \beta X_{\Phi}/2} \ket{0}_T \label{eq:bhtrace2}
 \end{align}
 where $\scalebox{.6}{$\Delta$} \beta = \beta_{CH} - \beta_{BH}$, $\ket{\Phi} = \ket{\Phi_{BH}}\ket{\Phi_{CH}}$ and $\Lambda_{\Phi,\ell}$ was defined in \eqref{eq:lambda} and $t$ is an arbitrary choice of time as in \eqref{eq:inflaton1sttracedef}. 

Since \eqref{eq:bhtr} is somewhat complicated, and the rigorous (for physicists!) arguments that it defines a trace are fairly involved, we will first give a more intuitive but not at all trustworthy explanation for why it should act as a trace for $\widetilde{\mathcal{A}}_R$. Using \eqref{eq:lambda} and \eqref{eq:ptinterpretation}, we can formally write
\begin{align}
    e^{-\beta_{BH} \Lambda_{\Phi,\ell}/2} e^{\scalebox{.6}{$\Delta$} \beta X_{\Phi}/2} &= \exp\left( \frac{\scalebox{.6}{$\Delta$} \beta (\boldsymbol{\hat A}_{CH} - A_{CH,0})}{8G \beta_{CH}} + \frac{\beta_{CH}}{2} \boldsymbol{h}_{\Phi_{CH},\ell} +\frac{\beta_{BH}}{2} \boldsymbol{h}_{\Phi_{BH},\ell} - \frac{\beta_{BH}}{2} \boldsymbol{h}_{r,0}\right)
\end{align}
But the formal density matrices $\boldsymbol{\rho}_{\Phi_{BH},\ell}$ and $\boldsymbol{\rho}_{\Phi_{CH},\ell}$ on $\mathcal{A}_{BH,\ell}$ and $\mathcal{A}_{CH,\ell}$ satisfy
\begin{align}
    \boldsymbol{\rho}_{\Phi_{BH},\ell} = \exp(-\beta_{BH} \boldsymbol{h}_{\Phi_{BH},\ell} + {\rm const})~~~\text{and}~~~\boldsymbol{\rho}_{\Phi_{CH},\ell} = \exp(-\beta_{CH} \boldsymbol{h}_{\Phi_{CH},\ell} + {\rm const}),
\end{align}
where the constants are as usual divergent. So
\begin{align}
    \ket{\Phi_{BH}}\ket{\Phi_{CH}} = \exp(-\beta_{BH} \boldsymbol{h}_{\Phi_{BH},\ell}/2-\beta_{CH} \boldsymbol{h}_{\Phi_{CH},\ell}/2 + {\rm const}) \ket{\rm \textbf{MAX}},
\end{align}
where $\ket{\textbf{MAX}}$ is (very very formally!) a maximally entangled state on $\mathcal{A}_r$. We therefore have
\begin{align}
    e^{-\beta_{BH} \Lambda_{\Phi,\ell}/2} e^{\scalebox{.6}{$\Delta$} \beta X_{\Phi}/2}\ket{\Phi}\ket{0}_T= \exp\left(\frac{\scalebox{.6}{$\Delta$} \beta(\boldsymbol{\hat A}_{CH} - A_{CH,0})}{8G \beta_{CH}} - \frac{\beta_{BH}}{2} \boldsymbol{h}_{r,0}+ {\rm const}\right)\ket{\rm \textbf{MAX}}\ket{0}_T.
\end{align}
Our formal manipulations therefore seem to suggest that the state \eqref{eq:bhtr} should have a density matrix $\boldsymbol{\rho}_R$ on $\mathcal{A}_R$ given by
\begin{align}
\boldsymbol{\rho}_R = \exp\left(\frac{\scalebox{.6}{$\Delta$} \beta(\boldsymbol{\hat A}_{CH} - A_{CH,0})}{4G \beta_{CH}} - \beta_{BH} \boldsymbol{h}_{r,0}+ {\rm const}\right).
\end{align}
Since $\log \boldsymbol{\rho}_R$ generates the same modular flow for $\mathcal{A}_R$ as $H$, the state \eqref{eq:bhtr} should satisfy the KMS condition for the boost Hamiltonian $H$ on the algebra $\mathcal{A}_R$. But operators in $\widetilde{\mathcal{A}}_R$ are boost invariant, so, as in Section \ref{sec:inflatonalgebra}, \eqref{eq:bhtr} should define a trace on that subalgebra.

 We now turn to more rigorous analysis of the properties of \eqref{eq:bhtr}. We first show that \eqref{eq:bhtrace2} is boost invariant and hence independent of the arbitrary choice $t$. We have
 \begin{align}\nonumber
 H  e^{-\beta_{BH} \Lambda_{\Phi,\ell}/2}\ket{\Phi} e^{\scalebox{.6}{$\Delta$} \beta X_{\Phi}/2} \ket{0}_T &= e^{\scalebox{.6}{$\Delta$} \beta X_{\Phi}/2} \ket{0}_T \left(\Lambda_{\Phi,\ell} - \Delta_{\Phi_{CH}}^{-\scalebox{.6}{$\Delta$} \beta/2\beta_{CH}}\scalebox{.8}{$\Delta$} H_{\Phi,\ell} \Delta_{\Phi_{CH}}^{\scalebox{.6}{$\Delta$} \beta/2\beta_{CH}}\right)e^{-\beta_{BH} \Lambda_{\Phi,\ell}/2}\ket{\Phi}
 \\&=e^{\scalebox{.6}{$\Delta$} \beta X_{\Phi}/2} \ket{0}_T e^{-\beta_{BH} \Lambda_{\Phi,\ell}/2}\left(\Lambda_{\Phi,\ell} - \Delta_{\Phi}^{-1/2}\scalebox{.8}{$\Delta$} H_{\Phi,\ell} \Delta_{\Phi}^{1/2}\right)\ket{\Phi}\nonumber
 \\&=e^{\scalebox{.6}{$\Delta$} \beta X_{\Phi}/2} \ket{0}_T e^{-\beta_{BH} \Lambda_{\Phi,\ell}/2}\left(\scalebox{.8}{$\Delta$} H_{\Phi,r} - \Delta_{\Phi}^{-1/2}\scalebox{.8}{$\Delta$} H_{\Phi,\ell}\right)\ket{\Phi},\nonumber
 \\&=e^{\scalebox{.6}{$\Delta$} \beta X_{\Phi}/2} \ket{0}_T e^{-\beta_{BH} \Lambda_{\Phi,\ell}/2}\left(\scalebox{.8}{$\Delta$} H_{\Phi,r} - J \scalebox{.8}{$\Delta$} H_{\Phi,\ell} \right)\ket{\Phi},\label{eq:traceboostfinalstep}
 \end{align}
 where $\Delta_\Phi = \Delta_{\Phi_{BH}} \Delta_{\Phi_{CH}}$ is the modular operator for $\ket{\Phi} = \ket{\Phi_{BH}}\ket{\Phi_{CH}}$ on $\mathcal{A}_r$ and $J$ is the corresponding modular conjugation operator, which we chose to be the discrete symmetry generator defined in \eqref{eq:Jdef}. In the second step we used the fact that $\Lambda_{\Phi,\ell}$ generates modular flows of $\mathcal{A}_{BH,\ell}$ and $\mathcal{A}_{CH,\ell}$. In the third step, we used
 \begin{align}
 \Delta_{\Phi_{CH}}\ket{\Phi} = \Delta_{\Phi_{BH}} \ket{\Phi} = \ket{\Phi}.
 \end{align}
 Finally in the last step we used
 \begin{align}
     \Delta_{\Phi}^{-1/2}\scalebox{.8}{$\Delta$} H_{\Phi,\ell}\ket{\Phi}=J S'_\Phi \scalebox{.8}{$\Delta$} H_{\Phi,\ell}\ket{\Phi} = J  \scalebox{.8}{$\Delta$} H_{\Phi,\ell}^\dagger\ket{\Phi} =J  \scalebox{.8}{$\Delta$} H_{\Phi,\ell}\ket{\Phi}. 
 \end{align}
 However, since $H$ and $\log \Delta_{\Phi_{CH}}$ and $\log \Delta_{\Phi_{BH}}$ all have odd parity under conjugation by $J$, we have
 \begin{align}
     0 = H + JHJ = \scalebox{.8}{$\Delta$} H_{\Phi,r} - J \scalebox{.8}{$\Delta$} H_{\Phi,\ell} J - \scalebox{.8}{$\Delta$} H_{\Phi,\ell} + J \scalebox{.8}{$\Delta$} H_{\Phi,r} J.
 \end{align}
 Since conjugation by $J$ exchanges $\mathcal{A}_\ell$ and $\mathcal{A}_r$, the first two terms are contained in $\mathcal{A}_r$, while the last two are contained in $\mathcal{A}_\ell$. It follows that
 \begin{align}\label{eq:exchangedeltaH}
    \scalebox{.8}{$\Delta$} H_{\Phi,r} - J \scalebox{.8}{$\Delta$} H_{\Phi,\ell} J = 0.
 \end{align}
 The possibility of a $c$-number on the right-hand side of \eqref{eq:exchangedeltaH} is ruled out by the fact that $J^2 = 1$. We conclude that \eqref{eq:traceboostfinalstep} vanishes and hence that \eqref{eq:bhtr} is boost invariant.\footnote{Note that  \eqref{eq:bhtr} would in fact still be boost invariant even if we had not chosen $\ket{\Phi}$ to be canonically purified with respect to $J$. However the proof would be considerably more involved.} By an analogous argument to \eqref{eq:equivalencetraceinflaton}, \eqref{eq:bhtr} is also independent of the choice of expansion \eqref{eq:tildeAexp}.

 An intuitive physical explanation for the boost invariance of \eqref{eq:bhtr} is the following. At finite $G$, there is an exponentially small possibility 
 \begin{align}
     p_{BH} \sim \exp((A_{BH} + A_{CH} - 4 \pi \ell_{dS}^2)/4G)
 \end{align}
 that the no-boundary Hartle-Hawking state contains a black hole. In the limit $G \to 0$, multiplying expectation values in the Hartle-Hawking state of operators that project onto the presence of a black hole by a divergent constant should define a semifinite weight on $\mathcal{A}_R$. By construction, this weight should be boost invariant and should have a probability distribution proportional to $p_{BH}$, which becomes $\exp(\scalebox{.6}{$\Delta$} \beta X_\Phi)$ in the $G \to 0$ limit. In fact, we expect that it is exactly \eqref{eq:bhtr}.

 To show that $\Tr$ is indeed a trace on $\widetilde{\mathcal{A}}_R$, we first show that, for any operators $a,b \in \mathcal{A}_{R}$, we have the KMS condition
 \begin{align}
 \label{eq:BHKMS}
       \int d X_{\Phi}e^{\scalebox{.6}{$\Delta$} \beta X_{\Phi}} &\braket{\Phi|e^{-\beta_{BH} \Lambda_{\Phi,\ell}/2} a(t)  b(t') e^{-\beta_{BH} \Lambda_{\Phi,\ell}/2} |\Phi}\\&\stackrel{?}{=}  \int d X_{\Phi}e^{\scalebox{.6}{$\Delta$} \beta X_{\Phi}} \braket{\Phi|e^{-\beta_{BH} \Lambda_{\Phi,\ell}/2}   b(t') a(t+ i\beta_{BH})e^{-\beta_{BH} \Lambda_{\Phi,\ell}/2} |\Phi}.
 \end{align}
 Using the boost invariance of \eqref{eq:bhtr} and \eqref{eq:dthCHR}, we have 
 \begin{align}
     \int d X_{\Phi} e^{\scalebox{.6}{$\Delta$} \beta X_{\Phi}} &\braket{\Phi|e^{-\beta_{BH} \Lambda_{\Phi,\ell}/2} a(t)  b(t') e^{-\beta_{BH} \Lambda_{\Phi,\ell}/2} |\Phi} \\&= \int d X_{\Phi}e^{\scalebox{.6}{$\Delta$} \beta X_{\Phi}} \braket{\Phi|e^{-\beta_{BH} \Lambda_{\Phi,\ell}/2} a(t- i\beta_{BH}/2)  b(t'-i\beta_{BH}/2) e^{-\beta_{BH} \Lambda_{\Phi,\ell}/2} |\Phi}
     \\&= \int d X_{\Phi}e^{\scalebox{.6}{$\Delta$} \beta X_{\Phi}} \braket{\Phi|a(t)  b(t')e^{-\beta_{BH} \Lambda_{\Phi,\ell}} |\Phi}
 \end{align}
 and 
 \begin{align}
     \int d X_{\Phi}e^{\scalebox{.6}{$\Delta$} \beta X_{\Phi}} &\braket{\Phi|e^{-\beta_{BH} \Lambda_{\Phi,\ell}/2}   b(t') a(t+i\beta_{BH})e^{-\beta_{BH} \Lambda_{\Phi,\ell}/2} |\Phi} 
     \\&= \int d X_{\Phi}e^{\scalebox{.6}{$\Delta$} \beta X_{\Phi}} \braket{\Phi|e^{-\beta_{BH} \Lambda_{\Phi,\ell}/2}   b(t'-i\beta_{BH}/2) a(t+i\beta_{BH}/2)e^{-\beta_{BH} \Lambda_{\Phi,\ell}/2} |\Phi}
     \\&= \int d X_{\Phi}e^{\scalebox{.6}{$\Delta$} \beta X_{\Phi}} \braket{\Phi|b(t')e^{-\beta_{BH} \Lambda_{\Phi,\ell}}  a(t) |\Phi}.
 \end{align}
 It follows that the KMS condition \eqref{eq:BHKMS} is equivalent to
 \begin{align}\label{eq:toshow}
    \int d X_{\Phi}\,e^{\scalebox{.6}{$\Delta$} \beta X_{\Phi}} \braket{\Phi| b \,e^{-\beta_{BH} \Lambda_{\Phi,\ell}} \,a |\Phi} \stackrel{?}{=} \int d X_{\Phi}\,e^{\scalebox{.6}{$\Delta$} \beta X_{\Phi}} \braket{\Phi|a\,  b
    \,e^{-\beta_{BH} \Lambda_{\Phi,\ell}} |\Phi},
 \end{align}
for all $a,b \in \mathcal{A}_R$. To check \eqref{eq:toshow}, it suffices to consider operators of the form
 \begin{align} \label{eq:simpleform}
     a =   \int ds \,\,a_{BH}  \otimes a_{CH,s} \,e^{i X_{\Phi , R} s}
 \end{align}
 with $a_{BH} \in \mathcal{A}_{BH,r}$ and $a_{CH,s} \in \mathcal{A}_{CH,r}$ holomorphic in the strip $0 < \mathrm{Im}(s) < \beta_{CH} -\beta_{BH}$. This is because the linear span of operators of the form \eqref{eq:simpleform} is s.o.t. dense in  $\mathcal{A}_{R}$. 
   Let $a_{BH}(\alpha) = \Delta_{\Phi_{BH}}^{i\alpha/\beta_{BH}} a_{BH} \Delta_{\Phi_{BH}}^{-i\alpha/\beta_{BH}}$ and $a_{CH,s}(\alpha) = \Delta_{\Phi_{CH}}^{i\alpha/\beta_{CH}} a_{CH,s} \Delta_{\Phi_{CH}}^{-i\alpha/\beta_{CH}}$. The left hand side of \eqref{eq:toshow} then becomes
 \begin{align}\label{eq:dStraceIdLHS}
  \int d X_\Phi ds ds'&\,e^{\scalebox{.6}{$\Delta$} \beta+ i s + is') X_\Phi}   \braket{\Phi|b_{BH} b_{CH,s} \Delta_{\Phi_{CH}}^{-is/\beta_{CH}}e^{-\beta_{BH} \Lambda_{\Phi,\ell}} a_{BH} a_{CH,s'}|\Phi}\nonumber
  \\& = 2 \pi \int ds \, \braket{\Phi| b_{BH} b_{CH,s}(s) e^{-\beta_{CH} \Lambda_{\Phi,\ell}} a_{BH} a_{CH, (i \scalebox{.6}{$\Delta$} \beta - s)}|\Phi}
 \end{align}
 while the right hand side becomes
 \begin{align}\label{eq:dStraceIdRHS}
  \int d X_\Phi ds' ds&\,e^{(\scalebox{.6}{$\Delta$} \beta+ i s + is') X_\Phi}  \braket{\Phi| a_{BH} a_{CH,s'} \Delta_{\Phi_{BH}}^{-is'/\beta_{BH}} b_{BH} b_{CH,s} \Delta_{\Phi_{CH}}^{-is/\beta_{CH}}e^{-\beta_{BH} \Lambda_{\Phi,\ell}} |\Phi}\nonumber
  \\& = 2 \pi \int ds \braket{\Phi| a_{BH} a_{CH,(i\scalebox{.6}{$\Delta$} \beta -s)}(i \scalebox{.6}{$\Delta$} \beta)  b_{BH}b_{CH,s}(s) e^{-\beta_{BH} \Lambda_{\Phi,\ell}} |\Phi}
 \end{align}
 
 Let $A' = J A J$. By the definition of the Tomita operator $S_{\Phi_{BH}} =  J_{BH} \Delta_{\Phi_{BH}}^{1/2}$, we have
 \begin{align}
     a_{BH} \ket{\Phi_{BH}} = S_{\Phi_{BH}} a_{BH}^\dagger \ket{\Phi_{BH}} =  J_{BH} a_{BH}^\dagger(-i\beta_{BH}/2) \ket{\Phi_{BH}} = a_{BH}^\dagger(-i\beta_{BH}/2)' \ket{\Phi_{BH}}
 \end{align}
 Similarly $a_{CH,s} \ket{\Phi_{CH}}  =  a_{CH,s}^\dagger(-i \beta_{CH}/2)' \ket{\Phi_{CH}}$.
 Furthermore, since $a_{BH}' \in \mathcal{A}_{BH,\ell}$, we have
 \begin{align}
[a_{BH}', \beta_{BH} \Lambda_{\Phi}] = [a_{BH}', - \log \Delta_{\Phi_{BH}}] = -\left([a_{BH}, \log \Delta_{\Phi_{BH}}]\right)'
 \end{align}
 and similarly $[a_{CH,s}', \beta_{CH} \Lambda_{\Phi}] = -\left([a_{CH,s}, \log \Delta_{\Phi_{CH}}]\right)'$. It follows that 
 \begin{align}
   &\braket{\Phi| b_{BH} b_{CH,s}(s) e^{-\beta_{CH} \Lambda_{\Phi,\ell}} a_{BH} a_{CH,(i\scalebox{.6}{$\Delta$} \beta -s)}|\Phi} \\
   &= \braket{\Phi| b_{BH} b_{CH,s}(s) e^{-\beta_{BH} \Lambda_{\Phi,\ell}} a_{BH}^\dagger(-i \beta_{BH}/2)' a_{CH,(i\scalebox{.6}{$\Delta$} \beta -s)}^\dagger(-i \beta_{CH}/2)'|\Phi}\nonumber
   \\&= \braket{\Phi| a_{BH}^\dagger(i \beta_{BH}/2)' a_{CH,(i\scalebox{.6}{$\Delta$} \beta -s)}^\dagger(i \beta_{BH}-i \beta_{CH}/2)'b_{BH} b_{CH,s}(s) e^{-\beta_{BH} \Lambda_{\Phi,\ell}} |\Phi}\nonumber
   \\&= \braket{\Phi| a_{BH} a_{CH,(i\scalebox{.6}{$\Delta$} \beta -s)}(i \scalebox{.6}{$\Delta$} \beta) b_{BH} b_{CH,s}(s) e^{-\beta_{BH} \Lambda_{\Phi,\ell}} |\Phi}
 \end{align}
 This proves the equivalence of \eqref{eq:dStraceIdLHS} and \eqref{eq:dStraceIdRHS} and hence of \eqref{eq:toshow} and the KMS condition \eqref{eq:BHKMS}. 
 
 It is then comparatively straightforward to complete the proof that \eqref{eq:bhtr} is tracial. As in the analogous argument in Section \ref{sec:inflatonalgebra}, for any $\tilde a, \tilde b \in \widetilde{\mathcal{A}}_{R}$ we have 
 \begin{align}
     \widetilde{a} \widetilde{b} = \int dt dt' a(t') b(t) = \int dt dt'' a(t + t'') b(t).
 \end{align}
 Hence
 \begin{align}
     \Tr[\widetilde{a} \widetilde{b}] &=  \int dt'' d X_{\Phi}\,e^{\scalebox{.6}{$\Delta$} \beta X_{\Phi}} \braket{\Phi|e^{-\beta_{BH} \Lambda_{\Phi,\ell}/2} a(t + t'') b(t) e^{-\beta_{BH} \Lambda_{\Phi,\ell}/2} |\Phi}
     \\&=  \int dt'' d X_{\Phi}\,e^{\scalebox{.6}{$\Delta$} \beta X_{\Phi}} \braket{\Phi|e^{-\beta_{BH} \Lambda_{\Phi,\ell}/2}  b(t) a(t + t'' + i\beta_{BH}) e^{-\beta_{BH} \Lambda_{\Phi,\ell}/2} |\Phi}
     \\&=  \int dt' d X_{\Phi}\,e^{\scalebox{.6}{$\Delta$} \beta X_{\Phi}} \braket{\Phi|e^{-\beta_{BH} \Lambda_{\Phi,\ell}/2}  b(t) a(t + t') e^{-\beta_{BH} \Lambda_{\Phi,\ell}/2} |\Phi}
     \\&= \Tr[\widetilde{b}\widetilde{a}],\label{eq:bhtracialproofend}
 \end{align}
In the third equality, we  defined $t' = t''+ i \beta_{BH}$ and shifted the contour of integration to the real $t'$ line. 

As in Section \ref{sec:inflatonalgebra}, the proof of \eqref{eq:bhtracialproofend} given \eqref{eq:BHKMS} would have been even simpler if we could have directly replaced $a(t)$ by $\tilde a$ in \eqref{eq:bhtr}. However because \eqref{eq:bhtr} is boost invariant, doing so leads to a universal divergent factor for any $\tilde a \in \widetilde{\mathcal{A}}_R$ because of the integral over $t$.

As in \eqref{eq:inflatontrace2}, we can remove this divergent factor and define \eqref{eq:bhtr} explicitly as a linear functional of $\tilde a$ by writing
\begin{align}
    \Tr(\tilde a) = C \lim_{X_{\rm max} \to + \infty} \frac{1}{X_{\rm max}} \int_{- \infty}^{X_{\rm max}} d X_{\Phi}\,e^{\scalebox{.6}{$\Delta$} \beta X_{\Phi}} \braket{\Phi|e^{-\beta_{BH} \Lambda_{\Phi,\ell}/2} a(t) e^{-\beta_{BH} \Lambda_{\Phi,\ell}/2}|\Phi}
\end{align}
Here $C$ is a constant that depends on the first and second moments of the net energy flux between the horizons in the Unruh state. This is because the late-time behaviour of $X_\Phi$ and $X_{\Phi,R}$ is determined, like $\phi_{\rm cl}$, by a Wiener process, this time with positive, rather than negative, drift. It follows that, as $X_{\rm max} \to + \infty$, the probability $p(X_\Phi,t)$ for finite $X_\Phi$ and large $t$ remains roughly constant until $t \sim X_{\rm max}/ 2C$ for some $C >0$, whereupon it quickly drops to zero.

 Because the trace $\Tr$ exists and has $\Tr(\mathds{1}) = \infty$ and, to the best of our knowledge, the algebra $\mathcal{A}_R$ has trivial center, $\mathcal{A}_R$ must be either a Type I$_\infty$ or Type II$_\infty$ von Neumann factor. As in the inflaton case, we rule out Type I$_\infty$ because of the existence of a trace-rescaling automorphism for the algebra. In this case the automorphism is $X_\Phi \to X_\Phi + \kappa$ for some constant $\kappa$, which rescales the trace by a factor of $\exp(\scalebox{.6}{$\Delta$} \beta \kappa)$. This generates an automorphism of $\widetilde{\mathcal{A}}_R$ because $H$ is invariant under $X_\Phi \to X_\Phi + \kappa$.

 Since increasing $X_\Phi \to X_\Phi + \kappa$ increases $\boldsymbol{\hat A_{CH}}/4G$ by $\beta_{CH} \kappa$ while decreasing $\boldsymbol{\hat A}_{BH}/4G$ by $\beta_{BH} \kappa$, we see that this rescaling correctly captures the change in the number of total number of horizon microstates $\exp((A_{BH} + A_{CH})/4G)$.

 To make a more precise match between entropies of the algebra $\widetilde{\mathcal{A}}_R$ and generalised entropy, we follow \cite{CPW} in considering states with a semiclassical timeshift $T \approx 0$. Specifically, we consider states of the form
 \begin{align}
     \ket{\hat\Psi} &= \varepsilon^{-1/2}\int dT f(T/\varepsilon) \ket{\Psi}\ket{T}
     \\& = \varepsilon^{1/2}\int dX_\Phi F(\varepsilon X_\Phi) \ket{\Psi}\ket{X_\Phi}
 \end{align}
 where $\varepsilon > 0$ is small, $f$ is a square-integrable function with Fourier transform $F$, $\ket{\Psi} \in \mathcal{H}_0$ is an arbitrary state and $\ket{T}$ and $\ket{X_\Phi}$ are respectively delta-function normalised position and momentum eigenstates for $L^2(\R)$. For simplicity of notation, we assume as in Section \ref{sec:bddensity}, that $\ket{\Psi}$ is already rotation invariant. If not, we simply integrate over rotations of $\ket{\Psi}$ and purify the resulting state on $\mathcal{A}_r$.

 We then claim that the density matrix $\rho_{\hat\Psi}$ for $\ket{\hat\Psi}$ on $\widetilde{\mathcal{A}}_R$ can be written as
 \begin{align}\label{eq:approxrhobhds}
     \rho_{\hat\Psi} \approx \int dt \,\varepsilon e^{iHt}\overline{F}(\varepsilon X_{\Phi,R}) e^{\beta_{BH} \Lambda_{\Phi,\ell}/2} \Delta_{\Psi|\Phi} e^{\beta_{BH} \Lambda_{\Phi,\ell}/2}  e^{-\scalebox{.6}{$\Delta$} \beta X_\Phi} F(\varepsilon X_{\Phi,R}) e^{-iHt}.
 \end{align}
 Here $\Delta_{\Psi|\Phi}$ is the relative modular operator of $\ket{\Psi}$ relative to $\ket{\Phi} = \ket{\Phi_{BH}}\ket{\Phi_{CH}}$ on $\mathcal{A}_r$. To verify this, we need to show that the right-hand side  of \eqref{eq:approxrhobhds} is affiliated to $\widetilde{\mathcal{A}}_R$ and that
 \begin{align}\label{eq:bhdensityworks}
     \Tr(\rho_{\hat\Psi}\tilde a) \stackrel{?}{\approx} \braket{\hat\Psi|a|\hat\Psi}
 \end{align}
 for all $\tilde a \in \mathcal{A}_R$. 

 To show the former, since \eqref{eq:approxrhobhds} is manifestly boost invariant, it suffices to show that
 \begin{align}\label{eq:isitinAR}
     [e^{\beta_{BH} \Lambda_{\Phi,\ell}/2} \Delta_{\Psi|\Phi} e^{\beta_{BH} \Lambda_{\Phi,\ell}/2}  e^{-\scalebox{.6}{$\Delta$} \beta X_\Phi}, a'] \stackrel{?}{=} 0
 \end{align}
 for all $a' \in \mathcal{A}_L = \mathcal{A}_R'$. Clearly \eqref{eq:isitinAR} commutes with $X_\Phi$. It remains to check
 \begin{align}
     a' = a'_{BH} a'_{CH}
 \end{align}
 with $a'_{BH} \in \mathcal{A}_{BH,\ell}$ and 
 \begin{align}
     a'_{CH} \in \Delta_{\Phi_{CH}}^{iT/\beta_{CH}} \,\mathcal{A}_{CH, \ell}\, \Delta_{\Phi_{CH}}^{-iT/\beta_{CH}}.
 \end{align}
 We have
 \begin{align}
     \beta_{CH} [X_\Phi,a'_{CH}] = [\log \Delta_{\Phi_{CH}}, a'_{CH}] =  -\beta_{CH} [\Lambda_{\Phi,\ell},a'_{CH}] =   [\log \Delta_{\Psi|\Phi}, a'_{CH}]
 \end{align}
 and
 \begin{align}
     \beta_{BH} [\Lambda_{\Phi,\ell},a'_{BH}] = - [\log \Delta_{\Psi|\Phi}, a'_{BH}].
 \end{align}
 The desired result \eqref{eq:isitinAR} follows immediately.

 To show \eqref{eq:bhdensityworks}, we have
 \begin{align}
     \Tr(\rho_{\hat\Psi}\tilde a) &\approx \int dX_\Phi |F(\varepsilon X_\Phi)|^2 \braket{\Phi| \Delta_{\Psi|\Phi} e^{\beta_{BH} \Lambda_{\Phi,\ell}/2} \tilde a e^{-\beta_{BH} \Lambda_{\Phi,\ell}/2}|\Phi}
     \\&\approx \int dX_\Phi |F(\varepsilon X_\Phi)|^2 \braket{\Phi| \Delta_{\Psi|\Phi} \tilde a |\Phi}
     \\&\approx \int dX_\Phi |F(\varepsilon X_\Phi)|^2 \braket{\Psi| \tilde a |\Psi}.
     \\&\approx \braket{\hat\Psi|\tilde a |\hat\Psi}
 \end{align}
 In the first step we used the fact that $F(\varepsilon X_{\Phi,R})$ is slowly varying as a function of $X_{\Phi,R}$ to commute it past $\tilde a$ and $e^{-\beta_{BH} \Lambda_{\Phi,\ell}/2}$ while incurring only $O(\varepsilon)$ error. In the second step we used 
 \begin{align}
     [\Lambda_{\Phi,\ell},\tilde a] = [H, \tilde a] = 0.
 \end{align}
 Finally, in the third step we used (\ref{eq:appxeq}).

 To compute the entropy of $\rho_{\hat\Psi}$, we write
 \begin{align}
     \rho_{\hat\Psi} &\approx \int dt \,\varepsilon e^{i\Lambda_{\Phi,\ell} t}\overline{F}(\varepsilon X_{\Phi,R}) e^{\beta_{BH} \Lambda_{\Phi,\ell}/2} \Delta_{\Psi|\Phi} e^{\beta_{BH} \Lambda_{\Phi,\ell}/2}  e^{-\scalebox{.6}{$\Delta$} \beta X_\Phi} F(\varepsilon X_{\Phi,R}) e^{-i\Lambda_{\Phi,\ell}t}
     \\&\approx \int dt \,\varepsilon \overline{F}(\varepsilon X_{\Phi,R}) e^{\beta_{BH} \Lambda_{\Phi,\ell}/2} e^{i\Lambda_{\Phi,\ell} t}\Delta_{\Psi|\Phi}e^{-i\Lambda_{\Phi,\ell}t} e^{\beta_{BH} \Lambda_{\Phi,\ell}/2}  e^{-\scalebox{.6}{$\Delta$} \beta X_\Phi} F(\varepsilon X_{\Phi,R}) 
     \\&\approx \varepsilon \overline{F}(\varepsilon X_{\Phi,R}) e^{\beta_{BH} \Lambda_{\Phi,\ell}/2} \Delta_{\tilde \Psi|\Phi} e^{\beta_{BH} \Lambda_{\Phi,\ell}/2}  e^{-\scalebox{.6}{$\Delta$} \beta X_\Phi} F(\varepsilon X_{\Phi,R})
 \end{align}
 where the semifinite weight $\ket{\tilde\Psi}$ on $\mathcal{A}_r$ is defined by
 \begin{align}
     \braket{\tilde\Psi|a|\tilde\Psi} = \int dt\braket{\Psi|e^{i\Lambda_{\Psi,\ell}t} a e^{-i\Lambda_{\Psi,\ell}t}|\Psi} = \int dt\braket{\Psi|e^{i H_0t} a e^{-i H_0 t}|\Psi}.
 \end{align}
 Since $\ket{\tilde\Psi}$ is invariant under conjugation by $e^{-it\Lambda_{\Phi,\ell}}$ as a weight on $\mathcal{A}_r$ and the weight on $\mathcal{A}_\ell$ defined by $\ket{\Phi}$ is also invariant under
 conjugation by $e^{-it\Lambda_{\Phi,\ell}}$, we have
 \begin{align}
     [\Delta_{\tilde \Psi|\Phi}, \Lambda_{\Phi,\ell} ] = 0
 \end{align}
 and hence
 \begin{align}
     \log \left(e^{\beta_{BH} \Lambda_{\Phi,\ell}/2} \Delta_{\tilde \Psi|\Phi} e^{\beta_{BH} \Lambda_{\Phi,\ell}/2}\right) = \log  \Delta_{\tilde \Psi|\Phi} + \beta_{BH} \Lambda_{\Phi,\ell}.
 \end{align}
 Finally, we obtain 
 \begin{align}
     \log \rho_{\hat\Psi} &\approx \log\left( \varepsilon |F(\varepsilon X_{\Phi,R})|^2\right) - \scalebox{.6}{$\Delta$} \beta X_\Phi + \log  \Delta_{\tilde \Psi|\Phi} - \beta_{BH} \Lambda_{\Phi,\ell},
     \end{align}
 where we again used the fact that $F(\varepsilon X_{\Phi,R})$ commutes with everything up to $O(\varepsilon)$ corrections. So
 \begin{align}
    S(\rho_{\hat\Psi}) &=  -\braket{\hat\Psi|\log \rho_{\hat\Psi}|\hat\Psi}
    \\&\approx -\braket{\log\left( \varepsilon |F(\varepsilon X_{\Phi,R})|^2\right)} + \scalebox{.6}{$\Delta$} \beta\braket{X_\Phi} - \braket{\log \Delta_{\tilde\Psi|\Phi}} + \beta_{BH} \braket{\Lambda_{\Phi,\ell}}
    \\&\approx-\braket{\log\left( \varepsilon |F(\varepsilon X_{\Phi,R})|^2\right)} + \frac{\scalebox{.6}{$\Delta$} \beta}{\beta_{CH}} \braket{\frac{\boldsymbol{\hat A}_{CH} - A_{CH,0}}{4G}} + S_{\rm ren}(\tilde\Psi) - \beta_{BH} \braket{\boldsymbol{h}_r}\label{eq:substitute}
    \\&\approx-\braket{\log\left( \varepsilon |F(\varepsilon X_{\Phi,R})|^2\right)} +  \braket{\frac{\boldsymbol{\hat A}_{CH} - A_{CH,0}}{4G}} +S_{\rm ren}(\tilde\Psi) +\braket{\frac{\boldsymbol{\hat A}_{BH} - A_{BH,0}}{4G}}, \label{eq:komarsubstitute}
 \end{align}
 where, as with \eqref{eq:formalgenentinflaton},
 \begin{align} 
     S_{\rm ren}(\tilde\Psi) = \braket{\Psi| \log \Delta_{\tilde\Psi|\Phi} + \beta_{BH} h_{\Phi_{BH},\ell} + \beta_{CH} h_{\Phi_{CH},\ell} |\Psi}
 \end{align}
 can be formally interpreted (up to a divergent constant) as the entropy of the boost-invariant weight $\ket{\tilde\Psi}$ divided by its divergent normalisation. In \eqref{eq:substitute}, we substituted in the formulas \eqref{eq:modopsplit}, \eqref{eq:deltaphiBHsplit}, \eqref{eq:lambda} and \eqref{eq:ptinterpretation} for $\log \Delta_{\Phi}$, $\Lambda_{\Phi,\ell}$ and $X_\Phi$ in terms of sesquilinear forms. Then, in \eqref{eq:komarsubstitute}, we used \eqref{eq:nomissingoperators}.

 The first term in \eqref{eq:komarsubstitute} describes the entropy of fluctuations in the horizon areas, or equivalently in the timeshift. The second term is the entropy of the $\ket{\tilde \Psi}$ at fixed timeshift. Finally we have area terms for the black hole and cosmological horizons. Up to the usual divergent constant, the full formula \eqref{eq:komarsubstitute} therefore describes the generalised entropy for the state $\ket{\hat\Psi}$, with the qualification that the state $\ket{\Psi}$ of the quantum fields needs to be formally averaged over boosts.

 \section{Generic extremal surfaces} \label{sec:generic}
 In this section, we conclude by briefly discussing gravitational algebras in a generic classical background. Our set up is similar to that described in \cite{Jensen_2023}, but our point of view and conclusions will differ somewhat from theirs.
 
 A classical background is defined by a classical Lorentzian metric $g_{\rm cl}$ and a classical matter field configuration $\phi_{\rm cl}$, that form a solution to the classical Einstein equations
 \begin{align}
     R_{\mu \nu} - \frac{1}{2} g_{\mu\nu} R = 8\pi G T_{\mu \nu}.
 \end{align}
 Here, the classical matter fields $\phi_{\rm cl}$ need to be scaled so that their stress-energy tensor $T_{\mu\nu} \sim O(1/G)$ in the $G \to 0$ limit. We can then consider the Hilbert space $\mathcal{H}_{\rm QFT}$ of quantum fluctuations of both the graviton field $h = (g - g_{\rm cl})/\sqrt{G}$ and quantum perturbations $\delta \phi$ of the matter fields (scaled to have $O(1)$ stress-energy tensor and hence vanishing backreaction as $G \to 0$). 

 Generically, such a background will have no isometries whose generators need to be treated as gauge constraints in the $G \to 0$ limit. We do need to impose gauge constraints associated to perturbative diffeomorphisms, but these act only on the graviton field in the strict $G \to 0$ limit and can be dealt with using standard techniques. It is therefore possible to specify subregions in a gauge-invariant manner. Moreover the algebra $\mathcal{A}_0$ of quasilocal QFT operators within a given subregion are already gauge invariant as quantum gravity observables.\footnote{Here, we define $\mathcal{A}_0$ to only include operators that are already gauge invariant with respect to any local matter gauge group and, in the case of graviton operators, with respect to perturbative diffeomorphisms.} There is no need to introduce any clock. By usual quantum field theory arguments, this algebra will be Type III$_1$ factor. 

 In particular, we can consider the algebra $\mathcal{A}_0$ associated to a domain of dependence, or wedge, in the classical background $g_{\rm cl}$ whose boundary is an extremal surface. We will assume for the moment that the wedge is compact, but will later discuss the case where it also has an asymptotic boundary in the conformal compactification of the spacetime. Except in special cases like the ones discussed in Sections \ref{sec:inflaton} and \ref{sec:blackholes}, such a wedge will not be a causal diamond associated to any particular worldline, and hence $\mathcal{A}_0$  will not be the algebra of observables associated to a single observer. But the algebra $\mathcal{A}$ nonetheless exists. 

 With this restriction in place, we can also consider classical geometries with a kink, or shift in boost angle, across the extremal surface, analogous to the introduced for SdS black holes in Section \ref{sec:sds}. Because the boundary surface is extremal, the resulting classical geometry will continue to satisfy the classical Einstein equations. For each shift in boost angle, there is a different Hilbert space $\mathcal{H}_T$ describing quantum fluctuations about this background.
 We can then consider the direct integral Hilbert space
 \begin{align}
     \mathcal{H} = \int^\oplus dT\, \mathcal{H}_T
 \end{align}
 that describes smooth superpositions over boost angles.

 There is a unambiguous and gauge-invariant natural action of the algebra $\mathcal{A}_0$ on the Hilbert space $\mathcal{H}_T$ for any boost angle $T$. But there are also additional operators localised within the wedge that change the boost angle and that have finite $G \to 0$ limits acting on the Hilbert space $\mathcal{H}$. In particular, for any state $\ket{\Phi} \in \mathcal{H}_0$, there is a densely defined operator $X_\Phi$ localised in the wedge defined by 
 \begin{align}
     X_{\Phi} = \frac{\boldsymbol{\hat A}_{\rm ext} - A_{{\rm ext},0}}{4 G} + \boldsymbol{h}_{\Phi} = H_{\Phi}' - \log \Delta_{\Phi}.
 \end{align}
 Here the operator $\boldsymbol{\hat A}_{\rm ext}$ describes the area of the extremal surface (including $O(G)$ perturbative corrections), $A_{{\rm ext},0}$ is the area of the same surface in the classical background geometry, $\boldsymbol{h}_{\Phi}$ is a formal one-sided modular Hamiltonian for the state $\ket{\Phi}$ on the algebra $\mathcal{A}$ (defined so that $\braket{\Phi|\boldsymbol{h}_{\Phi}|\Phi} = 0$) and 
 \begin{align}
     X_{\Phi}' = \frac{\boldsymbol{\hat A}_{\rm ext} - A_{{\rm ext},0}}{4 G} + \boldsymbol{h}_{\Phi}'
 \end{align} 
 commutes with both $\mathcal{A}_0$ and $X_{\Phi}$. As before, $\hat A_{\rm ext}$, $\boldsymbol{h}_{\Phi}$ and $\boldsymbol{h}_{\Phi}'$ exist only as sesquilinear forms, but $X_\Phi'$ and the modular operator $-\log \Delta_\Phi = \boldsymbol{h}_{\Phi} - \boldsymbol{h}_{\Phi}'$ for $\ket{\Phi}$ on $\mathcal{A}$ are densely defined operators. We can identify the Hilbert spaces $\mathcal{H}_t \cong \mathcal{H}_0$ in a way that identifies the natural actions of $\mathcal{A}$ on each and such that induced isomorphism
 \begin{align}\label{eq:genericident}
     \mathcal{H} \cong \mathcal{H}_0 \otimes L^2(\R)
 \end{align}
 allows us to identify $X_{\Phi}'$ with $-2\pi i \partial_T \in \mathcal{B}(L^2(\R))$. Together $\mathcal{A}_0$ and $X_{\Phi}$ generate the full wedge algebra $\mathcal{A}$, which is therefore just the modular crossed product. The commutant algebra $\widetilde{\mathcal{A}}'$ is generated by $X_{\Phi}'$ and $\Delta_{\Phi}^{-iT/2\pi} \mathcal{A}' \Delta_{\Phi}^{iT/2\pi}$. It is easy to check that the latter agrees the unique natural action of $\mathcal{A}'$ on $\mathcal{H}_T$ given the identification \eqref{eq:genericident}. The gravitational commutant algebra $\mathcal{A}_0'$ is can be identified as usual with the algebra of observables localised in the complementary wedge.

 For $a \in \mathcal{A}$, we can write as usual
 \begin{align}
     \Tr(a) = \int dX_{\Phi}'  \braket{\Phi|a|\Phi},
 \end{align}
 with $\braket{\Phi|a|\Phi}$ a function of $X_\Phi'$. For semiclassical states $\ket{\hat \Psi} \in \mathcal{H}$ of the form
 \begin{align}
     \ket{\hat \Psi} = \int dT \,\varepsilon^{-1/2}f(T/\varepsilon) \ket{\Psi}\ket{T}
 \end{align}
 with $\varepsilon >0$ small, $f$ a square-integrable function with Fourier transform $F$ and $\ket{\Psi} \in \mathcal{H}_0$,  the density matrix $\rho_{\hat \Psi}$ satisfies \cite{CPW}
 \begin{align} \label{eq:logrhoexpansion}
     -\log \rho_{\hat \Psi} \approx  X_\Phi' - \log \Delta_{\Psi|\Phi} - \log\left[\varepsilon |F(\varepsilon X_\Phi)|^2\right]
 \end{align}
 Since
 \begin{align}
     X_\Phi' - \log \Delta_{\Psi|\Phi} = \frac{\boldsymbol{\hat A}_{\rm ext} - A_{{\rm ext},0}}{4 G} + \boldsymbol{h}_{\Psi}
 \end{align}
 with $\boldsymbol{h}_{\Psi}$ the one-sided modular operator for the state $\ket{\Psi}$, the entropy $S(\hat \Psi) = -\braket{\hat\Psi|\log \rho_{\hat \Psi}|\hat \Psi}$ is equal to the generalised entropy of the wedge (up to the usual divergent constant).\footnote{The last term in \eqref{eq:logrhoexpansion} just gives the entropy of fluctuations in the area of the extremal surface; see \cite{CPW} for details.}

 So far, we have not talked about the possibly of the wedge containing an asymptotic boundary or localised observer of the form described in \cite{CLPW}. This was because, in the absence of any isometries of the classical background, there was no need to do so in order to obtain a nontrivial Type II algebra. If an asymptotically-flat or -AdS boundary exists, we can take $G \to 0$ limits where the fluctuations in the ADM mass of that boundary are either $O(1/\sqrt{G})$ or $O(1)$. However in both cases the algebra is the same. Because the time translation symmetry generated by the ADM mass only exists asymptotically and is broken by the classical background, the boundary time (relative to the background) and the ADM mass are both gauge-invariant observables can be measured within the wedge. As a result, the wedge algebra will contain a $\mathcal{B}(L^2(\R))$ tensor product subfactor on which the fluctuations in the ADM mass and the boundary time act as conjugate variables. If the fluctuations in the ADM mass are $O(1/\sqrt{G})$, the fluctuations in the boundary time will be $O(\sqrt{G})$. If the ADM mass fluctuations are $O(1)$, the fluctuations in the boundary time will also be $O(1)$. This affects the definition of operators that are dressed to the boundary, but not of operators that are dressed to the classical background (and hence not the overall structure of the algebra). Similarly, including an observer as in \cite{CLPW} just introduces an additional tensor product factor $\mathcal{B}(L^2(\R^+))$ to the wedge algebra, without otherwise affecting its structure. In particular, the time read by the observer's clock at a particular point in the classical background is a physical gauge-invariant observable.\footnote{Because the clock energy is bounded from below, the time shown by the clock can only be defined approximately. But there are no issues with gauge invariance.}

 \subsection*{Acknowledgements}
 We would like to thank Edward Witten, Juan Maldacena, Jonah Kudler-Flam, Sam Leutheusser and Gautam Satishchandran for valuable discussions. This work was supported by the Berkeley Center for Theoretical Physics, the Department of Energy through QuantISED Award DE-SC0019380 and an Early Career Award DE-FOA-0002563, by AFOSR award FA9550-22-1-0098 and by a Sloan Fellowship.

 \appendix
 \section{Glossary}
 \textbf{Note:} The bold font is reserved for the sesquilinear forms that are not operators. Tilded operators and algebras are gauge-invariant with respect to isometries of the background spacetime.

  \noindent\textbf{Section 2}
  \begin{enumerate}
      \item $\mathcal{H}$: The unique natural Hilbert space of quantum fields and free gravitons for the global de Sitter space in the strict $G\rightarrow 0$ limit. It is well-defined because the Cauchy slice in the global slicing is compact.
 \item $\mathcal{A}$: The static-patch algebra of QFT operators.
 \item $\widetilde{\mathcal{A}}$: The subalgebra of $\mathcal{A}$, after imposing the gauge constraints associated with the isometry of the static patch.
 \item $H:$ the boost Hamiltonian.
\item $\ket{\Psi_{BD}}:$ The Bunch-Davies weight. It is unnormalizable, annihilated by the boost $H$, satisfies the KMS condition for $\mathcal{A}$, is tracial for $\mathcal{A},$ and is \emph{not} in the natural Hilbert space $\mathcal{H}.$ We often abbreviate it as $\Psi$.
\item $\ket{\Phi}$: An arbitrary normalizable state in the Hilbert space $\mathcal{H}$.
  \end{enumerate}

 \noindent\textbf{Section 3}\\
  \textbf{Asymptotically-flat black holes}
  \begin{enumerate}
  \item $T$: The asymptotic right boundary time reached
by starting at $t=0$ on the left boundary and travelling along a spacelike geodesic orthogonal to the time-translation Killing vector. The mode is locally pure gauge. We will call $T$ the timeshift mode.
  \item $\mathcal{H}_{\rm QFT}$: The natural Hilbert space of quantum fields and free gravitons for an asymptotically flat black hole in the strict $G\rightarrow 0$ limit; see Figure \ref{fig:preferredmodes}. $\mathcal{H}$ consists of states that look like the Minkowski vacuum both near the black hole bifurcation surface and spatial infinity. The Hilbert space $\mathcal{H}$ does not contain any invariant states such as the Hartle-Hawking, Unruh or Boulwar\'{e} vacua. Can be written as $\mathcal{H}_{\rm QFT} \cong \mathcal{H}_0 \otimes L^2(\R)$.
 \item $\mathcal{H}_0$: The Hilbert space of quantum fields and free gravitons with the timeshift frozen at $T=0$. The full Hilbert space can be decomposed as $\mathcal{H} =\mathcal{H}_0\otimes L^2(\R),$ where the timeshift $T$ acts as the position operator on $L^2(\R)$.
  \item $\mathcal{H}$: The Hilbert space of quantum fields with $O(1)$ (rather than $O(\sqrt{G})$) timeshift fluctuations. Like for $\mathcal{H}_{\rm QFT}$, we have  $\mathcal{H}\cong\mathcal{H}_0 \otimes L^2(\R)$, but now the timeshift (rather than the timeshift rescaled by $O(G^{-1/2}$) acts as the position operator on $L^2(\R)$. See the paragraph surrounding (\ref{eq:hsemi}). This is the Hilbert space on which the Type II algebra $\mathcal{A}_R$ acts.
  \item $\Phi$: An arbitrary normalizable state in the Hilbert space $\mathcal{H}_0$.

  \item $H:$ The boost Hamiltonian on $\mathcal{H}_0$. 
  \item $h_R,$ $h_L:$ The renormalized ADM masses. They satisfy $H=h_R-h_L$ and are densely-defined and unbounded. They are canonically conjugated to the timeshift $T$, thus act as the momemtum operator on $L^2(\R).$ Notice that they are \emph{not} the one-sided boosts, which are only sesquilinear forms.
  \item $\scalebox{.8}{$\Delta$} H_{\Phi, r}$, $\scalebox{.8}{$\Delta$} H_{\Phi, \ell}:$ Densely-defined, unbounded operators that capture the difference between the modular Hamiltonian of $\Phi$ and the boost Hamiltonian. See (\ref{eq:tildesplitting}) and the surrounding paragraph.
  \item $X_{\Phi,R}$: Defined by  $X_{\Phi,R} = h_{R}-\scalebox{.8}{$\Delta$} H_{\Phi, r}$. Can be formally interpreted as the sum of the area operator and a one-sided modular Hamiltonian for $\ket{\Phi}$; see \eqref{eq:hphiR}.
\item $\mathcal{A}_r$, $\mathcal{A}_\ell$: Type III von Neumann factors for QFT observables localised in the right and left exteriors, respectively. They are commutant of each other, $\mathcal{A}_r=\mathcal{A}_\ell'$.

\item $\mathcal{A}_R$, $\mathcal{A}_L$: The full gravitational algebras for the right and left exteriors respectively, generated by $\mathcal{A}_{r/\ell}$ along with (bounded functions of) $h_{R/L}$. They are commutants $\mathcal{A}_L=\mathcal{A}_R'$. 

\end{enumerate}
  \textbf{Schwarzchild-de Sitter (SdS) black holes}
  \begin{enumerate}
   \item $T$: The shift in boost time when travelling through the black hole and cosmological horizons. The mode is locally pure gauge. We will call $T$ the timeshift mode. See the red slice in Figure \ref{fig:sds}.
  \item $\mathcal{H}$: The Hilbert space of quantum fields and free gravitons for a SdS black hole in the strict $G\rightarrow 0$ limit, where the timeshift mode has $O(1)$ fluctuation.
\item $\mathcal{H}_T$: The Hilbert space for a Schwarzschild-de Sitter black hole with fixed timeshift $T$. We have $\mathcal{H}\cong \int^{\otimes}_\R dt\, \mathcal{H}_t.$
\item $\mathcal{H}_{BH}, \mathcal{H}_{CH}$: Hilbert space subfactors of $\mathcal{H}_0 \cong \mathcal{H}_{BH} \otimes \mathcal{H}_{CH}$ defined so that all operators near the black hole bifurcation surface are contained in $\mathcal{H}_{BH}$ and all operators near the cosmological bifurcation surface are contained in $\mathcal{H}_{CH}$.
\item $\mathcal{H}_{CH,T}$: Defined so that the algebra of operators $\mathcal{B}(\mathcal{H}_{CH,T})$ is the commutant of $\mathcal{B}(\mathcal{H}_{BH})$ on $\mathcal{H}_T\cong \mathcal{H}_{BH}\otimes\mathcal{H}_{CH,T}$.
\item $U_r$, $U_\ell$: The right and left black hole exteriors in the SdS spacetime. They are invariant under the boosts. See Figure \ref{fig:sds}.
\item $U_{BH}$, $U_{CH}$: The black hole and cosmological wedges in the Penrose diagram, respectively. All operators in $U_{BH}$ act only on $\mathcal{B}(\mathcal{H}_{BH})$ while all operators in $U_{CH}$ act only on $\mathcal{B}(\mathcal{H}_{CH})$. They are not invariant under the boosts. See Figure \ref{fig:wedge}. 
\item $\mathcal{A}_{BH,r}$, $\mathcal{A}_{CH,r}$: Type III subfactors of $\mathcal{B}(\mathcal{H}_{BH})$ and $\mathcal{B}(\mathcal{H}_{CH})$ respectively, containing all operators in those algebras that are localised in the right black hole exterior. (Note that, somewhat confusingly, this is to the left of the cosmological bifurcation surface.) 
\item $\mathcal{A}_{BH,\ell}$, $\mathcal{A}_{CH,\ell}$: Commutants of $\mathcal{A}_{BH,r}$ and $\mathcal{A}_{CH,r}$ on $\mathcal{B}(\mathcal{H}_{BH}$ and $\mathcal{B}(\mathcal{H}_{CH}$. They contain all operators in those algebras that are localised in the left black hole exterior.
\item $\mathcal{A}_r$, $\mathcal{A}_\ell$: The algebras for the right and left exteriors, $\mathcal{A}_r \cong \mathcal{A}_{BH,r} \otimes \mathcal{A}_{CH,r}$ and $\mathcal{A}_\ell \cong \mathcal{A}_{BH,\ell} \otimes \mathcal{A}_{CH,\ell}$.
\item $H_0$: The boost Hamiltonian on $\mathcal{H}_0$. It generates automorphisms of $\mathcal{A}_\ell$ and $\mathcal{A}_r$ but mixes $\mathcal{H}_{BH}$ and $\mathcal{H}_{CH}$.
\item $H$: The boost Hamiltonian on $\mathcal{H}$. Defined by \eqref{eq:HdefSdS}.
\item $\boldsymbol{h}_{r,0}$, $\boldsymbol{h}_{\ell,0}$: One-sided boost Hamiltonians on the right and left wedges, respectively, at zero timeshift.
\item $\boldsymbol{h}_{\Phi_{CH},r}$, $\boldsymbol{h}_{\Phi_{CH},\ell}$: One-sided modular Hamiltonians of a state $\ket{\Phi_{CH}} \in \mathcal{H}_{CH}$ on the algebras $\mathcal{A}_{CH,r}$ and $\mathcal{A}_{CH,\ell}$ respectively.
\item $\boldsymbol{h}_{\Phi_{BH},r}$, $\boldsymbol{h}_{\Phi_{BH},\ell}$: One-sided modular Hamiltonians of a state $\ket{\Phi_{BH}} \in \mathcal{H}_{BH}$ on the algebras $\mathcal{A}_{BH,r}$ and $\mathcal{A}_{BH,\ell}$ respectively.
\item $\Lambda_{\Phi,\ell}$, $\Lambda_{\Phi,r}$: Densely-defined operators that generate boosts of $\mathcal{A}_r$ (resp. $\mathcal{A}_{\ell}$) and modular flows of $\mathcal{A}_{BH,\ell}$ and $\mathcal{A}_{CH,\ell}$ (resp. $\mathcal{A}_{BH,r}$ and $\mathcal{A}_{CH,r}$).
\item $\mathcal{A}_{R}$, $\mathcal{A}_{L}$: The full algebras of operators in the right and left exteriors before imposing gauge constraints.
\item $\widetilde{\mathcal{A}}_R$, $\widetilde{\mathcal{A}}_L$: The boost- and rotation-invariant subalgebras of $\mathcal{A}_{R}$ and $\mathcal{A}_{L}$.
\item $X_\Phi$: Densely-defined operator localised in the left wedge that is canonically conjugate to the timeshift mode $T$ given the identification \eqref{eq:H_CHtId}. The physical interpretation of $X_\Phi$ is explained in \eqref{eq:ptinterpretation}
\item $X_{\Phi , R}$: Defined by $X_{\Phi} - \log \Delta_{\Phi_{CH}}$. Localised in the right exterior.

\end{enumerate}

 \section{Some results from modular theory}\label{appx:a}
 We review some basic Tomita-Takesaki modular theory for quantum field theory. For more details, see, for example, \cite{Witten_2018,Haag1993LocalQP}. The central object of modular theory is the Tomita operator $S_\Psi$ which is an densely-defined, unbounded, antiunitary operator that satisfies
 \begin{align}
     S_\Psi\, a\ket{\Psi} = a^\dagger \ket{\Psi}
 \end{align} for some cyclic, separating states $\Psi.$ Clearly, $S_\Psi^2=1,$ so it is invertible. This implies that the Tomita operator has a unique polar decomposition,
 \begin{align}
     S_\Psi = J_\Psi \Delta_\Psi^{1/2},
 \end{align} where $J_\Psi$ is antiunitary and $\Delta_\Psi^{1/2}$ is Hermitian and positive-definite. They have some useful properties, {\it e.g.,} the modular operator preserves the state,
 \begin{align}\label{eq:prop1}
     \Delta_\Psi \ket{\Psi} = \ket{\Psi},
 \end{align} and the modular flow preserves the algebra,
 \begin{align}\label{eq:prop2}
     \Delta_\Psi^{\i s}\, \mathcal{A}\, \Delta_\Psi^{-\i s} = \mathcal{A},\quad s\in\R.
 \end{align} These will be useful below. Similarly, one can define the relative modular operators by 
 \begin{align}
     S_{\Phi|\Psi} a \ket{\Psi} = a^\dagger \ket{\Phi}
 \end{align} and 
 \begin{align}
     \Delta_{\Phi|\Psi} = S_{\Phi|\Psi}^\dagger S_{\Phi|\Psi},
 \end{align} etc.

 With these preparation, we prove a few lemmas that will be useful in the main text. The first one, also proved in \cite{CPW,CLPW}, says that the relative modular operator satisfies
 \begin{align}\label{eq:appxeq}
     \bra{\Psi}\Delta_{\Phi|\Psi} a \ket{\Psi} = \bra{\Phi} a\ket{\Phi}.
 \end{align} To prove, we write out the relative Tomita operators explicitly,
 \begin{align}
     \bra{\Psi}\Delta_{\Phi|\Psi} a \ket{\Psi}  = \bra{\Psi}S_{\Phi|\Psi}^\dagger S_{\Phi|\Psi} a \ket{\Psi} =\bra{\Psi}S_{\Phi|\Psi}^\dagger a^\dagger \ket{\Phi} =\overline{\bra{S_{\Phi|\Psi}\Psi}a^\dagger\ket{\Phi}} = \overline{\bra{\Phi}a^\dagger\ket{\Phi}} = \bra{\Phi}a\ket{\Phi},
 \end{align} as desired.
 
 Second, given a reference system $\mathcal{H}_R$ and a state $\ket{\Phi} = \ket{\Phi_0} \ket{0} + \ket{\Phi_1} \ket{1} \in \mathcal{H} \otimes \mathcal{H}_R$, the relative modular operator is $\Delta_{\Phi|\Psi} = \Delta_{\Phi_0|\Psi} + \Delta_{\Phi_1|\Psi}.$ We claim that the relative Tomita operator is 
 \begin{align}
     S_{\Phi|\Psi} = S_{\Phi_0|\Psi}\otimes \ket{0} +  S_{\Phi_1|\Psi}\otimes \ket{1},
 \end{align} where the kets $\ket{0}$ and $\ket{1}$ are meant to be the embedding maps that add an ancilla qubit in the designated state. This can be checked straightforwardly: for any $a\in\mathcal{A},$
 \begin{align}
     S_{\Phi|\Psi} \,a\otimes \mathbb{1} \ket{\Psi} =\big(S_{\Phi_0|\Psi}\otimes \ket{0} +  S_{\Phi_1|\Psi}\otimes \ket{1}\big)\, a\otimes \mathbb{1} \ket{\Psi} = a^\dagger \otimes\mathbb{1} \big(\ket{\Phi_0} \ket{0} + \ket{\Phi_1} \ket{1}\big).
  \end{align} Then, the relative modular operator can be easily computed,
  \begin{align}\label{eq:modopsadd}
      \Delta_{\Phi|\Psi} &=  S_{\Phi|\Psi} ^\dagger  S_{\Phi|\Psi} = S_{\Phi_0|\Psi}^\dagger S_{\Phi_0|\Psi}\cdot \braket{0|0}+S_{\Phi_1|\Psi}^\dagger S_{\Phi_1|\Psi}\cdot \braket{1|1} \\
      &=\Delta_{\Phi_0|\Psi} + \Delta_{\Phi_1|\Psi},
  \end{align} as desired. We have used that $\ket{0}$ and $\ket{1}$ are orthonormal. The generalisation of \eqref{eq:modopsadd} to continuous integrals over states is straightforward.
 
 The last lemma is as follows. If we define the time-evolved state $\ket{\Phi(s)} = \Delta_\Psi^{-\i s} \ket{\Phi}$, then the relative modular operator becomes
\begin{align}
    \Delta_{\Phi(s)|\Psi} = \Delta_{\Psi}^{-\i s}\,\Delta_{\Phi|\Psi}\, \Delta_{\Psi}^{\i s}.
\end{align} To show this, we claim that the time-evolved relative Tomita operator is
\begin{align}
    S_{\Phi(s)|\Psi} = \Delta_{\Psi}^{-\i s}\, S_{\Phi|\Psi}\, \Delta_{\Psi}^{\i s}.
\end{align} This can be checked straightforwardly: for any $a\in \mathcal{A},$
\begin{align}
    S_{\Phi(s)|\Psi} a \ket{\Psi} &= \Delta_{\Psi}^{-\i s}\, S_{\Phi|\Psi}\, \Delta_{\Psi}^{\i s} a \ket{\Psi} \\
    &= \Delta_{\Psi}^{-\i s}\, S_{\Phi|\Psi}\, \big( \Delta_{\Psi}^{\i s} a \Delta_{\Psi}^{-\i s}\big) \ket{\Psi}\\
    &=\Delta_{\Psi}^{-\i s}\big( \Delta_{\Psi}^{\i s} a^\dagger \Delta_{\Psi}^{-\i s}\big) \ket{\Phi}\\
    & = a^\dagger \ket{\Phi(s)}.
\end{align} We used (\ref{eq:prop1}) in the second equality and (\ref{eq:prop2}) in the third one.
 
 \section{Spherical harmonics on $S^3$}\label{sh}
 We briefly summarize some facts about (hyper)spherical harmonics on the three-sphere that will be useful for us in the main text. For more details, see, for example, the book \cite{sphere}, especially their (3.86) and (3.87). For our purposes, the most important results are (\ref{eq:formulasphere1}) and (\ref{ortho}) that will be used for calculating the boost Hamiltonian in quasi-de Sitter space.

 The spherical harmonics and their derivatives are normalized as
 \begin{align}
     \int d^3x \sin^2\chi\,\sin\theta\; Y^{k\ell m}\,Y^\ast {}^{k'\ell' m'} = \delta^{k k'}\delta^{\ell\ell'}\delta^{mm'}
 \end{align} and 
 \begin{align}\label{norm}
     \int d^3x \sin^2\chi\,\sin\theta \;\nabla_a Y^{k\ell m}\nabla^a \,Y^\ast {}^{k'\ell' m'} = k(k+2)\delta^{k k'}\delta^{\ell\ell'}\delta^{mm'},
 \end{align} where the complex conjugate is (evaluated at antipodal points)
 \begin{align}
     Y^\ast {}^{k'\ell' m'} = Y^{k'\ell' -m'}.
 \end{align}
 
Explicitly, the spherical harmonics take the form
 \begin{align}
     Y^{k\ell m}=N_{k\ell}\,\sin^\ell\chi \,C^{\ell+1}_{k-\ell}(\cos\chi)\,Y^{\ell m}(\theta,\phi),
 \end{align} where the normalization is
 \begin{align}
     N_{k\ell} = (-1)^k(2\ell)!!\sqrt{\frac{2(k+1)(k-\ell)!}{\pi(k+\ell+1)!}}
 \end{align} and $C^{\ell+1}_{k-\ell}$ are the Gegenbauer polynomials. With this, one can explicitly evaluate the following integrals,
 \begin{align}\label{eq:formulasphere1}
 \int \cos\chi  Y^{k\ell m}\,Y^\ast {}^{k'\ell' m'} &= -\frac{1}{2}\sqrt{\frac{(k_\text{min}+\ell+2)(k_\text{min}-\ell+1)}{(k_\text{min}+1)(k_\text{min}+2)}}\delta^{k\pm 1 k'}\delta^{\ell\ell'}\delta^{mm'},\\
     \int \cos\chi \nabla_a Y^{k\ell m}\nabla^a \,Y^\ast {}^{k'\ell' m'} &= -\frac{k_\text{min}(k_\text{min}{+3})}{2}\sqrt{\frac{(k_\text{min}+\ell+2)(k_\text{min}-\ell+1)}{(k_\text{min}+1)(k_\text{min}+2)}}\delta^{k\pm 1 k'}\delta^{\ell\ell'}\delta^{mm'},\label{ortho}
 \end{align} where $k_\text{min}:=\min\{k,k'\}.$ Notice that the additive constant is ``+3'' instead of ``+2'' as in (\ref{norm}). 

\bibliographystyle{unsrt}
\bibliography{ref}

\end{document}